\documentclass[12pt]{article}
\usepackage[usenames,dvipsnames,svgnames,table]{xcolor} 
\usepackage{kotex}
\usepackage{graphicx}
\usepackage{epsfig}
\usepackage{amssymb, amsmath, mathtools}
\usepackage{verbatim}
\usepackage{natbib}
\usepackage[affil-it]{authblk}
\usepackage{mathrsfs}
\usepackage{here}
\usepackage{makecell}
\usepackage{tikz}
\usepackage{enumerate}
\usepackage[shortlabels]{enumitem}
\usepackage{yfonts}
\usepackage{comment}
\usepackage{longtable}
\usepackage{tikz,amsmath, amssymb,bm,color}
\usetikzlibrary{circuits.logic.US,circuits.logic.IEC,fit}
\usepackage{xr}
\usepackage{setspace}
\usepackage{booktabs,threeparttable,multirow,array,longtable}

\usepackage{algorithm2e}
\usepackage{multirow}
\usepackage{physics}
\usepackage{amsthm}
\usepackage{makecell}
\usepackage{booktabs}
\usepackage{multirow}
\usepackage{subcaption}

\usepackage[normalem]{ulem}

\usepackage[colorlinks]{hyperref} 

\usepackage[framemethod=TikZ]{mdframed}
\mdfdefinestyle{JL}{%
	linecolor=blue,
	outerlinewidth=1pt,
	roundcorner=2pt,
	innertopmargin=0.1\baselineskip,
	innerbottommargin=0.1\baselineskip,
	innerrightmargin=2pt,
	innerleftmargin=2pt,
	backgroundcolor=orange!100!white}

\hypersetup{
	colorlinks,
	citecolor=Black,
	linkcolor=Red,
	urlcolor=Blue}

\includecomment{note} 

\newtheorem{theorem}{Theorem}[section]
\newtheorem{lemma}[theorem]{Lemma}
\newtheorem{definition}[theorem]{Definition}
\newtheorem{proposition}[theorem]{Proposition}
\newtheorem{corollary}[theorem]{Corollary}

\newtheorem{remark}[theorem]{Remark}

\everymath{\displaystyle}

\newcommand{\rmU}{\mathrm{U}}

\newcommand{\be}{\begin{equation}}
	\newcommand{\ee}{\end{equation}}
\newcommand{\bea}{\begin{eqnarray*}}
	\newcommand{\eea}{\end{eqnarray*}}
\newcommand{\bean}{\begin{eqnarray}}
	\newcommand{\eean}{\end{eqnarray}}
\newcommand{\ben}{\begin{enumerate}}
	\newcommand{\een}{\end{enumerate}}
\newcommand{\bi}{\begin{itemize}}
	\newcommand{\ei}{\end{itemize}}
\newcommand{\brem}{\begin{remark}}
	\newcommand{\erem}{\end{remark}}
\newcommand{\bcen}{\begin{center}}
	\newcommand{\ecen}{\end{center}}
\newcommand{\bsv}{\begin{semiverbatim}}
	\newcommand{\esv}{\end{semiverbatim}}

\newcommand{\bt}{\begin{theorem}}
	\newcommand{\et}{\end{theorem}}
\newcommand{\bl}{\begin{lemma}}
	\newcommand{\el}{\end{lemma}}
\newcommand{\bd}{\begin{definition}}
	\newcommand{\ed}{\end{definition}}
\newcommand{\bc}{\begin{corollary}}
	\newcommand{\ec}{\end{corollary}}
\newcommand{\bp}{\begin{proposition}}
	\newcommand{\ep}{\end{proposition}}

\newcommand{\bfY}{ \mathbf{Y}}

\newcommand{\bbR}{ \mathbb{R}}

\usepackage{xr}
\makeatletter
\newcommand*{\addFileDependency}[1]{
	\typeout{(#1)}
	\@addtofilelist{#1}
	\IfFileExists{#1}{}{\typeout{No file #1.}}
}
\makeatother

\usepackage[textwidth=16cm, textheight=22cm, centering]{geometry}

\title{Bayesian Estimation of the Eigenstructure in High-Dimensional Approximate Factor Models}
\author[1]{Seongmin Kim}
\author[2]{Jaeyong Lee}
\affil[1]{Human-Centered Artificial Intelligence Research Institute, Ewha Womans University}
\affil[2]{Department of Statistics, Seoul National University}

\allowdisplaybreaks
\begin{document}
	
\maketitle

\begin{abstract}
High-dimensional economic datasets often display strong co-movement driven by a small number of latent factors, which are typically modeled using approximate factor models. When the number of variables is large relative to the sample size, the eigenvalues and eigenvectors of the sample covariance matrix are severely distorted, which in turn makes principal component based estimators of the factor structure unstable. To address the high-dimensional problem, we propose a Bayesian model for approximate factor structures. We show that the posterior convergence rate is of the same order as benchmark results for high-dimensional spiked covariance models. Simulation studies show that the proposed method more accurately recovers the factor structure in approximate factor models than existing methods. Real data analyses on macro--financial datasets illustrate that the proposed method provides interpretable estimates of latent factor structure and performs competitively in forecasting exercises.
\end{abstract}

\section{Introduction}

High-dimensional economic and financial datasets often exhibit strong co-movement across variables.
A small number of common macroeconomic and market-wide factors—such as business cycle conditions, monetary policy, and aggregate risk—account for a large share of this co-movement \citep{repec:eee:jetheo:v:13:y:1976:i:3:p:341-360,forni2000generalized}. In equity returns, the Fama--French three-factor model \citep{fama1993common} explains a substantial portion of cross-sectional variation through market, size, and value factors.
In many empirical applications, however, the factors are unobserved and the idiosyncratic errors exhibit mild cross-sectional dependence, which motivates the approximate factor model of \cite{chamberlain1983arbitrage}.

The approximate factor model assumes that each observation
$Y_i \in \bbR^p$ can be written as
\[
Y_i = B f_i + \epsilon_i,\quad i=1,\ldots,n,
\]
where $f_i \in \mathbb{R}^k$ denotes the $k$-dimensional vector of latent factors, $B \in \mathbb{R}^{p\times k}$ is the loading matrix, $\epsilon_i \in \bbR^p$ denotes the idiosyncratic error, and $p$ and $n$ are positive integers. We let $Cov(f_i) = I_k$ and $Cov(\epsilon_i) = \Sigma_e$, so that the population covariance of $Y_i$ is
\[
\Sigma_0 = B B^T + \Sigma_e.
\]
A strict factor model assumes that the idiosyncratic errors are cross-sectionally uncorrelated, which is equivalent to requiring that $\Sigma_e$ be diagonal. In contrast, the approximate factor model permits weak cross-sectional dependence in the idiosyncratic errors. The column space of $B$ defines the $k$-dimensional factor subspace spanned by the latent common factors, while the rank-$k$ matrix $B B^T$ captures the corresponding systematic co-movement among the $p$ variables.

A large literature on approximate factor models analyzes principal component based estimators, develops information criteria for determining the number of factors, and studies weak factors and residual cross-sectional dependence \citep{bai2002determining,bai2003inferential,fan2013large,
wang2017asymptotics}. By contrast, Bayesian work that directly targets approximate factor models is relatively limited. Most existing Bayesian factor models are formulated as strict factor models and impose global--local shrinkage or sparsity priors on the loading matrix \citep{lopes2004bayesian,bhattacharya2011sparse}.

In this paper, we propose a Bayesian model for approximate factor structures that works directly with the covariance decomposition
$\Sigma_0 = B B^T + \Sigma_e$.
Writing the factor covariance as
\[
B B^T = \rmU \Lambda \rmU^T,
\]
where $\rmU \in \mathbb{R}^{p\times k}$ has orthonormal columns, we parameterize the factor component through
(i) $\rmU$, which identifies the orientation of the $k$-dimensional factor subspace $\mathrm{span}(\rmU)$, and
(ii) $\Lambda = \mathrm{diag}(\lambda_1,\ldots,\lambda_k)$, which quantifies the strength of each factor direction.
We place priors directly on $(\rmU,\Lambda,\Sigma_e)$ and treat $(\rmU,\Lambda)$ as the primary parameters of interest. This parameterization is rotation-free, in the sense that for any orthogonal matrix $Q \in \mathbb{R}^{k\times k}$, the factor representation
\[
Y_i = Bf_i + \epsilon_i
\]
is equivalent to
\[
Y_i = (BQ)(Q^T f_i) + \epsilon_i,
\]
while the induced pair $(\rmU,\Lambda)$ remains unchanged.
By focusing on the eigenstructure of $B B^T$, the model targets the low-dimensional factor subspace and the magnitudes of the associated spikes, while avoiding the rotational ambiguity of the loading matrix.

The decomposition $\Sigma_0 = B B^T + \Sigma_e$ naturally gives rise to a spiked covariance structure. When the nonzero eigenvalues of $B B^T$ diverge while the eigenvalues of $\Sigma_e$ remain bounded, the leading eigenvalues of $\Sigma_0$ are separated from the remaining noise spectrum. In this regime, the leading eigenspace of $\Sigma_0$ is asymptotically aligned with the factor subspace induced by $B$, so that inference on the leading eigenstructure of $\Sigma_0$ can be interpreted as inference on the common component of the approximate factor model. This places the approximate factor model within the spiked covariance framework of \cite{johnstone2001distribution}. Recent works have established detailed asymptotic results for eigenvalue separation and eigenspace recovery in such spiked covariance models \citep{wang2017asymptotics,cai2020limiting,lee2024bayesian,kim2025eigenstructure}, providing a useful foundation for studying factor recovery under high-dimensional asymptotics. Our Bayesian parameterization $(\rmU,\Lambda,\Sigma_e)$ allows us to bring these spiked covariance insights to the approximate factor model setting.

Our main contributions are threefold.
First, we develop theoretical guarantees for the proposed Bayesian approximate factor model by extending recent results on Bayesian spiked
covariance estimation to the factor-analytic setting. Under eigenvalue separation and weak idiosyncratic dependence, we establish posterior contraction rates for $(\rmU,\Lambda)$ that are of the same order as existing frequentist rates for high-dimensional spiked covariance estimation.
Second, we specify a computationally tractable prior on $(\rmU,\Lambda,\Sigma_e)$ that encodes rotation invariance via a rotation-invariant prior on $\rmU$ and shrinkage on $\Lambda$, and we describe an efficient posterior computation scheme.
Third, simulation studies and real data analyses suggest that our approach yields improved factor recovery, interpretable low-dimensional factor structure, and competitive medium-horizon to long-horizon forecasting performance in factor-augmented vector autoregressions (FAVAR) of \cite{bernanke2005measuring}, relative to principal component analysis and standard approximate factor model benchmarks.

The remainder of the paper is organized as follows. Section 2 introduces the model and the identification target. Section 3 presents the prior specification and posterior computation algorithm, and Section 4 provides the theoretical results. Section 5 reports simulation studies, and Section 6 presents real data analyses on macro--financial data. Section 7 concludes. Proofs of the lemmas and theorems are collected in the Appendix.

\section{Model and Identification Target}\label{sec:pre_afm}
\subsection{Notation}
Let $\mathbb{R}$ denote the set of real numbers. For a positive integer $p$, let $\mathbb{R}^p$ denote the $p$–dimensional Euclidean space.
All vectors are column vectors and all norms on $\mathbb{R}^p$ are Euclidean unless otherwise specified.

For a matrix $A = (a_{ij}) \in \mathbb{R}^{p\times q}$, $\|A\|_F := \Big( \sum_{i=1}^p \sum_{j=1}^q a_{ij}^2 \Big)^{1/2}$ denotes the Frobenius norm, $\|A\|$ denotes the spectral norm (that is, the largest singular value of $A$), and $\operatorname{tr}(A)$ denotes the trace.
We write $\lambda_1(A) \ge \cdots \ge \lambda_p(A)$ for the ordered eigenvalues of a
symmetric $p\times p$ matrix $A$.
For two symmetric matrices $A,B\in\mathbb{R}^{p\times p}$,
\[
A \preccurlyeq B
\quad\text{means}\quad
B-A \text{ is positive semidefinite},
\]
and $A \prec B$ means $B-A$ is positive definite.
We use $I_p$ for the $p\times p$ identity matrix and $0_{p\times q}$ for the $p\times q$ zero matrix.
For $a_1,\ldots,a_k\in\mathbb{R}$, $\mathrm{diag}(a_1,\ldots,a_k)$ denotes the diagonal matrix
with diagonal entries $a_1,\ldots,a_k$.

Let $\mathbb{V}_{p,k}$ denote the Stiefel manifold, i.e. the set of
$p\times k$ matrices with orthonormal columns:
\[
\mathbb{V}_{p,k} := \{\Gamma \in \mathbb{R}^{p\times k} : \Gamma^T \Gamma = I_k\},
\]
and $\mathrm{Unif}(\mathbb{V}_{p,k})$ denotes the uniform distribution on $\mathbb{V}_{p,k}$. The set of $p\times p$ orthogonal matrices is denoted by
\[
O_{p} := \{\Gamma \in \mathbb{R}^{p\times p} : \Gamma^T \Gamma = I_p\}.
\]
We write ${etr}(A) := \exp\{\operatorname{tr}(A)\}$ for a square matrix $A$.

For random variables, $X_1,\ldots,X_n$ are said to be i.i.d.\ if they are independent
and identically distributed. The uniform distribution on a set $A$ is denoted by $Unif(A)$, with density proportional to $1$ on $A$.
We write $X \sim N_p(\mu,\Sigma)$ for a $p$–variate normal distribution with mean $\mu$
and covariance matrix $\Sigma$.
The inverse-gamma distribution with shape $a>0$ and scale $b>0$ is denoted by
$\mathrm{IG}(a,b)$, with density proportional to $x^{-a-1}\exp(-b/x)$ for $x>0$.

For a positive-definite scale matrix $\Psi\in\mathbb{R}^{p\times p}$ and degrees of freedom $\nu>p$, the inverse-Wishart distribution is denoted by
$\mathrm{IW}_p(\nu,\Psi)$, with density proportional to
$|\Sigma|^{-\nu}etr(-\tfrac12\Psi\Sigma^{-1})$
for $\Sigma\succ 0$.
We use $I(\cdot)$ for the indicator function, so that $I(E)=1$ if the event $E$ holds and
$I(E)=0$ otherwise.

For positive sequences $a_n$ and $b_n$, we write $a_n \prec b_n$ if $a_n/b_n \to 0$,
$a_n \succ b_n$ if $b_n \prec a_n$, and $a_n \preceq b_n$ if $a_n/b_n$ is bounded
above by a constant. Similarly, $a_n \succeq b_n$ means $a_n/b_n$ is bounded below
by a positive constant.

\subsection{Approximate Factor Model}
Suppose $Y_1,\ldots,Y_n$ are independent $p$-dimensional random vectors that follow
\begin{align}\label{model:AFM}
    Y_i = B \eta_i + \epsilon_i,\quad \eta_i \overset{iid}{\sim} N_k(0,I_k),\ 
\epsilon_i \overset{iid}{\sim} N_p(0,\Sigma_e),\ 
\eta_i \perp \epsilon_i,
\end{align}
where $B$ is a $p\times k$ loading matrix, and $\Sigma_e$ is a $p\times p$ positive definite matrix. Equivalently, \eqref{model:AFM} can be expressed as
\begin{align*}
    Y_i \overset{iid}{\sim} N(0,BB^T + \Sigma_e),\quad i=1,\ldots,n,
\end{align*}
so that the population covariance matrix can be written as
$$\Sigma_0 = BB^T + \Sigma_e.$$

Consider the spectral decomposition of $BB^T$,
\begin{align*}
    BB^T = \rmU \Lambda \rmU^T,
\end{align*}
where $\rmU$ is a $p\times k$ matrix whose columns are the eigenvectors of $BB^T$, and $\Lambda={diag}(\lambda_1,\ldots,\lambda_k)$ is the diagonal matrix of its eigenvalues with
$\lambda_1 > \cdots > \lambda_k > 0$. The pair $(\rmU,\Lambda)$ determines the factor subspace and the strength of the common components, and will be the main object of inference in our Bayesian procedure.

\subsection{Common Component and Full Covariance Eigenstructure}

Under the approximate factor model,
$$\Sigma_0 = BB^T + \Sigma_e,$$
the rank-$k$ matrix $BB^T$ represents the common component, while $\Sigma_e$ captures the idiosyncratic covariance. Writing
$$BB^T = \rmU \Lambda \rmU^T,$$
the matrix $\rmU \in \mathbb{R}^{p\times k}$ spans the factor subspace and
$\Lambda = {diag}(\lambda_1,\ldots,\lambda_k)$ describes the strength of the common directions.

It is useful to distinguish the eigenstructure of the common component $BB^T$ from that of the full covariance matrix $\Sigma_0$. The pair $(\rmU,\Lambda)$ is the eigenstructure of the common component itself. By contrast, the eigenvalues and eigenvectors of $\Sigma_0$ are affected not only by $BB^T$ but also by the idiosyncratic covariance $\Sigma_e$. In particular, unless $\Sigma_e$ is proportional to the identity matrix or commutes with $BB^T$, the leading eigenvectors of $\Sigma_0$ need not coincide exactly with the columns of $\rmU$ in finite samples.

Nevertheless, under the standard approximate factor regime in which the nonzero eigenvalues of $BB^T$ are sufficiently large while the eigenvalues of $\Sigma_e$ remain bounded, the leading eigenstructure of $\Sigma_0$ is asymptotically driven by the common component. By Weyl's inequality, the leading eigenvalues of $\Sigma_0$ differ from those of $BB^T$ by at most $\|\Sigma_e\|$, and by Davis--Kahan type perturbation arguments, the leading eigenspace of $\Sigma_0$ is close to the column space of $\rmU$ when the spike-noise separation is sufficiently large \citep{horn2012matrix,davis1970rotation}.

For this reason, our Bayesian specification places priors directly on $(\rmU,\Lambda,\Sigma_e)$ and treats $(\rmU,\Lambda)$ as the primary object of inference. This parameterization directly targets the eigenstructure of the common component, while remaining connected to the leading eigenstructure of the full covariance matrix through the spiked covariance representation induced by
$$\Sigma_0 = \rmU \Lambda \rmU^T + \Sigma_e.$$
Accordingly, the theoretical results below are formulated in terms of recovery of the leading eigenstructure under this decomposition.

\section{Prior Specification and Posterior Computation}\label{sec:posterior_computation}
\subsection{Prior Specification}
The full model, placing priors on $\Lambda$, $\rmU$, and $\Sigma_e$, is formulated as follows
\begin{align}\label{model:bayes_AFM}
\begin{split}
    Y_i \vert \rmU,\Lambda,\Sigma_e &\overset{iid}{\sim} N(0,\rmU \Lambda \rmU^T + \Sigma_e)\\
    \rmU &\sim Unif(\mathbb{V}_{p,k})\\
    \lambda_i &\overset{ind}{\sim} IG (a_i-1,h/2),\quad i=1,\ldots,k\\
    \pi(\Sigma_e) & \propto IW(\Sigma_e\vert \nu_0,\Psi) I(b_0 I_p \preccurlyeq \Sigma_e \preccurlyeq b_1 I_p),
\end{split}
\end{align}
where $a_i>1$ for $i=1,\ldots,k$, $h>0$, $\nu_0>p$, and $\Psi = \psi I_p$ with $\psi>0$. These hyperparameters control the strength of shrinkage on the factor eigenvalues and the conditioning of the idiosyncratic covariance. Although the prior above is written with generic $a_1,\ldots,a_k$, in the results below these quantities are taken from the eigenvalues of the sample covariance matrix; see Assumption~(A5) in Section~\ref{sec:theoretical_results}.

The posterior distribution under \eqref{model:bayes_AFM} is given by
\begin{align*}
    &\pi(\rmU,\Lambda,\Sigma_e\vert \bfY)\\
    &\propto \prod_{i=1}^n p(Y_i\vert \rmU,\Lambda,\Sigma_e)\pi(\rmU)\pi(\Lambda)\pi(\Sigma_e)\\
    &\propto \abs{\rmU\Lambda\rmU^T+\Sigma_e}^{-n/2} etr(-\dfrac{n}{2}(\rmU\Lambda\rmU^T+\Sigma_e)^{-1}S)  I(\rmU^T\rmU = I_k)\\
    &\times \prod_{i=1}^k \lambda_i^{-a_i}\exp(-\dfrac{h}{2\lambda_i}) \times IW(\Sigma_e\vert \nu_0,\Psi) I(b_0 I_p \preccurlyeq \Sigma_e \preccurlyeq b_1 I_p),
\end{align*}
where $S = \dfrac{1}{n}\sum_{i=1}^n Y_i Y_i^T$ is the sample covariance.

\subsection{Posterior Sampling}
The sampling procedure is based on the Gibbs sampler. The approximate factor model can be expressed by 
\begin{align*}
    Y_i = \rmU\Lambda^{1/2}\eta_i + \epsilon_i,\quad \eta_i \overset{iid}{\sim} N_k(0,I_k),\ 
\epsilon_i \overset{iid}{\sim} N_p(0,\Sigma_e).
\end{align*}
Consider the following latent variable, $Z_i = \Lambda^{1/2}\eta_i$, then \eqref{model:bayes_AFM} can be written as follows
\begin{align}\label{model:AFM_sampling}
\begin{split}
    Y_i \vert Z_i,\rmU,\Sigma_e &\sim N_p(\rmU Z_i,\Sigma_e)\\
    Z_i \vert \Lambda &\sim N_k(0,\Lambda)\\
    \rmU &\sim Unif(\mathbb{V}_{p,k})\\
    \lambda_i &\overset{ind}{\sim} IG (a_i-1,h/2),\quad i=1,\ldots,k\\
    \pi(\Sigma_e) & \propto IW(\Sigma_e\vert \nu_0,\Psi) I(b_0 I_p \preccurlyeq \Sigma_e \preccurlyeq b_1 I_p).
\end{split}
\end{align}

Let $Z = (Z_1,\ldots,Z_n)\in\bbR^{k\times n}$ and $Y = (Y_1,\ldots,Y_n)\in\bbR^{p\times n}$.
The Gibbs sampling procedure of \eqref{model:AFM_sampling} is given by
\begin{enumerate}
    \item Sampling $Z$
        $$Z_i\vert Y_i,\rmU,\Lambda,\Sigma_e \sim N_k(\mu_i,V),$$
    where $V = (\Lambda^{-1}+\rmU^T\Sigma_e^{-1}\rmU)^{-1}$ and $\mu_i = V\rmU^T\Sigma_e^{-1}Y_i$.
    \item Sampling $\Lambda$
        $$\lambda_j\vert Z,Y,\rmU,\Sigma_e \sim IG(a_j+\dfrac{n}{2}-1,\dfrac{h}{2}+\dfrac{1}{2}\sum_{i=1}^nZ_{ji}^2).$$
    \item Sampling $\rmU$
    \begin{align*}
        [\rmU\vert Y,Z,\Lambda,\Sigma_e]\propto etr(\rmU^TF - \dfrac{1}{2}\rmU^T\Sigma_e^{-1}\rmU S_z),
    \end{align*}
    where $S_z= ZZ^T$ and $F= \Sigma_e^{-1}YZ^T$. This is a matrix Fisher--Bingham (MFB) distribution on the Stiefel manifold. We represent $\rmU$ as the first $k$ columns of a $p\times p$ orthogonal matrix and update it by applying pairwise column rotations, with each rotation angle sampled by univariate slice sampling; see Section~\ref{subsec:gamma_update} for details.

    \item Sampling $\Sigma_e$
    $$[\Sigma_e\vert Z,Y,\rmU,\Lambda]\propto \abs{\Sigma_e}^{-\nu_0-n/2}etr(-\dfrac{1}{2}\Sigma_e^{-1}(S_E+\Psi)) I(b_0 I_p \preccurlyeq \Sigma_e\preccurlyeq b_1 I_p),$$
    where $S_E= (Y-\rmU Z)(Y-\rmU Z)^T$. Direct sampling is difficult due to the eigenvalue constraints. We instead update the eigenvalues and eigenvectors of $\Sigma_e$ separately to efficiently handle the constraints; see Section \ref{subsec:sigmau_update} for details.
\end{enumerate}

The code for the Gibbs sampler is available on \url{https://github.com/zlatjdals/AFM}.

\subsubsection{Sampling $\rmU$}\label{subsec:gamma_update}
Let $\tilde R \in O_p$ be any orthogonal matrix whose first $k$ columns are given by $\rmU$. We utilize the singular value decomposition of the latent factor matrix $Z$. Let $S_z = ZZ^T$ and decompose it as $S_z = U_z D_z U_z^T$, where $D_z = \text{diag}(\delta_1, \ldots, \delta_k)$. We define the augmented matrices:
\begin{equation*}
\tilde F := F U_z,\qquad
\tilde D := \text{diag}(\delta_1,\ldots,\delta_k, 0,\ldots,0),
\end{equation*}
where $F =  \Sigma_e^{-1}Y Z^T$. The full conditional distribution of $\tilde R$ is given by
\begin{equation}
\pi(\tilde R \mid \cdots) \propto {etr}\left( \tilde R^T (\tilde F, 0_{p,p-k}) - \frac{1}{2} \tilde R^T \Sigma_e^{-1} \tilde R \tilde D \right).
\label{eq:R-density}
\end{equation}
We update $\tilde R$ via pairwise updates. We randomly partition the column indices $\{1, \ldots, p\}$ into disjoint pairs. For each pair $(u, v)$ with $u < v$, we perform the following three steps.

\begin{enumerate}[label=\textbf{Step \Roman*.}]
    \item \textbf{Rotation transformation via Signed Givens matrix.}\\
    Let $\tilde r_j$ denote the $j$th column of $\tilde R$, for $j=1,\ldots,p$. We update the pair of columns $(\tilde r_u, \tilde r_v)$ by applying a signed rotation. The new columns $(\tilde r_u', \tilde r_v')$ are obtained by post-multiplying the rotation matrix $G(\theta, \varepsilon)$:
    \begin{equation*}
    (\tilde r_u', \tilde r_v') = (\tilde r_u, \tilde r_v) R_\theta 
    \begin{pmatrix}
    \varepsilon & 0 \\
    0 & 1
    \end{pmatrix}
    = (\tilde r_u, \tilde r_v)
    \begin{pmatrix}
    \varepsilon \cos\theta & -\sin\theta \\
    \varepsilon \sin\theta & \cos\theta
    \end{pmatrix},
    \end{equation*}
    where $\theta \in (-\pi, \pi]$ is the rotation angle and $\varepsilon \in \{-1, 1\}$ represents the sign flip. The explicit update rule is:
    \begin{equation}
    \begin{aligned}
    \tilde r_u' &= \varepsilon (\cos\theta \, \tilde r_u + \sin\theta \, \tilde r_v), \\
    \tilde r_v' &= -\sin\theta \, \tilde r_u + \cos\theta \, \tilde r_v.
    \end{aligned}
    \label{eq:col_update}
    \end{equation}
    
   \item \textbf{Conditional distribution of $\theta$ and slice sampling.}\\
    Let
    \[
    \bar F := (\tilde F, 0_{p,p-k}),
    \]
    and write $\bar f_j$ for the $j$th column of $\bar F$. For $i,j \in \{u,v\}$, define
    \[
    B_{ij} := \tilde r_i^T \bar f_j,
    \qquad
    t_{ij} := \tilde r_i^T \Sigma_e^{-1} \tilde r_j.
    \]
    Then, after substituting \eqref{eq:col_update} into the log-density \eqref{eq:R-density}, the conditional log-density of $(\theta,\varepsilon)$ can be written as
    \[
    \ell(\theta,\varepsilon) \propto d_L(\theta,\varepsilon) - \frac12 d_K(\theta),
    \]
    where
    \begin{align*}
    d_L(\theta, \varepsilon)
    &=
    \cos\theta \, (\varepsilon B_{uu} + B_{vv})
    +
    \sin\theta \, (\varepsilon B_{vu} - B_{uv}), \\
    d_K(\theta)
    &=
    (\cos^2\theta \, t_{uu} + 2\sin\theta\cos\theta \, t_{uv} + \sin^2\theta \, t_{vv}) \tilde D_{uu} \\
    &\quad +
    (\sin^2\theta \, t_{uu} - 2\sin\theta\cos\theta \, t_{uv} + \cos^2\theta \, t_{vv}) \tilde D_{vv}.
    \end{align*}
    Integrating out $\varepsilon \in \{-1,1\}$ gives the marginal conditional density
    \[
    \pi(\theta \mid \cdots)
    \propto
    \exp\{\ell(\theta,1)\} + \exp\{\ell(\theta,-1)\},
    \qquad \theta \in (-\pi,\pi].
    \]
    Since this density does not belong to a standard parametric family, we update $\theta$ by univariate slice sampling \citep{neal2003slice}. 

    \item \textbf{Conditional distribution of $\varepsilon$ and Bernoulli Sampling.}\\
    Given $\theta$, the conditional probability of the sign $\varepsilon$ is:
    \begin{equation*}
    \pi(\varepsilon \mid \theta, \cdots) = \frac{\exp\{d_L(\theta, \varepsilon)\}}{\exp\{d_L(\theta, +1)\} + \exp\{d_L(\theta, -1)\}}, \quad \varepsilon \in \{-1, 1\}.
    \end{equation*}
    We draw $\varepsilon$ from this Bernoulli distribution and apply the update \eqref{eq:col_update}.
\end{enumerate}

\subsubsection{Sampling $\Sigma_e$}\label{subsec:sigmau_update}
Direct sampling from the truncated inverse-Wishart distribution via rejection sampling can be computationally inefficient, especially when the bounds $b_0$ and $b_1$ are tight or the dimension $p$ is large. To address this, we consider a reparameterization based on the spectral decomposition of $\Sigma_e$.

Let $\Sigma_e = P D P^T$ be the eigendecomposition of $\Sigma_e$, where $D = \mathrm{diag}(d_1, \ldots, d_p)$ is the diagonal matrix of eigenvalues and $P \in O_p$ is the orthogonal matrix of eigenvectors. The Jacobian of the transformation from $\Sigma_e$ to $(D, P)$ is given by
\[
(d\Sigma_e) = \prod_{1 \le i < j \le p} |d_i - d_j| (dD)(dP),
\]
where $(dP)$ represents the Haar measure on the orthogonal group $O_p$.
Substituting $\Sigma_e = P D P^T$ into the conditional posterior of $\Sigma_e$ derived in Section~\ref{sec:posterior_computation}, the joint posterior distribution of $(D, P)$ is proportional to:
\begin{align*}
    &\pi(D, P \mid Z, Y, \rmU, \Lambda) (dD)(dP) \\
    &\propto \left( \prod_{i<j} |d_i - d_j| \right) |D|^{-\nu_0 - n/2}  \times {etr}\left( -\frac{1}{2} P D^{-1} P^T (S_E + \Psi) \right) I(b_0 I_p \preccurlyeq D \preccurlyeq b_1 I_p) (dD)(dP),
\end{align*}
where $S_E = (Y - \rmU Z)(Y - \rmU Z)^T$ is the residual sum of squares matrix. Note that the constraint $b_0 I_p \preccurlyeq \Sigma_e \preccurlyeq b_1 I_p$ translates directly to the constraints on the eigenvalues: $b_0 \le d_i \le b_1$ for all $i=1,\ldots,p$.

Based on this factorization, we can construct a Gibbs sampler that alternates between updating the eigenvalues $D$ and the eigenvectors $P$:

\begin{enumerate}
    \item \textbf{Sampling Eigenvalues $D$ given $P$:}
    We update the eigenvalues $d_1, \ldots, d_p$ sequentially. Let $D_{-i} = \{d_j : j \neq i\}$ denote the set of eigenvalues excluding $d_i$. The full conditional density of $d_i$ is given by
    \begin{align*}
        [d_i \mid D_{-i}, P, \cdots] &\propto d_i^{-\nu_0 - n/2} \exp\left( -\frac{1}{2d_i} (P^T (S_E + \Psi) P)_{ii} \right) \nonumber \\
        &\quad \times \prod_{j \neq i} |d_i - d_j| \times I(b_0 \le d_i \le b_1).
    \end{align*}
    The first line corresponds to the kernel of an inverse-gamma distribution, while the second line includes the Jacobian term acting as a repulsive force between eigenvalues and the truncation constraints. Since this non-standard density is unimodal or locally well-behaved within the intervals defined by neighbors, we employ univariate slice sampling (\citealt{neal2003slice}) to update each $d_i$.

    \item \textbf{Sampling Eigenvectors $P$ given $D$:}
    The conditional density for the orthogonal matrix $P$ depends on the trace term involving $D^{-1}$:
    \begin{align*}
        [P \mid D, \cdots] &\propto {etr}\left( -\frac{1}{2} D^{-1} P^T (S_E + \Psi) P \right).
    \end{align*}
    This form corresponds to a matrix Bingham distribution (or a variant of the matrix Fisher-Bingham distribution with a zero linear term) on the manifold $O_p$. Similar to the update for $\rmU$ described in Section \ref{subsec:gamma_update}, we can sample $P$ by updating pairs of columns via random Givens rotations. The conditional distribution for the rotation parameters follows a structure analogous to the one derived for $\rmU$, simplified by the absence of the linear term.
\end{enumerate}

\section{Theoretical Results}\label{sec:theoretical_results}
In this section, we derive posterior convergence rates for $(\rmU,\Lambda)$ under a
spiked covariance structure on the true covariance matrix. Let $\Sigma_0$ denote the
covariance of $Y_i$, and let
$$
\lambda_{0,1} > \cdots > \lambda_{0,p},
$$
be the ordered eigenvalues of $\Sigma_0$. To demonstrate the posterior convergence rate, we consider the reparametrization of $\mathrm{U}$ and $\Sigma_e$ to diagonalize the sample covariance $S$. Consider the spectral decomposition of $nS=QWQ^T$ where $W=diag(n\hat{\lambda}_1,\ldots,n\hat{\lambda}_n,0,\ldots,0)$ and $Q$ is eigenvector matrix whose $i$th column is the eigenvector corresponding to the $i$th eigenvalue of $S$ which is denoted by $\hat{\lambda}_i$. Define $\Gamma = Q^T\mathrm{U}$ and $\Sigma_u = Q^T\Sigma_e Q$. The reparameterization aligns the basis with the sample eigenvectors, which facilitates the derivation of posterior contraction rates. Under the transformation, the posterior can be rewritten as
\begin{align}\label{eq:post_U_Lambda_Sigma}
\begin{split}
        &\pi(\Gamma,\Lambda,\Sigma_u\vert \bfY)\\
    &\propto \abs{\Gamma\Lambda{\Gamma}^T+\Sigma_u}^{-n/2} etr(-\dfrac{1}{2}(\Gamma\Lambda{\Gamma}^T+\Sigma_u)^{-1}W) I(\Gamma^T\Gamma= I_k)\\
    &\times \prod_{i=1}^k \lambda_i^{-a_i}\exp(-\dfrac{h}{2\lambda_i}) \times IW(\Sigma_u\vert \nu_0,\Psi) I(b_0 I_p \preccurlyeq \Sigma_u \preccurlyeq b_1 I_p).
\end{split}
\end{align}

Since the prior on $\Sigma_e$ is orthogonally invariant, i.e. $\pi(\Sigma_e) = \pi(P \Sigma_e P^T)$ for all orthogonal matrices $P$, and both $\Psi = \psi I_p$ and the constraint set $\{\Sigma_e: b_0 I_p \preccurlyeq \Sigma_e \preccurlyeq b_1 I_p\}$ are also orthogonally invariant, the prior has
the same functional form in the rotated coordinates $(\Gamma,\Lambda,\Sigma_u)$.

Following \citet{wang2017asymptotics} and \cite{kim2025eigenstructure}, we
impose the following high-dimensional conditions.

\begin{enumerate}[label={A\arabic*.}]
    \item High-dimensional regime: $p/n \to \infty$.
    \item For positive constants $c_0$ and $C_0$, $\lambda_{0,1},\ldots,\lambda_{0,p}$ satisfy the following inequality:
$$\lambda_{0,1}>\cdots>\lambda_{0,k}>C_0>\lambda_{0,k+1}>\cdots>\lambda_{0,p}>c_0.$$
    \item The $k$ spiked eigenvalues are separated by a constant $\delta_0>0$:
$$\dfrac{\lambda_{0,j}-\lambda_{0,j+1}}{\lambda_{0,j}}\geq\delta_0,\quad\forall j=1,\cdots ,k.$$
    \item The values $d_j=\dfrac{p}{n \lambda_{0,j}}$ are bounded above by a positive constant.
    \item Let $\hat{\lambda}_1 \ge \cdots \ge \hat{\lambda}_n>0$ denote the eigenvalues of the sample covariance matrix $S$. For $j = 1,\ldots,k$, the prior hyperparameters $a_j$ are given by
    $$
    a_j = \frac{n t}{2(\hat{\lambda}_j - t)} + 2,
    $$
    for some $t \in [\hat{\lambda}_{k+1}, \hat{\lambda}_n]$.
\end{enumerate}
The assumptions (A1)–(A4) ensure that the leading eigenvalues are separated from
the noise bulk and that the idiosyncratic covariance is well-conditioned.
Assumption (A5) makes explicit the choice of $a_1,\ldots,a_k$ mentioned in Section~\ref{sec:posterior_computation}, where these quantities are taken from the eigenvalues of the sample covariance matrix. We use this form in the arguments below, following \citet{kim2025eigenstructure}.

Recall the reparameterized parameter $\Gamma = Q^T \rmU$, which aligns the true loading matrix with the sample eigenvectors. Define the following subset of the Stiefel manifold:
\[
A_\epsilon = \Big\{ \Gamma\in \mathbb{V}_{p,k} : \big\|\Gamma -
\begin{pmatrix}
I_k\\ 0
\end{pmatrix}
\big\|_F < \epsilon \Big\}.
\]
Intuitively, $A_\epsilon$ collects all loading directions whose column space is within Frobenius distance $\epsilon$ of the canonical $k$-dimensional subspace.

\begin{lemma}\label{lem:shrink_post}
    Under model \eqref{model:AFM} and assumptions (A1)--(A4), suppose that the ratio $\lambda_{0,1}/\lambda_{0,k}$ is bounded by a positive constant. For any positive $\epsilon$ and $\epsilon_1$ satisfying $\sqrt{\frac{p}{n\lambda_{0,k}}} \prec \epsilon_1 < \epsilon$, we have
    \begin{align*}
        \pi(\Gamma\in A_\epsilon^c \mid \mathbf{Y})
        &\leq \exp\left(- \frac{n\epsilon_1^2\hat{\lambda}_k}{16b_1}\right) \\
        &\quad + \big[1+O(M_k)+ O(N)\big] \cdot \exp\left(-\frac{\eta }{4\sqrt{k}}\min_{l<k}(a_{l+1} - a_l) \right),
    \end{align*}
    where $n \prec \delta_n \prec \hat{\lambda}_k$, the parameter $\eta$ satisfies $\eta \min_{l<k}(a_{l+1} - a_l) \succ 1$, and
    \begin{align*}
        M_k &= O\left(\max\left\{\frac{a_k}{\delta_n},\frac{n\hat{\lambda}_1}{\delta_n^2}\right\} + (\hat{\lambda}_1)^{n} \exp\left(-\frac{n\hat{\lambda}_k}{6\delta_n}\right)\right), \\
        N &= O\left(\frac{n}{\delta_n}\right).
    \end{align*}
\end{lemma}

The proof relies on the covering number of the Stiefel manifold $\mathbb{V}_{p,k}$
and a comparison of the marginal posterior probability on $A_\epsilon$ and its
complement. In particular, the bound shows that the posterior probability of leaving the $\epsilon$-neighborhood $A_\epsilon$ decays exponentially fast in $n$, up to
factors $M_k$ and $N$. Full details are provided in the Appendix.

Here and below, \(\lambda_{(1)} \ge \cdots \ge \lambda_{(k)}\) denote the ordered values of \(\lambda_1,\ldots,\lambda_k\).

\begin{theorem}\label{thm:post_eigvals}
Assume that conditions (A1)--(A5) hold. Suppose that $\psi \succcurlyeq \sqrt{np}$, $\epsilon \prec n^{-1/2}$, and the ratio $\lambda_{0,1}/\lambda_{0,k}$ is bounded above by a positive constant. Additionally, assume that  $\max(\lambda_{0,k},p) \succ n^{3/2}$.
Then, for $i=1,\ldots,k$,
\[
    \mathbb{E}\Big[\frac{\lambda_{(i)} - \lambda_{0,i}}{\lambda_{0,i}} \,\Big\vert\, \bfY \Big]
    = O\Big(\lambda_{0,i}^{-1}\sqrt{\frac{p}{n}}\Big) + O(\beta_i),
\]
where $\beta_i \lesssim n^{-1/2+\delta}$ for any sufficiently small $\delta>0$.
\end{theorem}

Theorem~\ref{thm:post_eigvals} demonstrates that the posterior expectation of the $i$th spiked eigenvalue converges to the population spiked eigenvalue $\lambda_{0,i}$. The convergence rate is dominated by the term $\lambda_{0,i}^{-1}\sqrt{p/n}$. This rate is of the same order as the eigenvalue estimation rates established for high-dimensional spiked covariance models (e.g., \citealp{wang2017asymptotics,kim2025eigenstructure}).

For \(j=1,\ldots,k\), let \(\xi_{0,j}\) denote the population eigenvector associated with \(\lambda_{0,j}\), and let \(\xi_{(j)}\) denote the posterior eigenvector associated with the ordered eigenvalue \(\lambda_{(j)}\).

\begin{theorem}\label{thm:post_eigvecs}
Suppose that the assumptions of Theorem \ref{thm:post_eigvals} hold. For each $1\le j\le k$,
let $\xi_{0,j}$ be the $j$th population eigenvector and
$\xi_{(j)}$ the $j$th posterior eigenvector associated with $\lambda_{(j)}$.
Then
\[
\mathbb{E}\Big[1 - |\xi_{0,j}^T \xi_{(j)}|^2 \,\Big|\, \bfY\Big]
= O\Big(\frac{p}{n\lambda_{0,j}}\Big)
+ O_p(\zeta_j),
\]
where $\zeta_j = \lambda_{0,j}^{-1}\sqrt{p/n} + p/(n^{3/2}\lambda_{0,j}) + n^{-1}$.
\end{theorem}

The quantity $1 - |\xi_{0,j}^T \xi_{(j)}|^2$ is the squared sine of the angle between
the population eigenvector $\xi_{0,j}$ and its posterior counterpart $\xi_{(j)}$. 
Theorem~\ref{thm:post_eigvecs} establishes that the posterior distribution of eigenvectors concentrate around the true factor directions at the rate $p/(n\lambda_{0,j})$, up to smaller-order terms $\zeta_j$. Notably, this convergence rate matches the order of the estimation error for the sample eigenvectors derived from the sample covariance matrix \citep{wang2017asymptotics}.

\section{Simulation Studies}

We evaluate the performance of the proposed method using simulated data. Let $n \in \{30,40,50\}$ and $p \in \{300,500\}$. We control the signal strength by setting $d_j = p/(n \lambda_j)$ with $d_1 = 0.2$, $d_2 = 0.5$, and $d_3 = 1$. The observations are generated from the factor model:
\[
Y_i = B f_i + \epsilon_i,\qquad i=1,\ldots,n,
\]
where $B$ is a $p \times 3$ loading matrix. Each dataset is generated independently, and for each $(n, p)$ combination, we conduct 100 replications.

The data-generating process is as follows:
\begin{enumerate}
    \item The entries of the loading matrix are drawn as $B_{lj} \overset{iid}{\sim} N(0,1)$ for $l=1,\ldots,p$ and $j=1,\ldots,3$. Each column is then normalized to have a Euclidean norm of $\sqrt{\lambda_j}$.
    \item The latent factors are generated as $f_i \overset{iid}{\sim} N(0,I_3)$ for $i=1,\ldots,n$.
    \item The idiosyncratic errors are generated as $\epsilon_i \sim N(0, \Sigma_e)$, where $\Sigma_e = Q \mathrm{diag}(\sigma_1^2, \ldots, \sigma_p^2) Q^T$. The diagonal entries $\sigma_l^2$ are drawn independently from a $\mathrm{Gamma}(100, 100)$ distribution, and $Q$ is an orthogonal matrix drawn from the Haar measure.
\end{enumerate}
The resulting population covariance is $\Sigma_0 = B B^T + \Sigma_e$.

We compare the proposed Approximate Factor Model (AFM) with four alternatives: Penalized Maximum Likelihood (PML; \citealt{bai2016efficient}), generalized Shrinkage Inverse Wishart prior (gSIW; \citealt{kim2025eigenstructure}), Shrinkage Principal Orthogonal Complement Thresholding (S-POET; \citealt{wang2017asymptotics}), and the sample covariance matrix.

To evaluate the accuracy of the covariance estimation, we compute four metrics:
\begin{itemize}
    \item Relative spectral norm (RS):
    \[
     \bigl\|\Sigma_0^{-1/2} (\hat{\Sigma} - \Sigma_0) \Sigma_0^{-1/2}\bigr\|
    \]
    \item Relative Frobenius norm (RF):
    \[
     \dfrac{1}{\sqrt{p}}\bigl\|\Sigma_0^{-1/2} (\hat{\Sigma} - \Sigma_0) \Sigma_0^{-1/2}\bigr\|_F
    \]
    \item Spectral norm (Spec): $\|\hat{\Sigma} - \Sigma_0\|$
    \item Frobenius norm (Frob) :
    $\|\hat{\Sigma} - \Sigma_0\|_F$
\end{itemize}

\begin{table}[ht]
\centering
\scriptsize
\setlength{\tabcolsep}{4pt}
\renewcommand{\arraystretch}{1.15}
\begin{tabular}{cc lcccc}
\toprule
\textbf{p} & \textbf{n} & \textbf{Method} & \textbf{RS} & \textbf{RF} & \textbf{Spec} & \textbf{Frob}\\
\midrule
\multirow{15}{*}{300}
 & \multirow{5}{*}{30}
  & AFM    & \textbf{7.12(0.595)} & \textbf{0.593(0.0526)} & 28.7(4.70) & \textbf{36.8(3.03)} \\
  & & PML    & 13.3(0.900) & 1.22(0.0513) & 33.1(7.04) & 49.0(4.51) \\
  & & gSIW   & {8.33(0.782)} & 1.18(0.0345) & \textbf{27.2(4.12)} & 41.8(2.34) \\
  & & SPOET  & 9.60(0.738) & {0.859(0.0473)} & {27.5(4.71)} & {39.8(3.02)} \\
  & & Sample & 15.7(0.620) & 3.18(0.0457) & 31.8(6.24) & 69.9(2.94) \\
\cmidrule(lr){2-7}
 & \multirow{5}{*}{40}
  & AFM    & \textbf{5.59(0.460)} & \textbf{0.469(0.0322)} & 20.9(3.10) & \textbf{27.2(1.91)} \\
  & & PML    & 10.4(0.630) & 0.967(0.0401) & 24.6(4.53) & 36.8(2.72) \\
  & & gSIW   & {6.39(0.558)} & 1.07(0.0185) & \textbf{19.8(2.82)} & 32.8(1.43) \\
  & & SPOET  & 7.58(0.539) & {0.700(0.0322)} & {20.1(2.84)} & {29.9(1.81)} \\
  & & Sample & 12.5(0.419) & 2.75(0.0326) & 23.7(4.06) & 57.4(1.72) \\
\cmidrule(lr){2-7}
 & \multirow{5}{*}{50}
  & AFM    & \textbf{4.64(0.393)} & \textbf{0.398(0.0254)} & 16.6(2.39) & \textbf{21.7(1.41)} \\
  & & PML    & 8.66(0.525) & 0.806(0.0312) & 19.4(3.50) & 29.5(2.07) \\
  & & gSIW   & {5.15(0.479)} & 1.01(0.0131) & \textbf{15.7(2.26)} & 27.7(0.986) \\
  & & SPOET  & 6.35(0.460) & {0.599(0.0260)} & {16.0(2.29)} & {24.2(1.42)} \\
  & & Sample & 10.5(0.330) & 2.46(0.0311) & 18.8(3.21) & 49.7(1.29) \\
\bottomrule
\end{tabular}
\caption{Summary of results for $p=300$ with $n\in\{30,40,50\}$. Each entry reports the mean with the standard deviation in parentheses. For each metric column, the best method is shown in bold.}
\label{tbl:method_summary_p300}
\end{table}

\begin{table}[ht]
\centering
\scriptsize
\setlength{\tabcolsep}{4pt}
\renewcommand{\arraystretch}{1.15}
\begin{tabular}{cc lcccc}
\toprule
\textbf{p} & \textbf{n} & \textbf{Method} & \textbf{RS} & \textbf{RF} & \textbf{Spec} & \textbf{Frob}\\
\midrule
\multirow{15}{*}{500}
 & \multirow{5}{*}{30}
  & AFM    & \textbf{9.81(0.751)} & \textbf{0.672(0.0475)} & 52.6(6.81) & \textbf{63.0(4.83)} \\
  & & PML    & 35.1(97.1) & 2.21(5.08) & 97.0(190.0) & 122.0(195.0) \\
  & & gSIW   & {13.7(1.18)} & 1.32(0.0356) & \textbf{44.9(9.81)} & 67.5(6.19) \\
  & & SPOET  & 15.2(1.11) & {1.02(0.0478)} & {46.0(10.2)} & {65.5(7.14)} \\
  & & Sample & 24.1(0.634) & 4.09(0.0513) & 54.3(12.8) & 117.0(6.46) \\
\cmidrule(lr){2-7}
 & \multirow{5}{*}{40}
  & AFM    & \textbf{7.54(0.564)} & \textbf{0.519(0.0351)} & 39.2(4.70) & \textbf{47.1(3.41)} \\
  & & PML    & 15.8(0.714) & 1.15(0.0382) & 42.0(9.47) & 61.1(6.45) \\
  & & gSIW   & {10.3(0.836)} & 1.17(0.0259) & \textbf{32.6(5.54)} & 51.7(3.12) \\
  & & SPOET  & 11.6(0.730) & {0.800(0.0381)} & {32.9(6.12)} & {48.5(3.82)} \\
  & & Sample & 19.0(0.549) & 3.54(0.0377) & 39.2(7.79) & 95.4(3.42) \\
\cmidrule(lr){2-7}
 & \multirow{5}{*}{50}
  & AFM    & \textbf{6.15(0.417)} & \textbf{0.432(0.0253)} & 30.8(4.18) & \textbf{37.4(2.99)} \\
  & & PML    & 12.9(0.652) & 0.943(0.0326) & 33.7(7.58) & 49.0(4.70) \\
  & & gSIW   & {8.32(0.708)} & 1.09(0.0161) & \textbf{26.3(4.47)} & 43.3(2.14) \\
  & & SPOET  & 9.54(0.637) & {0.673(0.0278)} & {26.7(5.16)} & {39.1(2.91)} \\
  & & Sample & 15.9(0.456) & 3.16(0.0348) & 31.7(6.60) & 82.6(2.46) \\
\bottomrule
\end{tabular}
\caption{Summary of results for $p=500$ with $n\in\{30,40,50\}$. Each entry reports the mean with the standard deviation in parentheses. For each metric column, the best method is shown in bold.}
\label{tbl:method_summary_p500}
\end{table}

Tables~\ref{tbl:method_summary_p300} and \ref{tbl:method_summary_p500} summarize the estimation results for the overall covariance matrix. Across all settings, AFM consistently attains the lowest RS, RF, and Frobenius norm. Because RS and RF measure scale-adjusted estimation error, these results indicate that AFM accurately recovers the covariance structure in a global sense, effectively balancing the estimation of both the dominant spikes and the idiosyncratic component. In particular, its strong performance in Frobenius norm suggests that AFM achieves the most accurate overall reconstruction of the covariance matrix.

\begin{table}[ht]
\centering
\scriptsize
\setlength{\tabcolsep}{6pt}
\renewcommand{\arraystretch}{1.15}
\begin{tabular}{cc lccc}
\toprule
\textbf{p} & \textbf{n} & \textbf{Method} & \textbf{Spec} & \textbf{Frob} & \textbf{Max}\\
\midrule
\multirow{9}{*}{300}
 & \multirow{3}{*}{30}
  & AFM    & \textbf{0.393(0.0762)} & \textbf{2.01(0.0701)} & \textbf{0.0336(0.00411)} \\
  & & PML    & 0.847(0.117) & 5.16(0.157) & 0.748(0.0961) \\
  & & SPOET  & {0.768(0.0968)} & {5.12(0.139)} & {0.747(0.104)} \\
\cmidrule(lr){2-6}
 & \multirow{3}{*}{40}
  & AFM    & \textbf{0.394(0.0668)} & \textbf{2.09(0.0736)} & \textbf{0.0342(0.00376)} \\
  & & PML    & 0.718(0.107) & {4.40(0.159)} & 0.692(0.107) \\
  & & SPOET  & {0.681(0.0802)} & 4.46(0.149) & {0.655(0.0888)} \\
\cmidrule(lr){2-6}
 & \multirow{3}{*}{50}
  & AFM    & \textbf{0.402(0.0634)} & \textbf{2.17(0.0821)} & \textbf{0.0365(0.00466)} \\
  & & PML    & 0.629(0.0781) & {3.98(0.119)} & 0.605(0.0810) \\
  & & SPOET  & {0.607(0.0550)} & 4.03(0.120) & {0.574(0.0642)} \\
\bottomrule
\end{tabular}
\caption{Estimation error of the error covariance matrix for $p=300$. Results are reported as mean (standard deviation) over 100 replications. Best results are in bold.}
\label{tbl:cov_error_p300}
\end{table}

\begin{table}[ht]
\centering
\scriptsize
\setlength{\tabcolsep}{6pt}
\renewcommand{\arraystretch}{1.15}
\begin{tabular}{cc lccc}
\toprule
\textbf{p} & \textbf{n} & \textbf{Method} & \textbf{Spec} & \textbf{Frob} & \textbf{Max}\\
\midrule
\multirow{9}{*}{500}
 & \multirow{3}{*}{30}
  & AFM    & \textbf{0.618(0.136)} & \textbf{2.53(0.0934)} & \textbf{0.0273(0.00454)} \\
  & & PML    & 12.0(95.6) & 19.8(114.0) & 6.40(41.4) \\
  & & SPOET  & {0.826(0.110)} & {6.58(0.158)} & {0.810(0.115)} \\
\cmidrule(lr){2-6}
 & \multirow{3}{*}{40}
  & AFM    & \textbf{0.613(0.127)} & \textbf{2.59(0.101)} & \textbf{0.0280(0.00417)} \\
  & & PML    & 0.758(0.122) & {5.67(0.138)} & 0.717(0.120) \\
  & & SPOET  & {0.715(0.101)} & 5.76(0.139) & {0.694(0.107)} \\
\cmidrule(lr){2-6}
 & \multirow{3}{*}{50}
  & AFM    & \textbf{0.625(0.135)} & \textbf{2.67(0.111)} & \textbf{0.0290(0.00400)} \\
  & & PML    & 0.679(0.0877) & {5.13(0.126)} & 0.657(0.0929) \\
  & & SPOET  & {0.648(0.0806)} & 5.21(0.126) & {0.625(0.0881)} \\
\bottomrule
\end{tabular}
\caption{Estimation error of the error covariance matrix for $p=500$. Results are reported as mean (standard deviation) over 100 replications. Best results are in bold.}
\label{tbl:cov_error_p500}
\end{table}

Regarding estimation of the idiosyncratic covariance matrix, Tables~\ref{tbl:cov_error_p300} and \ref{tbl:cov_error_p500} report the estimation errors of $\Sigma_e$ under the Spec, Frob, and Max metrics. Since the sample covariance estimator and gSIW are designed for estimation of the full covariance matrix rather than the idiosyncratic covariance component, they are not included in this comparison. Among the methods that explicitly estimate $\Sigma_e$, AFM consistently outperforms the competing approaches across all settings and remains stable even in the challenging case $(p,n)=(500,30)$, where PML exhibits substantial variability.

Taken together, the simulation results show that AFM performs strongly in both full covariance estimation and idiosyncratic covariance estimation. In particular, its favorable performance under the relative spectral and relative Frobenius norms indicates accurate recovery of the main covariance structure in a scale-adjusted sense, while the error covariance results show that it also isolates the residual idiosyncratic component more effectively than competing methods. Overall, these findings suggest that AFM provides a reliable approach to recovering both the common factor structure and the remaining error covariance in high-dimensional approximate factor models.

\section{Real Data Analyses}
\subsection{S\&P 500 Returns}

We applied the proposed Approximate Factor Model (AFM) to the monthly log returns of S\&P 500 constituents from January 2015 to December 2023. The dataset comprises $p=462$ companies over a period of $n=108$ months. We examined the estimated factor loading matrices across different Global Industry Classification Standard (GICS) sectors. The number of factors, $k$, was determined using the $IC_2$ criterion proposed by \cite{bai2002determining}. The selected number of factors was $k=4$.

\begin{figure}[htbp]
    \centering
    \includegraphics[width=0.95\linewidth]{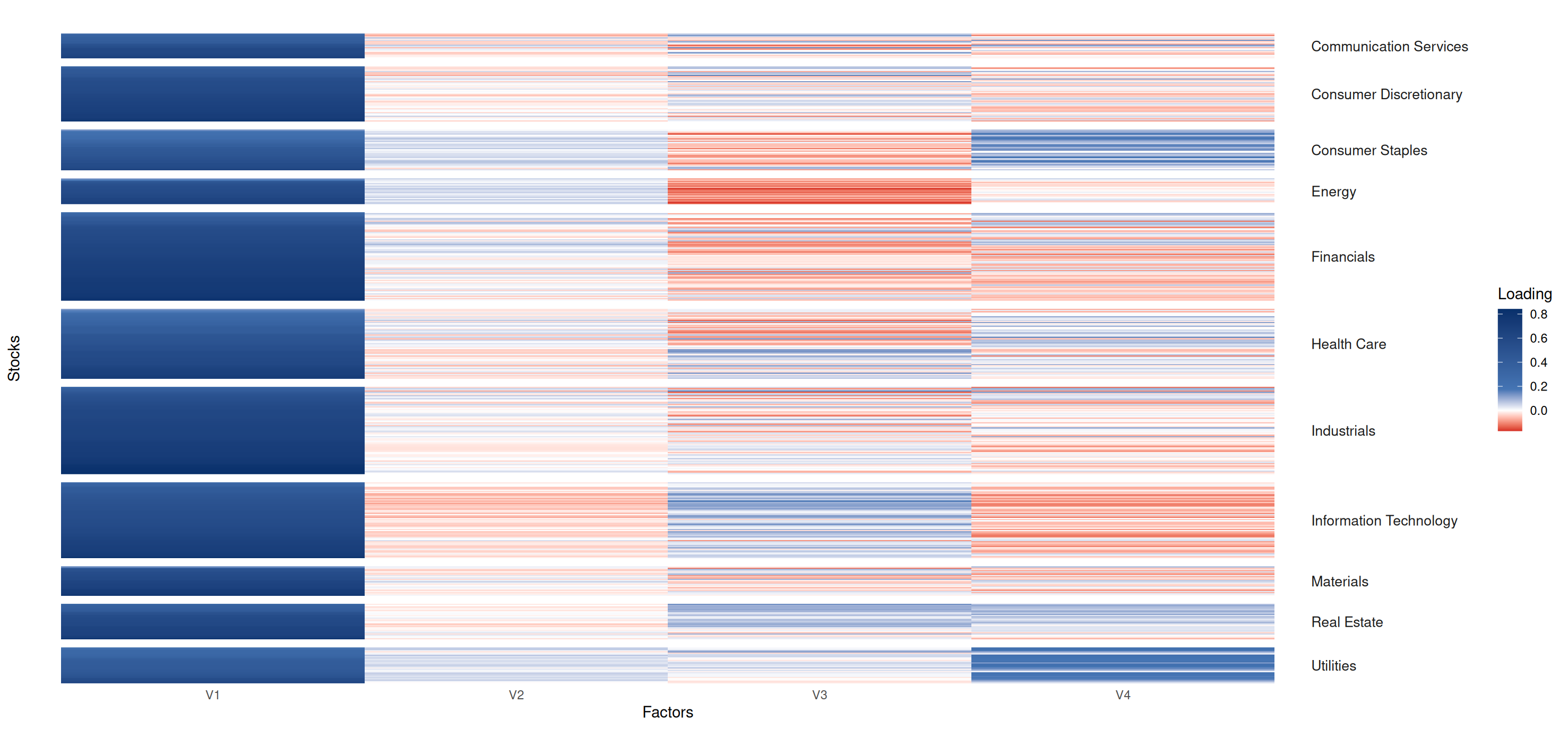}
    \caption{Factor loading matrix estimated by the proposed AFM.}
    \label{fig:SNP_AFM}
\end{figure}

\begin{figure}[htbp]
    \centering
    \includegraphics[width=0.95\linewidth]{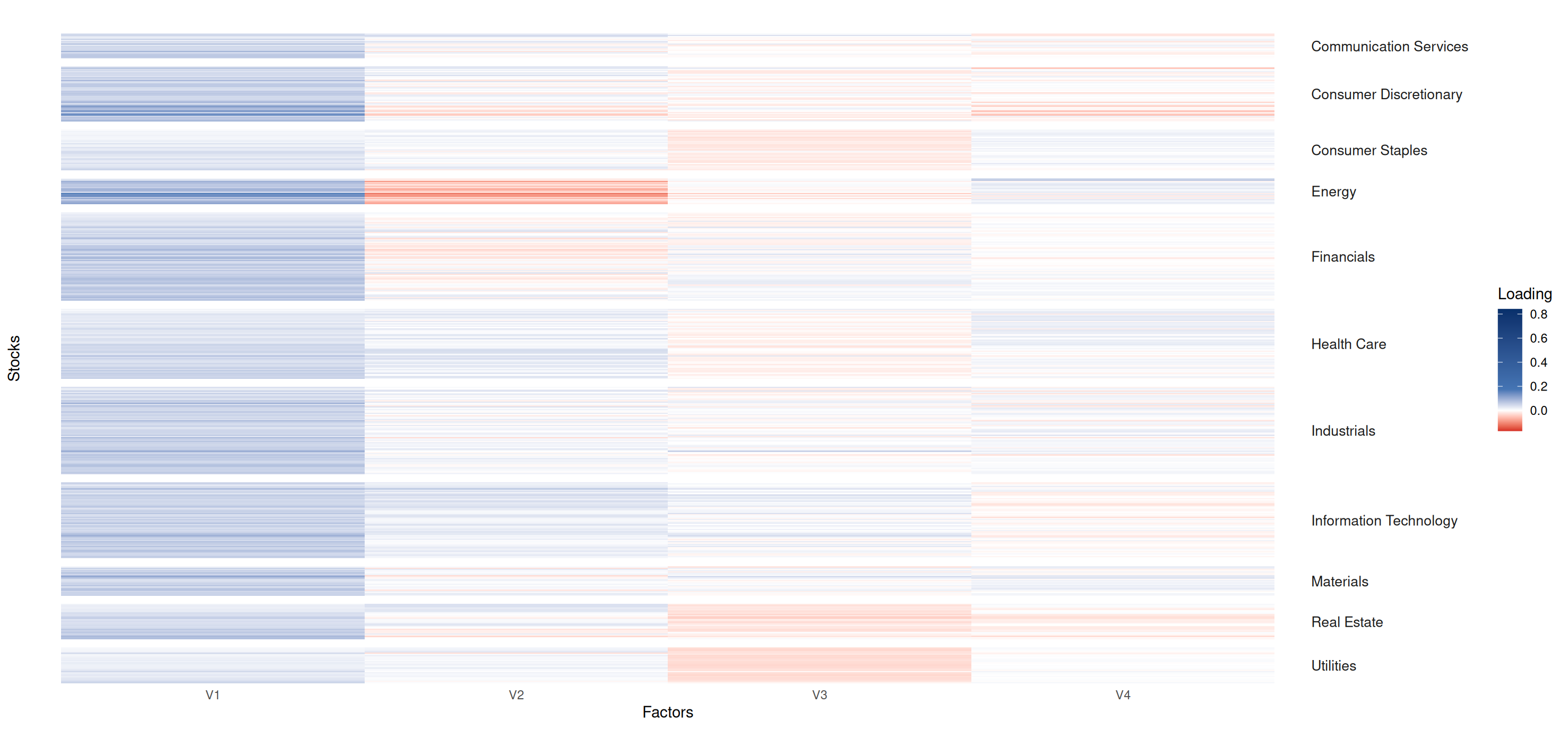}
    \caption{Factor loading matrix estimated by PCA.}
    \label{fig:SNP_PCA}
\end{figure}

Figure \ref{fig:SNP_AFM} displays the factor loading matrix estimated by the proposed AFM, while Figure \ref{fig:SNP_PCA} reports the corresponding loading structure from PCA. A clear difference is that the first AFM factor exhibits uniformly positive and substantially larger loadings across nearly all sectors, indicating a strong market-wide component. While the first principal component from PCA also captures comovement in the equity panel, the corresponding loading pattern is much weaker in magnitude and less sharply separated.

Beyond the leading factor, AFM reveals more distinct sector-level contrasts in the higher-order factors. In particular, several sectors display clearer concentration patterns in the third and fourth AFM factors, whereas the PCA loadings appear more diffuse and less structured. Because the figures are presented as heatmaps, we interpret them in terms of loading magnitude and sectoral concentration rather than direction. Overall, AFM provides a clearer decomposition of common market variation and sector-specific heterogeneity than PCA in this real data analysis.

\subsection{Real Data Forecasting Exercise: Factor-Augmented Vector Autoregression}
We apply the proposed method within the Factor-Augmented Vector Autoregression (FAVAR) framework of \cite{bernanke2005measuring}. The model is specified as
\begin{align*}
    \begin{pmatrix}
        X_t \\ \eta_t
    \end{pmatrix}
    &= 
    \sum_{j=1}^{\ell} \Phi_j
    \begin{pmatrix}
        X_{t-j} \\ \eta_{t-j}
    \end{pmatrix}
    + \nu_t, \qquad t=1,\ldots,n, \\
    Y_t &= B \eta_t + \epsilon_t, \qquad t=1,\ldots,n,
\end{align*}
where $t$ indexes time, $n$ is the number of observations in the estimation sample, $X_t \in \mathbb{R}^M$ denotes the target variables, namely the 12 industry portfolio excess returns, $\eta_t \in \mathbb{R}^k$ denotes the latent factor vector, and $Y_t \in \mathbb{R}^p$ is a large macroeconomic panel used to extract these factors. The lag order of the VAR is denoted by $\ell$.

Our estimation strategy follows a two-step procedure. First, we estimate the loading matrix $B$ and the latent factors $\eta_t$ from the high-dimensional panel $Y_t$ using the proposed AFM method (and competing alternatives). Second, conditioning on the estimated factors $\hat{\eta}_t$, we fit a Bayesian VAR to forecast the target series $X_t$. The Bayesian VAR estimation is implemented using the \texttt{BVAR} package \citep{kuschnig2021bvar}.

For the macroeconomic panel $Y_t$, we utilize the monthly FRED-MD dataset \citep{mccracken2020fred}. The number of latent factors is determined using the $IC_2$ criterion of \cite{bai2002determining}. For the target series $X_t$, we use the ``12 Industry Portfolios'' excess returns obtained from the \cite{french_data_library}. The estimation is performed using data from January 2018 to December 2021 (48 months), and forecasts are generated for the period of January–June 2022.

We compare the proposed AFM with three alternative factor estimation methods: Penalized Maximum Likelihood (PML; \citealt{bai2016efficient}), Shrinkage Principal Orthogonal Complement Thresholding (S-POET; \citealt{wang2017asymptotics}), and Principal Component Analysis (PCA). Additionally, we include a baseline VAR that utilizes only $X_t$ without factors as a benchmark.

For all factor-based models, once the factors $\hat{\eta}_t$ are estimated, we employ the \texttt{BVAR} package of \cite{kuschnig2021bvar} to generate forecasts and posterior credible intervals.

Forecast accuracy is measured by the Root Mean Squared Error (RMSE):
\[
\mathrm{RMSE}
=
\sqrt{
\frac{1}{M H}\sum_{j=1}^M\sum_{h=1}^{H}
\bigl(X_{n+h,j} - \hat{X}_{n+h,j}\bigr)^2
},
\]
where $M$ is the number of portfolios (variables in $X_t$) and $H$ is the forecast horizon.

\begin{table}[t!]
    \centering
    \setlength{\tabcolsep}{4pt}
    \begin{tabular}{cc ccccc}
    \hline
    H & lag & AFM & SPOET & PML & PCA & VAR \\
    \hline
    \multirow{3}{*}{1}
     & 4 & 0.0806 & 0.0816 & 0.0881 & \textbf{0.0799} & 0.1124 \\
     & 5 & \textbf{0.0796} & 0.0820 & 0.0867 & 0.0809 & 0.1066 \\
     & 6 & \textbf{0.0744} & 0.0840 & 0.0899 & 0.0834 & 0.1132 \\
    \hline
    \multirow{3}{*}{3}
     & 4 & \textbf{0.0676} & 0.0725 & 0.0689 & 0.0737 & 0.0791 \\
     & 5 & \textbf{0.0599} & 0.0693 & 0.0658 & 0.0708 & 0.0778 \\
     & 6 & \textbf{0.0581} & 0.0685 & 0.0661 & 0.0697 & 0.0797 \\
    \hline
    \multirow{3}{*}{6}
     & 4 & \textbf{0.0808} & 0.0856 & 0.0820 & 0.0879 & 0.0851 \\
     & 5 & \textbf{0.0749} & 0.0840 & 0.0836 & 0.0855 & 0.0895 \\
     & 6 & \textbf{0.0681} & 0.0812 & 0.0796 & 0.0831 & 0.0850 \\
    \hline
    \end{tabular}
    \caption{Forecasting RMSE for the 12 Industry Portfolios. The best performing method for each setting is highlighted in bold.}
    \label{tbl:err_H_lag}
\end{table}

Table~\ref{tbl:err_H_lag} reports the RMSE results for horizons $H \in \{1,3,6\}$ across lag orders $\ell \in \{4,5,6\}$. The baseline VAR without factors consistently exhibits the highest errors, confirming the benefit of incorporating latent macroeconomic information. From short-term to long-term forecasts, the proposed AFM demonstrates the best or second-best performance in all cases.

\begin{figure}[ht]
    \centering
    \includegraphics[scale=0.50]{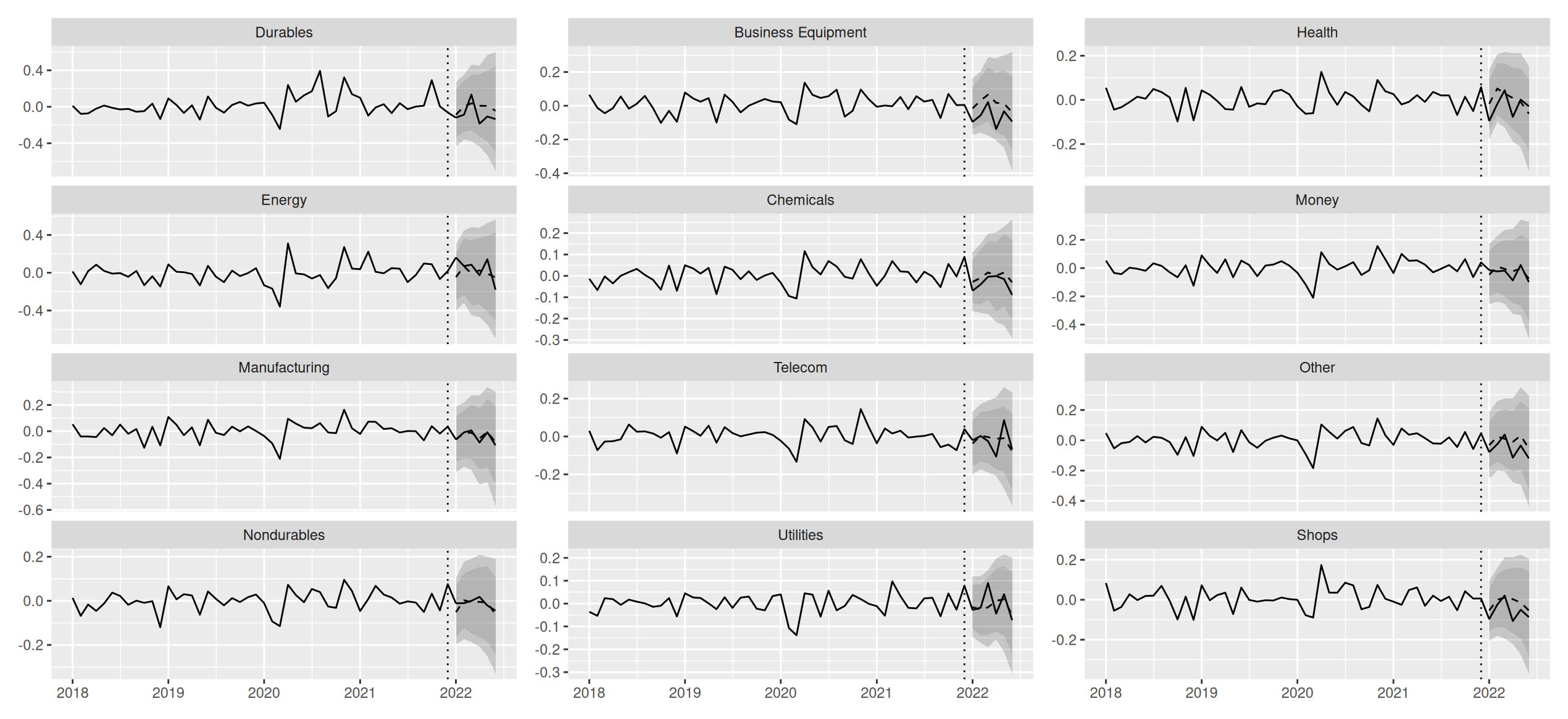}
    \caption{ Six-step-ahead forecasts ($H=6$) for the 12 industry portfolio excess returns using the AFM-based FAVAR. The dashed line denotes the posterior median forecast, while the dark and light shaded areas represent the 95\% and 99\% credible intervals, respectively.\label{fig:favar_afm}}
\end{figure} 

Figure~\ref{fig:favar_afm} displays the six-step-ahead forecasts ($H=6$) using the AFM-based model with $\ell=6$. The realized returns largely fall within the 95\% credible intervals, and the posterior median tracks the observed series closely. Notably, sectors such as {Money} and {Nondurables} show precise alignment between the forecast and actual data, illustrating the model's capability to capture sector-specific dynamics driven by common factors.

\subsection{Korean Macroeconomic Data: Dynamic Analysis}
For this application, we use a dynamic extension of the proposed model. We apply the method to a Korean macroeconomic dataset constructed from the Economic Statistics System (ECOS) of the Bank of Korea. The dataset covers 2010Q2--2026Q1 and contains $n=64$ quarterly observations on $p=149$ transformed variables. For descriptive convenience, the variables are organized into eight broad macroeconomic groups: interest rates, real activity and investment, sentiment and composite indicators, equity market variables, prices, employment, money and liquidity, and exchange rates and foreign reserves. This real data analysis is exploratory and descriptive: the theoretical results in Section~\ref{sec:theoretical_results} pertain to the static model, not to the dynamic specification used here.

Specifically, we use the dynamic factor specification
\begin{align*}
    Y_t &= \rmU \Lambda^{1/2}\eta_t + u_t, \qquad u_t \sim N_p(0,\Sigma_e), \\
    \eta_t &= A\eta_{t-1} + \xi_t, \qquad \xi_t \sim N_k(0,I_k-A^2),
\end{align*}
where $Y_t=(Y_{1t},\ldots,Y_{pt})^T \in \mathbb{R}^p$ is the transformed macroeconomic vector at time $t$, $\eta_t \in \mathbb{R}^k$ is the latent factor vector, $\rmU=(u_{ij}) \in \mathbb{R}^{p\times k}$ has orthonormal columns and spans the common factor subspace, $\Lambda=diag(\lambda_1,\ldots,\lambda_k)$ quantifies the strength of each factor direction, and $\Sigma_e=(\sigma_{e,ij}) \in \mathbb{R}^{p\times p}$ is the idiosyncratic covariance matrix. We set
\[
A=diag(\rho_1,\ldots,\rho_k), \qquad |\rho_j|<1 \ \text{for } j=1,\ldots,k,
\]
so that each factor follows its own AR(1) dynamics. To fix factor scale, we impose the stationary normalization $Var(\eta_t)=I_k$. Under the diagonal AR(1) specification above, this is achieved by setting $\xi_t \sim N_k(0,I_k-A^2)$, so that the covariance matrix of $\xi_t$ is chosen to produce the desired stationary covariance of the factor process. This is the same type of stationary state specification used in \citet{aguilar2000bayesian}, where the covariance matrix of the Gaussian state error is linked to the marginal covariance through the stationarity relation. Consequently, the scaled factors $f_t=\Lambda^{1/2}\eta_t$ are contemporaneously uncorrelated with covariance matrix $\Lambda$, and the contemporaneous covariance of the common component is $\rmU\Lambda\rmU^T$. The priors on $(\mathrm{U},\Lambda,\Sigma_e)$ are the same as in the static model. The number of factors is set to $k=2$ by the $IC_2$ criterion of \citet{bai2002determining}. Posterior computation updates the latent states, $\rmU$, $\Lambda$, $\Sigma_e$, and the factor-specific AR coefficients within a Gibbs sampler; additional details are given in the Appendix.

\subsubsection{Communality}

In the classical factor-analysis terminology of \citet{thurstone1931multiple}, communality refers to the portion of a variable's total variance that is shared with the common factors. In our dynamic factor specification, we use the same term for the model-implied share of variance attributable to the common component. Since the scaled factors are contemporaneously uncorrelated with covariance matrix $\Lambda = diag(\lambda_1,\ldots,\lambda_k)$, factor $j$ contributes $u_{ij}^2\lambda_j$ to the marginal variance of variable $i$. Hence, under the model,
\begin{equation}
    \mathrm{Var}(Y_{it}) = \sum_{j=1}^k u_{ij}^2\lambda_j + \sigma_{e,ii},
\end{equation}
where $\sigma_{e,ii}$ denotes the $i$th diagonal element of $\Sigma_e$. The common variance of variable $i$ is therefore
\begin{equation}
    \mathrm{Var}_{\mathrm{common}}(Y_{it}) = \sum_{j=1}^k u_{ij}^2\lambda_j,
\end{equation}
and its communality is
\begin{equation}
    h_i^2
    =
    \frac{\sum_{j=1}^k u_{ij}^2\lambda_j}
    {\sum_{j=1}^k u_{ij}^2\lambda_j + \sigma_{e,ii}}.
\end{equation}
Communality measures how well each variable is represented by the common component and provides a convenient summary for comparing factor strength across variable groups.

Table~\ref{tab:group_communality} reports average communalities by group. Interest-rate variables have the strongest common component, with mean communality 0.362. Employment, sentiment and composite, and equity-market variables form a middle tier, with mean communalities around 0.17--0.18. By contrast, the average communality is only modest for money and liquidity, real activity and investment, and prices, while exchange rates and reserves show the weakest common component. Thus, the common structure in this dataset is meaningful but moderate, and it is most clearly visible in interest-rate variables.

\begin{table}[t]
\centering
\small
\begin{tabular}{lccc}
\toprule
Variable group & \# Variables & Mean communality & Median communality \\
\midrule
Interest rates & 14 & 0.362 & 0.374 \\
Employment & 12 & 0.184 & 0.161 \\
Sentiment / composite & 12 & 0.171 & 0.183 \\
Equity market & 9 & 0.170 & 0.154 \\
Money / liquidity & 21 & 0.110 & 0.084 \\
Real activity / investment & 53 & 0.101 & 0.064 \\
Prices & 19 & 0.097 & 0.054 \\
Exchange rates / reserves & 9 & 0.041 & 0.037 \\
\bottomrule
\end{tabular}
\caption{Communality by variable group}
\label{tab:group_communality}
\end{table}

The same pattern appears at the level of individual variables. Table~\ref{tab:top_comm} reports selected high-communality variables. The largest values are observed mainly for short-term interest-rate variables, including the Korea Inter-Bank Offered Rate (KORIBOR) at maturities of 3, 6, and 12 months, together with a small number of employment and construction-related variables. Even for these variables, however, the idiosyncratic share remains substantial. The common factors are therefore economically informative, but they do not dominate the variation of most variables in the dataset.

\begin{table}[t]
\centering
\small
\begin{tabular}{lcc}
\toprule
Variable & Communality & Idiosyncratic share \\
\midrule
KORIBOR (6-month) & 0.519 & 0.481 \\
Employment: business-support / rental services & 0.496 & 0.504 \\
KORIBOR (3-month) & 0.493 & 0.507 \\
Employment: construction & 0.442 & 0.558 \\
KORIBOR (12-month) & 0.432 & 0.568 \\
Industrial bank bond (1-year) & 0.431 & 0.569 \\
Corporate bond (3-year, AA-, private assessment) & 0.426 & 0.574 \\
Corporate bond (3-year, AA-) & 0.420 & 0.580 \\
\bottomrule
\end{tabular}
\caption{Selected high-communality variables}
\label{tab:top_comm}
\end{table}

\subsubsection{Variance Decomposition and Factor Interpretation}

Since the scaled factors $f_t = \Lambda^{1/2}\eta_t$ are contemporaneously uncorrelated with covariance matrix $\Lambda = diag(\lambda_1,\ldots,\lambda_k)$, the share of the total variance of variable $i$ explained by factor $j$ is
\begin{equation}
    \mathrm{Share}_{ij}^{\mathrm{total}}
    =
    \frac{u_{ij}^2\lambda_j}
    {\sum_{\ell=1}^k u_{i\ell}^2\lambda_\ell + \sigma_{e,ii}},
\end{equation}
and the corresponding idiosyncratic share is
\begin{equation}
    \mathrm{Share}_{i}^{\mathrm{idio}}
    =
    \frac{\sigma_{e,ii}}
    {\sum_{\ell=1}^k u_{i\ell}^2\lambda_\ell + \sigma_{e,ii}}.
\end{equation}
This decomposition summarizes how the common and idiosyncratic components contribute to the marginal variance of each variable; see also \citet{forni2000generalized,stock2016dynamic} for dynamic-factor decompositions into common and idiosyncratic parts.

The estimated two-factor structure is asymmetric in interpretability. Factor 2 is the more clearly interpretable factor and is dominated by interest-rate variables, with large loadings on KORIBOR and several government and corporate bond yields. Factor 1 is broader and combines labor-market, construction and investment, equity-turnover, and consumer-price related variables. We therefore interpret Factor 2 as an interest-rate factor and Factor 1 more cautiously as a broad domestic conditions factor.

The estimated factors do not map cleanly onto the predefined macroeconomic categories. Instead, the results suggest one relatively distinct factor dominated by interest-rate variables and another broader factor with substantial loadings from multiple categories. Table~\ref{tab:factor_interp} reports representative variable codes with large loadings; full definitions of the variable codes are provided in the Appendix.

\begin{table}[t]
\centering
\small
\begin{tabular}{llll}
\toprule
Factor & Code & Group & Interpretation \\
\midrule
\multirow{4}{*}{Factor 1}
& RA12 & Real activity / investment & \multirow{4}{*}{Broad domestic conditions} \\
& RA11 & Real activity / investment & \\
& RA30 & Real activity / investment & \\
& EQ3  & Equity market & \\
\midrule
\multirow{4}{*}{Factor 2}
& IR3  & Interest rates & \multirow{4}{*}{Interest-rate factor} \\
& IR2  & Interest rates & \\
& IR13 & Interest rates & \\
& IR10 & Interest rates & \\
\bottomrule
\end{tabular}
\caption{Representative variable codes and tentative interpretations of the two factors}
\label{tab:factor_interp}
\end{table}

\subsubsection{Dynamic Behavior}

For dynamic interpretation, it is convenient to work with the scaled factors $f_t=\Lambda^{1/2}\eta_t$, which absorb the factor-specific scale and are therefore easier to interpret in terms of factor strength.

\begin{figure}[t]
    \centering
    \includegraphics[width=.90\textwidth]{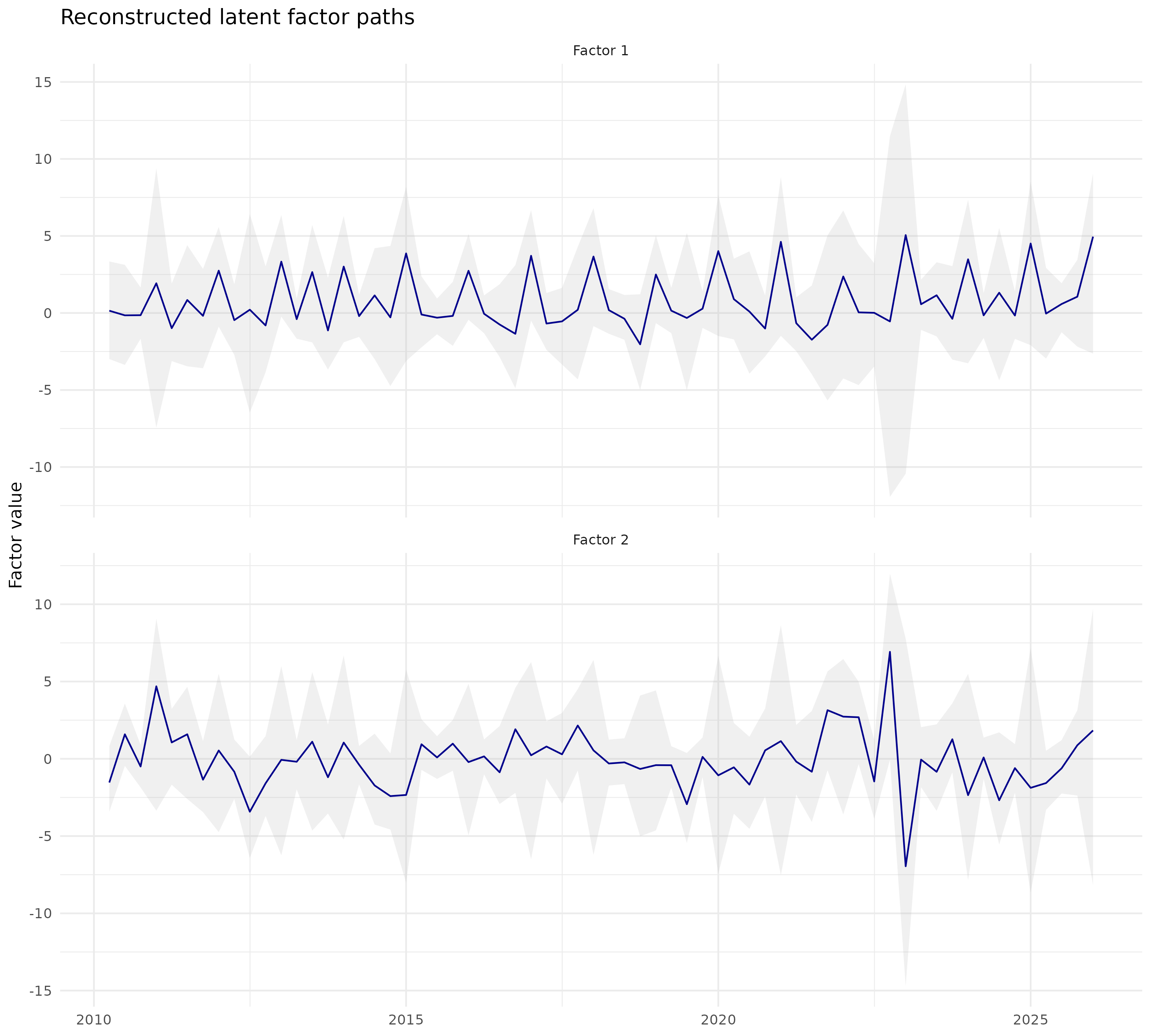}
    \caption{Posterior summaries of the smoothed latent factors $f_t=\Lambda^{1/2}\eta_t$. The solid lines denote posterior means, and the shaded bands denote pointwise 2.5\% and 97.5\% quantiles across posterior samples.}
    \label{fig:factor_paths}
\end{figure}

Figure~\ref{fig:factor_paths} plots the posterior mean paths of $f_t$ together with 95\% credible intervals. Both factors exhibit pronounced movements around 2022--2023, indicating a substantial common disturbance during that period. In addition, Factor~1 displays a clear recurrent pattern at roughly annual frequency, suggesting that some seasonal or other within-year variation may remain even after transformation. More broadly, both factor paths show short-run oscillations, although this feature is especially pronounced for Factor~1.

Overall, the Korean macroeconomic dataset exhibits a moderate common component, although its strength varies across variable groups. The common signal appears most clearly in interest-rate variables, while other categories show a more heterogeneous mix of common and idiosyncratic variation. Under the current specification with differenced and standardized quarterly variables, the factor paths are primarily useful for summarizing short-run common movements.

\section{Conclusion}

We introduce a novel Bayesian framework for approximate factor models. By placing priors directly on the eigenstructure of the common component, we address rotational non-identifiability while enabling accurate recovery of the factor subspace and associated eigenvalues. We derive posterior contraction rates for the leading eigenvalues and eigenspace, providing a theoretical foundation for the proposed procedure. Our simulation results confirm improved recovery of the underlying eigenstructure, particularly under relative error metrics. Furthermore, in real data analyses, the proposed method yields interpretable estimates of the common factor structure and delivers competitive forecasting performance at medium to long horizons in a factor-augmented VAR exercise, relative to standard shrinkage and PCA-based approaches. Extending the framework to dynamic factor models with time-varying eigenstructures is a promising direction for future research.

\section*{Acknowledgements}
Seongmin Kim was supported by Global - Learning \& Academic research institution for Master's·PhD students, and Postdocs(G-LAMP) Program of the National Research Foundation of Korea(NRF) grant funded by the Ministry of Education(No. RS-2025-25442252). Jaeyong Lee was supported by the National Research Foundation of Korea (NRF) grant funded by the Korea government (MSIT) (No. 2023R1A2C1003050 and RS-2024-00406127).


\bibliographystyle{dcu}
\bibliography{FA}

\end{document}


\maketitle


\section{Variable List in the Korean macroeconomic Panel}\label{app:full_variable_list}

The full variable list for the Korean macroeconomic application in Section~6.3 is reported below. The table corresponds to the variables that remain in the final transformed panel after preprocessing, coverage filtering, and quarterly subsampling. Variable codes are those used in the main text. Here, \textbf{Difference} refers to $\Delta x_t$, and \textbf{Log-difference} refers to $\Delta \log x_t$.

{\scriptsize
\begin{longtable}{p{1.5cm}p{3.3cm}p{7.2cm}p{2.2cm}}
\caption{Retained variable list for the final Korean macroeconomic panel}
\label{tab:full_variable_list}\\
\toprule
Code & Group & Variable label & Transformation \\
\midrule
\endfirsthead

\multicolumn{4}{c}{\tablename\ \thetable\ -- continued from previous page} \\
\toprule
Code & Group & Variable label & Transformation \\
\midrule
\endhead

\midrule
\multicolumn{4}{r}{Continued on next page} \\
\endfoot

\bottomrule
\endlastfoot

\multicolumn{4}{l}{\textit{Interest rates}} \\
IR1 & Interest rates & Korea Interbank Offered Rate (KORIBOR, 12-month) & Difference \\
IR2 & Interest rates & Korea Interbank Offered Rate (KORIBOR, 3-month) & Difference \\
IR3 & Interest rates & Korea Interbank Offered Rate (KORIBOR, 6-month) & Difference \\
IR4 & Interest rates & Government bond yield (10-year) & Difference \\
IR5 & Interest rates & Government bond yield (1-year) & Difference \\
IR6 & Interest rates & Government bond yield (20-year) & Difference \\
IR7 & Interest rates & Government bond yield (30-year) & Difference \\
IR8 & Interest rates & Government bond yield (3-year) & Difference \\
IR9 & Interest rates & Government bond yield (5-year) & Difference \\
IR10 & Interest rates & Industrial finance bond yield (1-year) & Difference \\
IR11 & Interest rates & Call rate (1-day, all transactions) & Difference \\
IR12 & Interest rates & Call rate (1-day, brokered transactions) & Difference \\
IR13 & Interest rates & Corporate bond yield (3-year, AA-) & Difference \\
IR14 & Interest rates & Corporate bond yield (3-year, AA-, private assessment) & Difference \\
\addlinespace[1mm]

\multicolumn{4}{l}{\textit{Real activity / investment}} \\
RA1 & Real activity / investment & Trading volume: individuals (sell) & Log-difference \\
RA2 & Real activity / investment & Trading volume: institutional investors (sell) & Log-difference \\
RA3 & Real activity / investment & Trading volume: institutional investors (buy) & Log-difference \\
RA4 & Real activity / investment & Trading volume: other corporations (sell) & Log-difference \\
RA5 & Real activity / investment & Trading volume: other corporations (buy) & Log-difference \\
RA6 & Real activity / investment & Trading volume: other corporations (net buy) & Log-difference \\
RA7 & Real activity / investment & Trading volume: other foreigners (sell) & Log-difference \\
RA8 & Real activity / investment & Trading volume: total sell & Log-difference \\
RA9 & Real activity / investment & Trading volume: total buy & Log-difference \\
RA10 & Real activity / investment & Trading volume: foreigners (sell) & Log-difference \\
RA11 & Real activity / investment & Employment: construction & Log-difference \\
RA12 & Real activity / investment & Employment: business facilities management, business support, and rental services & Log-difference \\
RA13 & Real activity / investment & Employment: professional, scientific, and technical services & Log-difference \\
RA14 & Real activity / investment & Service production: wholesale and retail trade (seasonally adjusted index) & Log-difference \\
RA15 & Real activity / investment & Service production: wholesale trade (seasonally adjusted index) & Log-difference \\
RA16 & Real activity / investment & Service production: wholesale of industrial agricultural and livestock products (seasonally adjusted index) & Log-difference \\
RA17 & Real activity / investment & Service production: wholesale of household goods (seasonally adjusted index) & Log-difference \\
RA18 & Real activity / investment & Service production: water supply, sewage, waste treatment, and materials recovery (seasonally adjusted index) & Log-difference \\
RA19 & Real activity / investment & Service production: water supply (seasonally adjusted index) & Log-difference \\
RA21 & Real activity / investment & Service production: wholesale of food, beverages, and tobacco (seasonally adjusted index) & Log-difference \\
RA22 & Real activity / investment & Service production: motor vehicle and parts sales (seasonally adjusted index) & Log-difference \\
RA23 & Real activity / investment & Service production: motor vehicle parts and motorcycle sales (seasonally adjusted index) & Log-difference \\
RA24 & Real activity / investment & Service production: motor vehicle sales (seasonally adjusted index) & Log-difference \\
RA25 & Real activity / investment & Service production: overall index (seasonally adjusted) & Log-difference \\
RA26 & Real activity / investment & Service production: waste collection, transport, treatment, and materials recovery (seasonally adjusted index) & Log-difference \\
RA27 & Real activity / investment & Service production: sewage, wastewater, and human waste treatment (seasonally adjusted index) & Log-difference \\
RA29 & Real activity / investment & Investment/construction: seasonally adjusted index & Log-difference \\
RA30 & Real activity / investment & Investment/construction: original index & Log-difference \\
RA31 & Real activity / investment & Industrial production: construction (seasonally adjusted) & Log-difference \\
RA32 & Real activity / investment & Industrial production: public administration (seasonally adjusted) & Log-difference \\
RA33 & Real activity / investment & Industrial production: mining and manufacturing (seasonally adjusted) & Log-difference \\
RA34 & Real activity / investment & Industrial production: services (seasonally adjusted) & Log-difference \\
RA35 & Real activity / investment & Industrial production index for all industries (excluding agriculture, forestry, and fisheries; seasonally adjusted) & Log-difference \\
RA36 & Real activity / investment & Manufacturing production: construction use (seasonally adjusted production index) & Log-difference \\
RA37 & Real activity / investment & Manufacturing production: other use (seasonally adjusted production index) & Log-difference \\
RA38 & Real activity / investment & Manufacturing production: agricultural use (seasonally adjusted production index) & Log-difference \\
RA39 & Real activity / investment & Manufacturing production: office use (seasonally adjusted production index) & Log-difference \\
RA40 & Real activity / investment & Manufacturing production: transport equipment (seasonally adjusted production index) & Log-difference \\
RA41 & Real activity / investment & Manufacturing production: capital goods (seasonally adjusted production index) & Log-difference \\
RA42 & Real activity / investment & Manufacturing production: electric power use (seasonally adjusted production index) & Log-difference \\
RA43 & Real activity / investment & Manufacturing production: manufacturing equipment use (seasonally adjusted production index) & Log-difference \\
RA44 & Real activity / investment & Manufacturing production: intermediate goods (seasonally adjusted production index) & Log-difference \\
RA45 & Real activity / investment & Manufacturing production: telecommunications and broadcasting use (seasonally adjusted production index) & Log-difference \\
RA46 & Real activity / investment & Shipment/inventory: grain mill products, starch, and starch products manufacturing (seasonally adjusted operation-rate index) & Log-difference \\
RA47 & Real activity / investment & Shipment/inventory: fruit and vegetable processing and preservation (seasonally adjusted operation-rate index) & Log-difference \\
RA48 & Real activity / investment & Shipment/inventory: other food products manufacturing (seasonally adjusted operation-rate index) & Log-difference \\
RA49 & Real activity / investment & Shipment/inventory: dairy products and edible ice products manufacturing (seasonally adjusted operation-rate index) & Log-difference \\
RA50 & Real activity / investment & Shipment/inventory: slaughtering, meat processing, and meat preservation (seasonally adjusted operation-rate index) & Log-difference \\
RA51 & Real activity / investment & Shipment/inventory: animal and vegetable fats and oils manufacturing (seasonally adjusted operation-rate index) & Log-difference \\
RA52 & Real activity / investment & Shipment/inventory: animal feed and prepared feed manufacturing (seasonally adjusted operation-rate index) & Log-difference \\
RA53 & Real activity / investment & Shipment/inventory: fish and seafood processing and preservation (seasonally adjusted operation-rate index) & Log-difference \\
RA54 & Real activity / investment & Shipment/inventory: food products manufacturing (seasonally adjusted operation-rate index) & Log-difference \\
RA55 & Real activity / investment & Shipment/inventory: manufacturing (seasonally adjusted operation-rate index) & Log-difference \\
\addlinespace[1mm]

\multicolumn{4}{l}{\textit{Sentiment / composite}} \\
SENT1 & Sentiment / composite & Composite index: Gyeonggi & Difference \\
SENT2 & Sentiment / composite & Composite index: Gwangju & Difference \\
SENT3 & Sentiment / composite & Composite index: Daegu & Difference \\
SENT4 & Sentiment / composite & Composite index: Daejeon & Difference \\
SENT5 & Sentiment / composite & Composite index: Busan & Difference \\
SENT6 & Sentiment / composite & Composite index: Seoul & Difference \\
SENT7 & Sentiment / composite & Composite index: Sejong & Difference \\
SENT8 & Sentiment / composite & Composite index: Ulsan & Difference \\
SENT9 & Sentiment / composite & Composite index: Incheon & Difference \\
SENT10 & Sentiment / composite & Composite index: nationwide & Difference \\
SENT11 & Sentiment / composite & CSI: Economic Sentiment Index (cyclical component) & Log-difference \\
SENT12 & Sentiment / composite & CSI: Economic Sentiment Index (original series) & Log-difference \\
\addlinespace[1mm]

\multicolumn{4}{l}{\textit{Equity market}} \\
EQ1 & Equity market & Equity: KOSDAQ index & Log-difference \\
EQ2 & Equity market & Equity: KOSPI index & Log-difference \\
EQ3 & Equity market & Equity: trading value (KOSPI market) & Log-difference \\
EQ4 & Equity market & Equity: trading value (KOSDAQ market) & Log-difference \\
EQ5 & Equity market & Equity: trading volume (KOSPI market) & Log-difference \\
EQ6 & Equity market & Equity: trading volume (KOSDAQ market) & Log-difference \\
EQ7 & Equity market & Equity: market capitalization (KOSPI market) & Log-difference \\
EQ8 & Equity market & Equity: foreign investors' net purchases (KOSPI market) & Log-difference \\
EQ9 & Equity market & Equity: foreign investors' net purchases (KOSDAQ market) & Log-difference \\
\addlinespace[1mm]

\multicolumn{4}{l}{\textit{Prices}} \\
P1 & Prices & Living prices: contract amount & Log-difference \\
P2 & Prices & Living prices: contract amount (daily average) & Log-difference \\
P3 & Prices & Living prices: number of contracts & Log-difference \\
P4 & Prices & Living prices: number of contracts (daily average) & Log-difference \\
P5 & Prices & CPI: noodles & Log-difference \\
P6 & Prices & CPI: vermicelli & Log-difference \\
P7 & Prices & CPI: peanuts & Log-difference \\
P8 & Prices & CPI: ramen & Log-difference \\
P9 & Prices & CPI: wheat flour & Log-difference \\
P10 & Prices & CPI: barley rice & Log-difference \\
P11 & Prices & CPI: bread and cereals & Log-difference \\
P12 & Prices & CPI: food & Log-difference \\
P13 & Prices & CPI: food and non-alcoholic beverages & Log-difference \\
P14 & Prices & CPI: rice & Log-difference \\
P15 & Prices & CPI: glutinous rice & Log-difference \\
P16 & Prices & CPI: overall index & Log-difference \\
P17 & Prices & CPI: beans & Log-difference \\
P18 & Prices & CPI: brown rice & Log-difference \\
P19 & Prices & CPI: mixed grains & Log-difference \\
\addlinespace[1mm]

\multicolumn{4}{l}{\textit{Employment}} \\
EMP1 & Employment & Employment: public administration, national defense, and social security administration & Log-difference \\
EMP2 & Employment & Employment: mining & Log-difference \\
EMP3 & Employment & Employment: finance and insurance & Log-difference \\
EMP4 & Employment & Employment: agriculture, forestry, and fisheries & Log-difference \\
EMP5 & Employment & Employment: wholesale and retail trade & Log-difference \\
EMP6 & Employment & Employment: real estate & Log-difference \\
EMP7 & Employment & Employment: water supply, sewage, waste treatment, and materials recovery & Log-difference \\
EMP8 & Employment & Employment: accommodation and food service activities & Log-difference \\
EMP9 & Employment & Employment: transportation and storage & Log-difference \\
EMP10 & Employment & Employment: electricity, gas, steam, and air conditioning supply & Log-difference \\
EMP11 & Employment & Employment: information and communications & Log-difference \\
EMP12 & Employment & Employment: manufacturing & Log-difference \\
\addlinespace[1mm]

\multicolumn{4}{l}{\textit{Money / liquidity}} \\
M1 & Money / liquidity & M2: CMA & Log-difference \\
M2 & Money / liquidity & M2: average balance, original series & Log-difference \\
M3 & Money / liquidity & M2: MMF shares & Log-difference \\
M4 & Money / liquidity & M2: money trusts with maturity less than 2 years & Log-difference \\
M5 & Money / liquidity & M2: time and savings deposits with maturity less than 2 years & Log-difference \\
M6 & Money / liquidity & M2: commercial paper & Log-difference \\
M7 & Money / liquidity & M2: savings deposits with free deposits and withdrawals & Log-difference \\
M8 & Money / liquidity & M2: certificates of deposit & Log-difference \\
M9 & Money / liquidity & M2: demand deposits & Log-difference \\
M10 & Money / liquidity & M2: currency in circulation & Log-difference \\
M11 & Money / liquidity & Lf: long-term financial instruments with maturity over 2 years & Log-difference \\
M12 & Money / liquidity & Lf: financial institution liquidity, by product (average balance, seasonally adjusted) & Log-difference \\
M13 & Money / liquidity & Lf: M2 & Log-difference \\
M14 & Money / liquidity & Lf: financial instruments issued by non-deposit-taking institutions & Log-difference \\
M15 & Money / liquidity & Lf: financial instruments issued by deposit-taking institutions & Log-difference \\
M16 & Money / liquidity & Deposits: time deposits, 2\% to less than 3\% & Difference \\
M17 & Money / liquidity & Deposits: time deposits, less than 2\% & Difference \\
M18 & Money / liquidity & Deposits: time deposits, 3\% to less than 4\% & Difference \\
M19 & Money / liquidity & Deposits: time deposits, 4\% to less than 5\% & Difference \\
M20 & Money / liquidity & Deposits: time deposits, 5\% to less than 6\% & Difference \\
M21 & Money / liquidity & Deposits: time deposits, 6\% to less than 7\% & Difference \\
\addlinespace[1mm]

\multicolumn{4}{l}{\textit{Exchange rates / reserves}} \\
FX1 & Exchange rates / reserves & Reserves: IMF reserve position & Log-difference \\
FX2 & Exchange rates / reserves & Reserves: gold & Log-difference \\
FX3 & Exchange rates / reserves & Reserves: foreign exchange & Log-difference \\
FX4 & Exchange rates / reserves & Reserves: special drawing rights & Log-difference \\
FX5 & Exchange rates / reserves & Reserves: total & Log-difference \\
FX6 & Exchange rates / reserves & FX: KRW/USD (market average exchange rate) & Log-difference \\
FX7 & Exchange rates / reserves & FX: KRW/CNY (market average exchange rate) & Log-difference \\
FX8 & Exchange rates / reserves & FX: KRW/EUR & Log-difference \\
FX9 & Exchange rates / reserves & FX: KRW/JPY (per 100 yen) & Log-difference \\
\addlinespace[1mm]

\end{longtable}
}

\section{Posterior Computation for the Dynamic Extension}\label{app:dynamic_sampler}

For the Korean macroeconomic application in Section~6.3, we use the dynamic extension
\begin{align}
Y_t &= \mathrm{U}\Lambda^{1/2} \eta_t + u_t,
\qquad
u_t \sim N_p(0,\Sigma_e), \label{eq:dyn_obs_app}\\
\eta_t &= A \eta_{t-1} + \xi_t,
\qquad
\xi_t \sim N_k(0, I_k - A^2),\quad\text{for}\;t=1,\ldots,n, \label{eq:dyn_state_app}
\end{align}
where
\[
A = \mathrm{diag}(\rho_1,\ldots,\rho_k),
\qquad |\rho_j|<1,\quad j=1,\ldots,k.
\]
The priors on $(\mathrm{U},\Lambda,\Sigma_e)$ are the same as in the static model.

For implementation, it is convenient to work with the scaled latent state
\[
Z_t := \Lambda^{1/2}\eta_t.
\]
Then the model can be written as
\begin{align}
Y_t &= \mathrm{U} Z_t + u_t,
\qquad
u_t \sim N_p(0,\Sigma_e), \label{eq:dyn_obs_Z}\\
Z_t &= A Z_{t-1} + \nu_t,
\qquad
\nu_t \sim N_k\!\bigl(0,\Lambda(I_k-A^2)\bigr), \label{eq:dyn_state_Z}
\end{align}
with stationary initial distribution
\[
Z_1 \sim N_k(0,\Lambda).
\]
This is equivalent to \eqref{eq:dyn_obs_app}--\eqref{eq:dyn_state_app} but matches the actual Gibbs implementation.

\paragraph{Sampling scheme.}
One MCMC iteration for the dynamic extension proceeds as follows.

\begin{enumerate}
   \item \textbf{Update the latent state path $Z_{1:n}$.}
    Conditional on $(\mathrm{U},\Lambda,\Sigma_e,A)$, the model
    \eqref{eq:dyn_obs_Z}--\eqref{eq:dyn_state_Z} is a linear Gaussian state-space model with
    \[
    Z_1 \sim N_k(0,\Lambda),\qquad
    Z_t \mid Z_{t-1} \sim N_k(AZ_{t-1},Q),\qquad
    Y_t \mid Z_t \sim N_p(\mathrm{U}Z_t,\Sigma_e),
    \]
    where
    \[
    Q := \Lambda(I_k-A^2).
    \]
    Hence the full conditional distribution of the latent path
    \[
    Z_{1:n}:=(Z_1^T,\ldots,Z_n^T)^T
    \]
    is Gaussian and can be sampled jointly by a standard forward-filtering backward-sampling (FFBS) step.
    
    In the forward-filtering pass, let
    \[
    m_{t|t-1} := \mathbb{E}(Z_t \mid Y_{1:t-1},\mathrm{rest}),
    \qquad
    P_{t|t-1} := \mathrm{Var}(Z_t \mid Y_{1:t-1},\mathrm{rest}),
    \]
    and
    \[
    m_{t|t} := \mathbb{E}(Z_t \mid Y_{1:t},\mathrm{rest}),
    \qquad
    P_{t|t} := \mathrm{Var}(Z_t \mid Y_{1:t},\mathrm{rest}).
    \]
    Starting from $m_{1|0}=0$ and $P_{1|0}=\Lambda$, the Kalman filter recursions are
    \[
    K_t
    =
    P_{t|t-1}\mathrm{U}^T
    \bigl(\mathrm{U}P_{t|t-1}\mathrm{U}^T+\Sigma_e\bigr)^{-1},
    \]
    \[
    m_{t|t}
    =
    m_{t|t-1}
    +
    K_t\bigl(Y_t-\mathrm{U}m_{t|t-1}\bigr),
    \qquad
    P_{t|t}
    =
    P_{t|t-1}
    -
    K_t\mathrm{U}P_{t|t-1},
    \]
    and, for $t=2,\ldots,n$,
    \[
    m_{t|t-1}=A\,m_{t-1|t-1},
    \qquad
    P_{t|t-1}=A\,P_{t-1|t-1}A^T+Q.
    \]
    
    In the backward-sampling pass, one first draws
    \[
    Z_n \sim N_k(m_{n|n},P_{n|n}),
    \]
    and then, for $t=n-1,\ldots,1$, draws
    \[
    Z_t \mid Z_{t+1},Y_{1:n},\mathrm{rest}
    \sim
    N_k(m_t^\ast,P_t^\ast),
    \]
    where
    \[
    J_t := P_{t|t}A^T(P_{t+1|t})^{-1},
    \]
    \[
    m_t^\ast
    =
    m_{t|t}
    +
    J_t\bigl(Z_{t+1}-A\,m_{t|t}\bigr),
    \qquad
    P_t^\ast
    =
    P_{t|t}
    -
    J_tP_{t+1|t}J_t^T.
    \]
    This yields a joint draw from the full conditional distribution of $Z_{1:n}$.

    \item \textbf{Update $\rho_1,\ldots,\rho_k$.}

    Conditional on $Z_{1:n}$ and $\Lambda$, the coordinates are decoupled because $A$ is diagonal. For each $j=1,\ldots,k$, the conditional log-posterior of $\rho_j$ is
    \begin{equation}
    \log \pi(\rho_j \mid Z_{j,1:n},\lambda_j,\text{rest})
    =
    \log \pi(\rho_j)
    - \frac{n-1}{2}\log(1-\rho_j^2)
    - \frac{1}{2\lambda_j(1-\rho_j^2)}
    \sum_{t=2}^n (Z_{j,t}-\rho_j Z_{j,t-1})^2
    + C,
    \label{eq:rho_logpost_Z}
    \end{equation}
    for $|\rho_j|<1$, where $C$ does not depend on $\rho_j$. In our implementation, each $\rho_j$ is updated by univariate slice sampling on $(-1,1)$ under a truncated Gaussian prior.

    \item \textbf{Update $\lambda_1,\ldots,\lambda_k$.}

    Conditional on $Z_{1:n}$ and $A$, each $\lambda_j$ has an inverse-gamma full conditional distribution. Writing
    \[
    S_j
    :=
    Z_{j1}^2
    +
    \frac{1}{1-\rho_j^2}
    \sum_{t=2}^n (Z_{j,t}-\rho_j Z_{j,t-1})^2,
    \]
    the update takes the form
    \[
    \lambda_j \mid Z_{j,1:n},\rho_j,\text{rest}
    \sim
    IG\!\left(
    a_j - 1 + \frac{n}{2},
    \frac{q}{2} + \frac{S_j}{2}
    \right),
    \]
    where $(a_j,q)$ denote the corresponding prior hyperparameters.

    \item \textbf{Update $(\mathrm{U},\Sigma_e)$.}

    Conditional on the sampled state path $Z_{1:n}$, the observation equation becomes
    \[
    Y_t = \mathrm{U}Z_t + u_t,\qquad u_t\sim N_p(0,\Sigma_e),
    \]
    so $(\mathrm{U},\Sigma_e)$ are updated using the same posterior blocks as in the static model, with $Z_t$ playing the role of the latent factor scores.
\end{enumerate}

\paragraph{Recovered scaled factors.}
For interpretation in Section~6.3.3, we report the scaled factors
\[
f_t := \Lambda^{1/2}\eta_t = Z_t.
\]
Thus, posterior summaries of $\{f_t\}_{t=1}^n$ are obtained directly from the retained draws of the latent state path $Z_{1:n}$.

\section{Asymptotic properties of sample covariance}
For $j = 1,\ldots,p$, let $\hat{\lambda}_j$ and $\lambda_{0,j}$ denote the $j$-th eigenvalues of the sample covariance matrix and the true covariance matrix,
respectively.

Recall the assumptions (A1)–(A5):
\begin{enumerate}[label={A\arabic*.}]
    \item High-dimensional regime: $p/n \to \infty$.
    \item There exist positive constants $c_0$ and $C_0$ such that
    the eigenvalues $\lambda_{0,1},\ldots,\lambda_{0,p}$ satisfy
    \[
        \lambda_{0,1} > \cdots > \lambda_{0,k} > C_0
        > \lambda_{0,k+1} > \cdots > \lambda_{0,p} > c_0 .
    \]
    \item The $k$ spiked eigenvalues are separated by a constant
    $\delta_0>0$:
    \[
        \frac{\lambda_{0,j} - \lambda_{0,j+1}}{\lambda_{0,j}}
        \;\geq\; \delta_0,
        \quad \forall j = 1,\ldots,k .
    \]
    \item For $j=1,\ldots,k$, the quantities
    \[
        d_j \;=\; \frac{p}{n \lambda_{0,j}}
    \]
    are bounded above by a positive constant.
    \item For $j=1,\ldots,k$, the hyperparameter of prior $a_j$ are given by
    \[
        a_j = \dfrac{nt}{2(\hat{\lambda}_j - t)} + 2,
    \]
    for some $t\in [\hat{\lambda}_{k+1},\hat{\lambda}_n]$.
\end{enumerate}
Assumptions $A1-A4$ are adopted from \cite{wang2017asymptotics}. Assumption A5 is adopted from \cite{kim2025eigenstructure} to obtain an improved posterior convergence rate.

Suppose that Assumptions $A1-A5$ hold.  Define
$$
\bar{d}
=\frac{1}{p-k} \sum_{i = k+1}^p \lambda_{0,i}.
$$
Then, by Lemma S1.2 of \cite{kim2025eigenstructure},
the eigenvalues of the sample covariance matrix satisfy
\[
    \frac{\hat{\lambda}_j}{\lambda_{0,j}}
    =
    \begin{cases}
        1 + \bar{d} d_j
        + \alpha_j \lambda_{0,j}^{-1} \sqrt{\dfrac{p}{n}}
        + \beta_j,
        & j = 1,\ldots,k, \\[0.5em]
        \bar{d} d_j
        + \alpha_j \lambda_{0,j}^{-1} \sqrt{\dfrac{p}{n}},
        & j = k+1,\ldots,n,
    \end{cases}
\]
where $\alpha_j$ satisfy $|\alpha_j|\le C$ for some positive constant
$C$, and $\beta_j \lesssim n^{-1/2+\delta}$ for any small $\delta>0$
and all sufficiently large $n$.

\section{Asymptotic properties of eigenstructure of posterior}

Before proceeding to the proof, we define the following subsets of the Stiefel manifold. For convenience, we assume that $\Gamma_{ii} \ge 0$.
\begin{align*}
    A_\epsilon &= \Big\{\Gamma\in \mathbb{V}_{p,k}: \abs{\abs{\Gamma- \
    \begin{pmatrix}
        I_k \\ 0
    \end{pmatrix}
    }}_F<\epsilon \Big\}\\
    B_\epsilon &= \Big\{ \Gamma\in \mathbb{V}_{p,k}: \inf_{Q\in O_{k}}\abs{\abs{\Gamma- \
    \begin{pmatrix}
        Q \\ 0
    \end{pmatrix}
    }}_F<\epsilon  \Big\}
\end{align*}

\begin{lemma}\label{lem:union_covering}
For $\epsilon_1, \epsilon_0 > 0$, the following inclusion holds:
\begin{align*}
    \Big\{
        \Gamma \in \mathbb{V}_{p,k}:
        \inf_{Q \in O_{k}}
        \Big\lVert
            \Gamma -
            \begin{pmatrix}
                Q \\[0.2em] 0
            \end{pmatrix}
        \Big\rVert_F
        < \epsilon_1
    \Big\}
    \subset
    \bigcup_{i=1}^{N\big(O_k,\lVert\cdot\rVert_F,\epsilon_0\big)}
    \Big\{
        \Gamma \in \mathbb{V}_{p,k}:
        \Big\lVert
            \Gamma -
            \begin{pmatrix}
                S_i \\[0.2em] 0
            \end{pmatrix}
        \Big\rVert_F
        < \epsilon_1 + \epsilon_0
    \Big\},
\end{align*}
where $O_k := O_{k}$ denotes the orthogonal group in $\mathbb{R}^{k \times k}$ and
$N(O_k,\lVert\cdot\rVert_F,\epsilon_0)$ is the $\epsilon_0$-covering number of $O_k$
with respect to the Frobenius norm $\lVert\cdot\rVert_F$.
\end{lemma}

\begin{proof}
Let $S_1,\ldots,S_{N(O_k,\lVert\cdot\rVert_F,\epsilon_0)}$ form an $\epsilon_0$-covering of
$O_k$ with respect to the distance $\lVert\cdot\rVert_F$ for $\epsilon_0 > 0$; that is, for
every $Q \in O_k$ there exists an index $i(Q)$ such that
\[
    \big\lVert S_{i(Q)} - Q \big\rVert_F < \epsilon_0.
\]

Consider any $\Gamma \in \mathbb{V}_{p,k}$ satisfying
\[
    \inf_{Q \in O_{k}}
    \Big\lVert
        \Gamma -
        \begin{pmatrix}
            Q \\[0.2em] 0
        \end{pmatrix}
    \Big\rVert_F
    < \epsilon_1.
\]
Then there exists $Q_\Gamma \in O_{k}$ such that
\[
    \Big\lVert
        \Gamma -
        \begin{pmatrix}
            Q_\Gamma \\[0.2em] 0
        \end{pmatrix}
    \Big\rVert_F
    < \epsilon_1.
\]
By the covering property of $\{S_i\}$, there exists an index $i(\Gamma)$ with
\[
    \big\lVert S_{i(\Gamma)} - Q_\Gamma \big\rVert_F < \epsilon_0.
\]

Applying the triangle inequality, we obtain
\begin{align*}
    \Big\lVert
        \Gamma -
        \begin{pmatrix}
            S_{i(\Gamma)} \\[0.2em] 0
        \end{pmatrix}
    \Big\rVert_F
    &\le
    \Big\lVert
        \Gamma -
        \begin{pmatrix}
            Q_\Gamma \\[0.2em] 0
        \end{pmatrix}
    \Big\rVert_F
    +
    \Big\lVert
        \begin{pmatrix}
            Q_\Gamma \\[0.2em] 0
        \end{pmatrix}
        -
        \begin{pmatrix}
            S_{i(\Gamma)} \\[0.2em] 0
        \end{pmatrix}
    \Big\rVert_F \\
    &=
    \Big\lVert
        \Gamma -
        \begin{pmatrix}
            Q_\Gamma \\[0.2em] 0
        \end{pmatrix}
    \Big\rVert_F
    +
    \big\lVert Q_\Gamma - S_{i(\Gamma)} \big\rVert_F \\
    &<
    \epsilon_1 + \epsilon_0.
\end{align*}
Hence $\Gamma$ belongs to
\[
    \bigcup_{i=1}^{N(O_k,\lVert\cdot\rVert_F,\epsilon_0)}
    \Big\{
        \Gamma \in \mathbb{V}_{p,k}:
        \Big\lVert
            \Gamma -
            \begin{pmatrix}
                S_i \\[0.2em] 0
            \end{pmatrix}
        \Big\rVert_F
        < \epsilon_1 + \epsilon_0
    \Big\},
\]
which proves the claimed inclusion.
\end{proof}

\hfill\break

\begin{lemma}[Modification of Lemma S1.10 in \citealt{kim2025eigenstructure}]\label{lem:block_approx}
Let $\bar{\Gamma} \in O_p$ be an orthogonal matrix whose first $k$ columns are given by
$\Gamma \in \mathbb{V}_{p,k}$. If $\Gamma \in A_\eta$, then
\[
    \inf_{Q_2 \in O_{p-k}}
    \left\lVert
        \begin{pmatrix}
            I_k & 0 \\
            0   & Q_2
        \end{pmatrix}
        - \bar{\Gamma}
    \right\rVert_F
    < 2\eta,
    \qquad \text{for all } \eta \in (0,1).
\]
Further, if $\Gamma \in B_\eta$, then
\[
    \inf_{Q_1 \in O_{k},\, Q_2 \in O_{p-k}}
    \left\lVert
        \begin{pmatrix}
            Q_1 & 0 \\
            0   & Q_2
        \end{pmatrix}
        - \bar{\Gamma}
    \right\rVert_F
    < 2\eta,
    \qquad \text{for all } \eta \in (0,1).
\]
\end{lemma}

\begin{proof}
Write the block decomposition
\[
    \bar{\Gamma}
    =
    \begin{pmatrix}
        \Gamma_{11} & \Gamma_{12} \\
        \Gamma_{21} & \Gamma_{22}
    \end{pmatrix},
    \qquad
    \Gamma
    =
    \begin{pmatrix}
        \Gamma_{11} \\
        \Gamma_{21}
    \end{pmatrix},
\]
where $\Gamma_{11} \in \bbR^{k \times k}$ and $\Gamma_{22} \in \bbR^{(p-k)\times(p-k)}$.

Let $A = I_{p-k}$ and $\tilde{A} = I_{p-k} - \Gamma_{21}\Gamma_{21}^T$.  
For all $\Gamma \in A_\eta \cup B_\eta$, Lemma S1.9 of \cite{kim2025eigenstructure}
(or the Hoffman--Wielandt theorem; see \cite{stewart1990matrix}) implies
\[
    \sum_{i=1}^{p-k} (\tilde{\lambda}_i - 1)^2
    \;\le\;
    \big\lVert \Gamma_{12}^T \Gamma_{12} \big\rVert_F^2
    \;<\; \eta^4,
\]
where $\tilde{\lambda}_i$ denotes the $i$th eigenvalue of $\tilde{A}$.

Consider the singular value decomposition
\[
    \Gamma_{22} = U D V^T,
\]
where $U, V \in O_{p-k}$ and
\[
    D = \operatorname{diag}\big(\sqrt{\tilde{\lambda}_1},\ldots,\sqrt{\tilde{\lambda}_{p-k}}\big).
\]
By orthogonal invariance of the Frobenius norm, we obtain
\begin{align*}
    \inf_{Q \in O_{p-k}}
    \big\lVert \Gamma_{22} - Q \big\rVert_F^2
    &=
    \inf_{Q \in O_{p-k}}
    \big\lVert D - Q \big\rVert_F^2                                                   \\
    &= \sum_{i=1}^{p-k} \Big[\big(\sqrt{\tilde{\lambda}_i} - q_i\big)^2 + (1-q_i)^2\Big]\\
    &=
    \sum_{i=1}^{p-k} \big(\sqrt{\tilde{\lambda}_i} - 1\big)^2                         \\
    &\le
    \max_{1 \le i \le p-k}
    \frac{1}{\big(\sqrt{\tilde{\lambda}_i} + 1\big)^2}
    \sum_{i=1}^{p-k} (\tilde{\lambda}_i - 1)^2                                       \\
    &\le \eta^4,
\end{align*}
where $q_{i}$ denotes $(i,i)$ element of $Q$, and hence
\[
    \inf_{Q \in O_{p-k}}
    \big\lVert \Gamma_{22} - Q \big\rVert_F
    \;\le\; \eta^2.
\]

\medskip
\noindent\textbf{Case 1: $\Gamma \in A_\eta$.}
By definition of $A_\eta$, we have
\[
    \big\lVert
        \Gamma -
        \begin{pmatrix}
            I_k \\
            0
        \end{pmatrix}
    \big\rVert_F^2
    =
    \big\lVert I_k - \Gamma_{11} \big\rVert_F^2
    +
    \big\lVert \Gamma_{21} \big\rVert_F^2
    < \eta^2.
\]
For any $Q_2 \in O_{p-k}$,
\begin{align*}
    &\left\lVert
        \bar{\Gamma}
        -
        \begin{pmatrix}
            I_k & 0 \\
            0   & Q_2
        \end{pmatrix}
    \right\rVert_F^2 \\[0.4em]
    &=
    \big\lVert I_k - \Gamma_{11} \big\rVert_F^2
    +
    \big\lVert \Gamma_{21} \big\rVert_F^2
    +
    \big\lVert \Gamma_{12} \big\rVert_F^2
    +
    \big\lVert \Gamma_{22} - Q_2 \big\rVert_F^2                                      \\
    &=
    \big( \big\lVert I_k - \Gamma_{11} \big\rVert_F^2
        +
        \big\lVert \Gamma_{21} \big\rVert_F^2
    \big)
    +
    \big\lVert \Gamma_{22} - Q_2 \big\rVert_F^2
    +
    \big\lVert \Gamma_{12} \big\rVert_F^2                                            \\
    &\le
    2\big(
        \big\lVert I_k - \Gamma_{11} \big\rVert_F^2
        +
        \big\lVert \Gamma_{21} \big\rVert_F^2
    \big)
    +
    \big\lVert \Gamma_{22} - Q_2 \big\rVert_F^2                                      \\
    &\le
    2\eta^2 + \eta^4
    \;\le\; 4\eta^2,
\end{align*}
for all $\eta \in (0,1)$. Taking the infimum over $Q_2 \in O_{p-k}$ yields
\[
    \inf_{Q_2 \in O_{p-k}}
    \left\lVert
        \bar{\Gamma}
        -
        \begin{pmatrix}
            I_k & 0 \\
            0   & Q_2
        \end{pmatrix}
    \right\rVert_F^2
    \le 4\eta^2,
\]
and hence
\[
    \inf_{Q_2 \in O_{p-k}}
    \left\lVert
        \bar{\Gamma}
        -
        \begin{pmatrix}
            I_k & 0 \\
            0   & Q_2
        \end{pmatrix}
    \right\rVert_F
    < 2\eta.
\]

\medskip
\noindent\textbf{Case 2: $\Gamma \in B_\eta$.}
By definition of $B_\eta$,
\[
    \inf_{Q_1 \in O_{k}}
    \big\lVert
        \Gamma -
        \begin{pmatrix}
            Q_1 \\
            0
        \end{pmatrix}
    \big\rVert_F^2
    =
    \inf_{Q_1 \in O_{k}}
    \big\{
        \big\lVert Q_1 - \Gamma_{11} \big\rVert_F^2
        +
        \big\lVert \Gamma_{21} \big\rVert_F^2
    \big\}
    < \eta^2.
\]
For any $Q_1 \in O_{k}$ and $Q_2 \in O_{p-k}$,
\begin{align*}
    &\left\lVert
        \bar{\Gamma}
        -
        \begin{pmatrix}
            Q_1 & 0 \\
            0   & Q_2
        \end{pmatrix}
    \right\rVert_F^2 \\[0.4em]
    &=
    \big\lVert Q_1 - \Gamma_{11} \big\rVert_F^2
    +
    \big\lVert \Gamma_{21} \big\rVert_F^2
    +
    \big\lVert \Gamma_{12} \big\rVert_F^2
    +
    \big\lVert \Gamma_{22} - Q_2 \big\rVert_F^2                                      \\
    &=
    \big(
        \big\lVert Q_1 - \Gamma_{11} \big\rVert_F^2
        +
        \big\lVert \Gamma_{21} \big\rVert_F^2
    \big)
    +
    \big\lVert \Gamma_{22} - Q_2 \big\rVert_F^2
    +
    \big\lVert \Gamma_{12} \big\rVert_F^2                                            \\
    &\le
    2\big(
        \big\lVert Q_1 - \Gamma_{11} \big\rVert_F^2
        +
        \big\lVert \Gamma_{21} \big\rVert_F^2
    \big)
    +
    \big\lVert \Gamma_{22} - Q_2 \big\rVert_F^2                                      \\
    &\le
    2\eta^2 + \eta^4
    \;\le\; 4\eta^2,
\end{align*}
for all $\eta \in (0,1)$. Taking the infimum over $Q_1 \in O_{k}$ and $Q_2 \in O_{p-k}$ gives
\[
    \inf_{Q_1 \in O_{k},\, Q_2 \in O_{p-k}}
    \left\lVert
        \bar{\Gamma}
        -
        \begin{pmatrix}
            Q_1 & 0 \\
            0   & Q_2
        \end{pmatrix}
    \right\rVert_F
    < 2\eta,
\]
which completes the proof.
\end{proof}

\begin{lemma}[Probability of a subset of the Stiefel manifold]\label{lem:prob_stiefel}
Define the subset of the Stiefel manifold by
$$
A_\epsilon
=
\Bigg\{
    \Gamma \in \mathbb{V}_{p,k}:
    \left\lVert
        \Gamma -
        \begin{pmatrix}
            I_k \\[0.2em] 0
        \end{pmatrix}
    \right\rVert_F
    < \epsilon
\Bigg\}.
$$
Then, for $\epsilon \le 2\sqrt{k}$,
$$
\mathbb{P}(A_\epsilon)
\;\ge\;
\Bigg(
    \frac{c\sqrt{k}}{\epsilon}
\Bigg)^{-\,k\big(p - \frac{k}{2} - \frac{1}{2}\big)},
$$
where $\mathbb{P}$ denotes the uniform (Haar) probability measure on $\mathbb{V}_{p,k}$ and
$c > 0$ is a universal constant.
\end{lemma}

\begin{proof}
We first relate $A_\epsilon$ to an event defined on the full orthogonal group $O_p$.
Let $\bar{\Gamma} \in O_p$ be an orthogonal matrix whose first $k$ columns are equal
to $\Gamma \in \mathbb{V}_{p,k}$. Then,
\begin{align*}
    \mathbb{P}(A_\epsilon)
    &\;\ge\;
    \mathbb{P}\Bigg(
        \Big\{
            \bar{\Gamma} \in O_p:
            \inf_{Q_2 \in O_{p-k}}
            \left\lVert
                \begin{pmatrix}
                    I_k & 0 \\
                    0   & Q_2
                \end{pmatrix}
                - \bar{\Gamma}
            \right\rVert_F
            < \epsilon
        \Big\}
    \Bigg).
\end{align*}

By Lemma~S1.6 of \cite{kim2025eigenstructure}, there exists a constant $c > 0$ such that
for all $\epsilon \le 2\sqrt{k}$,
\begin{align*}
    \mathbb{P}\Bigg(
        \Big\{
            \bar{\Gamma} \in O_p:
            \inf_{Q_2 \in O_{p-k}}
            \left\lVert
                \begin{pmatrix}
                    I_k & 0 \\
                    0   & Q_2
                \end{pmatrix}
                - \bar{\Gamma}
            \right\rVert_F
            < \epsilon
        \Big\}
    \Bigg)
    \;\ge\;
    \Bigg(
        \frac{c\sqrt{k}}{\epsilon}
    \Bigg)^{-\,k\big(p - \frac{k}{2} - \frac{1}{2}\big)}.
\end{align*}
Combining the two displays yields the claimed lower bound for $\mathbb{P}(A_\epsilon)$.
\end{proof}

\hfill\break

\begin{lemma}\label{lem:upp_bound_ci_diff}
Under model \eqref{main-model:AFM}, assume that conditions $A1-A4$ holds. Let $\epsilon,\epsilon_1,\epsilon_3 > 0$ with $\epsilon_3 > \epsilon > (\epsilon_1 + \epsilon_3)/2$, and suppose that
$a_1 < \cdots < a_k$. Then
\[
    \frac{
        \displaystyle \sup_{\Gamma \in A_{\epsilon_3}} \prod_{i=1}^k \Big(\frac{c_i}{n}\Big)^{a_i + n/2 - 1}
    }{
        \displaystyle \inf_{\Gamma \in A_{\epsilon}^c \cap B_{\epsilon_1}}
        \prod_{i=1}^k \Big(\frac{c_i}{n}\Big)^{a_i + n/2 - 1}
    }
    \;\preccurlyeq\;
    \frac{
        \exp\!\Big(4n\epsilon_3^2 \,\frac{\hat{\lambda}_1}{\hat{\lambda}_k}\Big)
    }{
        \exp\!\Big(
            \frac{\eta}{2\sqrt{k}}
            \min_{l < k} (a_{l+1} - a_l)
            \cdot
            \min_{l < k} \log\Big(\frac{\hat{\lambda}_l}{\hat{\lambda}_{l+1}}\Big)
        \Big)
    },
\]
where $\eta = 2\epsilon - \epsilon_1 - \epsilon_3 > 0$ and $A_\epsilon$, $B_\epsilon$ are as defined in Lemma~\ref{lem:union_covering}.
\end{lemma}

\begin{proof}
We first analyze the set
\[
    A_{\epsilon}^c \cap B_{\epsilon_1}
    =
    \bigg\{
        \Gamma \in \mathbb{V}_{p,k}:
        \inf_{Q_1 \in O_{k}}
        \left\lVert
            \begin{pmatrix}
                Q_1 \\[0.2em] 0
            \end{pmatrix}
            - \Gamma
        \right\rVert_F
        < \epsilon_1,\quad
        \left\lVert
            \begin{pmatrix}
                I_k \\[0.2em] 0
            \end{pmatrix}
            - \Gamma
        \right\rVert_F
        \ge \epsilon
    \bigg\}.
\]

Let $\epsilon_4 = \epsilon_3 - \epsilon > 0$.  
Suppose that $\{R_1,\ldots,R_{M(O_{k},\lVert\cdot\rVert_F,\epsilon_4)}\}$ forms a maximal
$\epsilon_4$-packing of $O_{k}$ with respect to the Frobenius norm, where we index the
points so that $R_{M(O_{k},\lVert\cdot\rVert_F,\epsilon_4)} = I_k$.  
Such a maximal packing is also an $\epsilon_4$-covering, so for every $Q_1 \in O_{k}$ there
exists $R_i$ such that $\lVert R_i - Q_1 \rVert_F < \epsilon_4$. By the triangle
inequality,
\begin{align}\label{ineq:tri_lem3.1}
    \left\lVert
        \begin{pmatrix}
            R_i \\[0.2em] 0
        \end{pmatrix}
        - \Gamma
    \right\rVert_F
    &\le
    \left\lVert
        \begin{pmatrix}
            R_i \\[0.2em] 0
        \end{pmatrix}
        -
        \begin{pmatrix}
            Q_1 \\[0.2em] 0
        \end{pmatrix}
    \right\rVert_F
    +
    \left\lVert
        \begin{pmatrix}
            Q_1 \\[0.2em] 0
        \end{pmatrix}
        - \Gamma
    \right\rVert_F \notag\\
    &\le
    \epsilon_4
    +
    \left\lVert
        \begin{pmatrix}
            Q_1 \\[0.2em] 0
        \end{pmatrix}
        - \Gamma
    \right\rVert_F.
\end{align}

Hence, the following inclusion holds:
\begin{align}\label{eq:subseteq}
    A_{\epsilon}^c \cap B_{\epsilon_1}
    &\subseteq
    \bigcup_{i=1}^{M(O_{k},\lVert\cdot\rVert_F,\epsilon_4)}
    \bigg\{
        \Gamma \in \mathbb{V}_{p,k}:
        \inf_{Q_2 \in O_{p-k}}
        \left\lVert
            \begin{pmatrix}
                R_i \\[0.2em] 0
            \end{pmatrix}
            - \Gamma
        \right\rVert_F
        < \epsilon_1 + \epsilon_4, \notag\\
    &\hspace{4.0cm}
        \inf_{Q_2 \in O_{p-k}}
        \left\lVert
            \begin{pmatrix}
                I_k \\[0.2em] 0
            \end{pmatrix}
            - \Gamma
        \right\rVert_F
        \ge \epsilon
    \bigg\}.
\end{align}

Let $\{S_1,\ldots,S_m\}$ be the subcollection of $\{R_1,\ldots,R_{M(O_{k},\lVert\cdot\rVert_F,\epsilon_4)}\}$ satisfying
\begin{equation}\label{eq:cond_S}
    \bigg\{
        \Gamma \in \mathbb{V}_{p,k}:
        \inf_{Q_2 \in O_{p-k}}
        \left\lVert
            \begin{pmatrix}
                S_i \\[0.2em] 0
            \end{pmatrix}
            - \Gamma
        \right\rVert_F
        < \epsilon_1 + \epsilon_4,\quad
        \left\lVert
            \begin{pmatrix}
                I_k \\[0.2em] 0
            \end{pmatrix}
            - \Gamma
        \right\rVert_F
        \ge \epsilon
    \bigg\}
    \neq \varnothing.
\end{equation}
Because $\epsilon > \epsilon_1 + \epsilon_4$ by assumption ($\epsilon > (\epsilon_1 + \epsilon_3)/2$ and $\epsilon_4 = \epsilon_3 - \epsilon$), we necessarily have $I_k \notin \{S_1,\ldots,S_m\}$.

For any $\Gamma \in \mathbb{V}_{p,k}$ and $i = 1,\ldots,m$, the triangle inequality yields
\[
    \left\lVert
        \begin{pmatrix}
            S_i \\[0.2em] 0
        \end{pmatrix}
        -
        \begin{pmatrix}
            I_k \\[0.2em] 0
        \end{pmatrix}
    \right\rVert_F
    \ge
    \left\lVert
        \begin{pmatrix}
            I_k \\[0.2em] 0
        \end{pmatrix}
        - \Gamma
    \right\rVert_F
    -
    \left\lVert
        \begin{pmatrix}
            S_i \\[0.2em] 0
        \end{pmatrix}
        - \Gamma
    \right\rVert_F.
\]
By the definition of $S_i$, there exists $\Gamma \in \mathbb{V}_{p,k}$ such that
\eqref{eq:cond_S} holds, and evaluating the above inequality at such $\Gamma$ gives
\begin{align}\label{eq:S_diff_I}
    \lVert S_i - I_k \rVert_F
    >
    \epsilon - (\epsilon_1 + \epsilon_4),
    \qquad i = 1,\ldots,m.
\end{align}

Using the definition of $S_i$, the \eqref{eq:subseteq} can be written as
\begin{align*}
    \eqref{eq:subseteq}
    &=
    \bigcup_{i=1}^m
    \bigg\{
        \Gamma \in \mathbb{V}_{p,k}:
        \inf_{Q_2 \in O_{p-k}}
        \left\lVert
            \begin{pmatrix}
                S_i \\[0.2em] 0
            \end{pmatrix}
            - \Gamma
        \right\rVert_F
        < \epsilon_1 + \epsilon_4, \\
    &\hspace{4.0cm}
        \inf_{Q_2 \in O_{p-k}}
        \left\lVert
            \begin{pmatrix}
                I_k \\[0.2em] 0
            \end{pmatrix}
            - \Gamma
        \right\rVert_F
        \ge \epsilon
    \bigg\} \\
    &\subseteq
    \bigcup_{i=1}^m
    \bigg\{
        \Gamma \in \mathbb{V}_{p,k}:
        \inf_{Q_2 \in O_{p-k}}
        \left\lVert
            \begin{pmatrix}
                S_i \\[0.2em] 0
            \end{pmatrix}
            - \Gamma
        \right\rVert_F
        < \epsilon_1 + \epsilon_4
    \bigg\},
\end{align*}
for some positive $\epsilon_4$ satisfying $\epsilon > \epsilon_1 + \epsilon_4$. (The first equality follows from \eqref{eq:cond_S}.)

Define
\[
    C_{\epsilon,i}
    =
    \bigg\{
        \Gamma \in \mathbb{V}_{p,k}:
        \inf_{Q_2 \in O_{p-k}}
        \left\lVert
            \begin{pmatrix}
                S_i \\[0.2em] 0
            \end{pmatrix}
            - \Gamma
        \right\rVert_F
        < \epsilon
    \bigg\}.
\]
Then we obtain
\begin{align}\label{eq:ratio}
    \frac{
        \displaystyle \sup_{\Gamma \in A_{\epsilon_3}} \prod_{j=1}^k \Big(\frac{c_j}{n}\Big)^{a_j + n/2 - 1}
    }{
        \displaystyle \inf_{\Gamma \in A_{\epsilon}^c \cap B_{\epsilon_1}} \prod_{j=1}^k \Big(\frac{c_j}{n}\Big)^{a_j + n/2 - 1}
    }
    &\le
    \frac{
        \displaystyle \sup_{\Gamma \in A_{\epsilon_3}} \prod_{j=1}^k \Big(\frac{c_j}{n}\Big)^{a_j + n/2 - 1}
    }{
        \displaystyle \inf_{\Gamma \in \cup_{i=1}^m C_{\epsilon + \epsilon_4,i}}
        \prod_{j=1}^k \Big(\frac{c_j}{n}\Big)^{a_j + n/2 - 1}
    } \notag\\
    &=
    \max_{1 \le i \le m}
    \bigg[
        \frac{
            \displaystyle \sup_{\Gamma \in A_{\epsilon_3}} \prod_{j=1}^k \Big(\frac{c_j}{n}\Big)^{a_j + n/2 - 1}
        }{
            \displaystyle \inf_{\Gamma \in C_{\epsilon + \epsilon_4,i}}
            \prod_{j=1}^k \Big(\frac{c_j}{n}\Big)^{a_j + n/2 - 1}
        }
    \bigg] \notag\\
    &=
    \max_{1 \le i \le m}
    \bigg[
        \frac{
            \displaystyle \sup_{\Gamma \in A_{\epsilon_3}} \prod_{j=1}^k \Big(\frac{c_j}{n}\Big)^{a_j + n/2 - 1}
        }{
            \displaystyle \inf_{\Gamma \in C_{\epsilon_3,i}}
            \prod_{j=1}^k \Big(\frac{c_j}{n}\Big)^{a_j + n/2 - 1}
        }
    \bigg],
\end{align}
where the last equality uses $\epsilon_4 = \epsilon_3 - \epsilon$.

Next, let $\bar{\Gamma} \in O_p$ be an orthogonal matrix whose first $k$ columns are
given by $\Gamma \in \mathbb{V}_{p,k}$, and define
\[
    D_{S,\eta}
    =
    \bigg\{
        \bar{\Gamma} \in O_p:
        \inf_{Q_2 \in O_{p-k}}
        \left\lVert
            \begin{pmatrix}
                S & 0 \\
                0 & Q_2
            \end{pmatrix}
            - \bar{\Gamma}
        \right\rVert_F
        < \eta
    \bigg\}.
\]
By Lemma~\ref{lem:block_approx}, we can bound \eqref{eq:ratio} as
\begin{align*}
    \eqref{eq:ratio}
    &\le
    \max_{1 \le i \le m}
    \bigg[
        \frac{
            \displaystyle \sup_{\bar{\Gamma} \in D_{I_k,2\epsilon_3}}
            \prod_{j=1}^k \Big(\frac{c_j}{n}\Big)^{a_j + n/2 - 1}
        }{
            \displaystyle \inf_{\bar{\Gamma} \in D_{S_i,2\epsilon_3}}
            \prod_{j=1}^k \Big(\frac{c_j}{n}\Big)^{a_j + n/2 - 1}
        }
    \bigg] \\
    &\preccurlyeq
    \frac{
        \exp\!\Big(4n\epsilon_3^2 \,\frac{\hat{\lambda}_1}{\hat{\lambda}_k}\Big)
    }{
        \exp\!\Big(
            \frac{\eta}{2\sqrt{k}}
            \min_{l < k} (a_{l+1} - a_l)
            \cdot
            \min_{l < k} \log\Big(\frac{\hat{\lambda}_l}{\hat{\lambda}_{l+1}}\Big)
        \Big)
    },
\end{align*}
where
\[
    \eta
    =
    \epsilon - (\epsilon_1 + \epsilon_4)
    =
    2\epsilon - \epsilon_1 - \epsilon_3 > 0,
\]
and the final inequality follows from \eqref{eq:S_diff_I} and Lemma~S1.13 of
\cite{kim2025eigenstructure}. This completes the proof.
\end{proof}

\hfill\break

\hfill\break

\begin{lemma}[Chernoff bound for the Gamma distribution]\label{lem:chernoff_gamma}
Let $X \sim \mathrm{Gam}(\alpha,\beta)$ with shape $\alpha > 0$ and rate $\beta > 0$.
Then, for any $\delta > \alpha/\beta$,
\[
    \mathbb{P}(X > \delta)
    \;\le\;
    \exp\!\Big( -\big(\beta - \tfrac{\alpha}{\delta}\big)\delta \Big)
    \Big( \tfrac{\alpha}{\delta \beta} \Big)^{-\alpha}.
\]
\end{lemma}

\begin{proof}
By the Chernoff bound, for any $t > 0$,
\[
    \mathbb{P}(X > \delta)
    \;\le\;
    \frac{\mathbb{E}[e^{tX}]}{e^{t\delta}}
    =
    \exp\!\Big( \log \mathbb{E}[e^{tX}] - t\delta \Big).
\]
Since $X \sim \mathrm{Gam}(\alpha,\beta)$ with rate $\beta$, its moment generating function is
\[
    \mathbb{E}[e^{tX}]
    =
    \Big(1 - \frac{t}{\beta}\Big)^{-\alpha},
    \qquad t < \beta.
\]
Define
\[
    g(t)
    =
    \log \mathbb{E}[e^{tX}] - t\delta
    =
    -\alpha \log\!\Big(1 - \frac{t}{\beta}\Big) - t\delta,
    \qquad 0 < t < \beta.
\]
Then
\[
    \frac{d g(t)}{dt}
    =
    -\delta
    + \frac{\alpha}{\beta - t}.
\]
Setting $g'(t) = 0$ yields
\[
    -\delta + \frac{\alpha}{\beta - t} = 0
    \quad\Longrightarrow\quad
    t^* = \beta - \frac{\alpha}{\delta}.
\]
Under $\delta > \alpha/\beta$, we have $0 < t^* < \beta$, so $t^*$ is a valid minimizer in $(0,\beta)$.
Evaluating the bound at $t = t^*$ gives
\begin{align*}
    \mathbb{P}(X > \delta)
    &\le
    \frac{\mathbb{E}[e^{t^* X}]}{e^{t^* \delta}}
    =
    \exp\!\big(-t^* \delta\big)
    \Big(1 - \frac{t^*}{\beta}\Big)^{-\alpha}.
\end{align*}
Noting that
\[
    1 - \frac{t^*}{\beta}
    =
    1 - \frac{\beta - \alpha/\delta}{\beta}
    =
    \frac{\alpha}{\delta \beta},
\]
we obtain
\[
    \mathbb{P}(X > \delta)
    \le
    \exp\!\Big( -\big(\beta - \tfrac{\alpha}{\delta}\big)\delta \Big)
    \Big( \tfrac{\alpha}{\delta \beta} \Big)^{-\alpha},
\]
which proves the claim.
\end{proof}

\hfill\break

\begin{lemma}\label{lem:approx_int_post}
Under model \eqref{main-model:AFM}, assume that conditions $A1-A4$ holds. Suppose $\Gamma \in B_\epsilon$ and consider
\[
    \frac{
        \displaystyle \int_{0}^\infty
        (\lambda + \sigma_1^2)^{-n/2}\lambda^{-a}
        \exp\!\Big(
            -\frac{[\Gamma^T W \Gamma]_{ii}}{2(\lambda + \sigma_2^2)}
            -\frac{h}{2\lambda}
        \Big)\,d\lambda
    }{
        \displaystyle \int_{0}^\infty
        \lambda^{-n/2-a}
        \exp\!\Big(
            -\frac{[\Gamma^T W \Gamma]_{ii}}{2\lambda}
            -\frac{h}{2\lambda}
        \Big)\,d\lambda
    },
\]
where $\sigma_1^2,\sigma_2^2 \in [b_0,b_1]$ may depend on $(\Lambda,\Gamma)$, $h>0$, and $k$ is fixed.  
Assume that $\sqrt{n\hat{\lambda}_1} \prec \delta_n \prec \hat{\lambda}_k$ and $h \prec n$. Then
\[
    1 + O\!\Big(\frac{n}{\delta_n}\Big)
    \;\le\;
    \frac{
        \displaystyle \int_{0}^\infty
        (\lambda + \sigma_1^2)^{-n/2}\lambda^{-a}
        \exp\!\Big(
            -\frac{[\Gamma^T W \Gamma]_{ii}}{2(\lambda + \sigma_2^2)}
            -\frac{h}{2\lambda}
        \Big)\,d\lambda
    }{
        \displaystyle \int_{0}^\infty
        \lambda^{-n/2-a}
        \exp\!\Big(
            -\frac{[\Gamma^T W \Gamma]_{ii}}{2\lambda}
            -\frac{h}{2\lambda}
        \Big)\,d\lambda
    }
    \;\le\;
    1 + O\!\Bigg(
        \max\Big(\frac{a}{\delta_n},\frac{n\hat{\lambda}_1}{\delta_n^2}\Big)
        + \hat{\lambda}_1^{\,n}\exp\!\Big(-\frac{n\hat{\lambda}_k}{6\delta_n}\Big)
    \Bigg).
\]
\end{lemma}

\begin{proof}
We first derive a lower bound for $[\Gamma^T W \Gamma]_{ii}$ for $i=1,\ldots,k$.
Note that
\begin{align*}
    [\Gamma^T W \Gamma]_{ii}
    &= n \sum_{j=1}^p \Gamma_{ji}^2 \hat{\lambda}_j \\
    &\ge n \sum_{j=1}^k \Gamma_{ji}^2 \hat{\lambda}_j \\
    &\ge n\Big(\sum_{j=1}^k \Gamma_{ji}^2\Big)\hat{\lambda}_k.
\end{align*}
Since $\Gamma \in B_\epsilon$ implies
\[
    \sum_{j=k+1}^p \sum_{i=1}^k \Gamma_{ji}^2 < \epsilon^2,
\]
we have $\sum_{j=1}^k \Gamma_{ji}^2 \ge 1 - \epsilon^2$, and therefore
\[
    [\Gamma^T W \Gamma]_{ii}
    \ge (1-\epsilon^2)\,n\hat{\lambda}_k.
\]

Let
\begin{align*}
    f(\lambda,\sigma_1^2,\sigma_2^2)
    &=
    (\lambda + \sigma_1^2)^{-n/2}\lambda^{-a}
    \exp\!\Big(
        -\frac{b}{2(\lambda + \sigma_2^2)} - \frac{h}{2\lambda}
    \Big), \\
    g(\lambda,\sigma_1^2,\sigma_2^2)
    &=
    (\lambda + \sigma_1^2)^{-n/2-a}
    \exp\!\Big(
        -\frac{b + h}{2(\lambda + \sigma_2^2)}
    \Big),
\end{align*}
where $b = [\Gamma^T W \Gamma]_{ii} > (1-\epsilon^2)n\hat{\lambda}_k$.

For a diverging sequence $\delta_n$, we split the integral as
\[
    \int_0^\infty f(\lambda,\sigma_1^2,\sigma_2^2) \, d\lambda
    =
    \int_0^{\delta_n} f(\lambda,\sigma_1^2,\sigma_2^2)\,d\lambda
    +
    \int_{\delta_n}^\infty
        g(\lambda,\sigma_1^2,\sigma_2^2)
        \frac{f(\lambda,\sigma_1^2,\sigma_2^2)}{g(\lambda,\sigma_1^2,\sigma_2^2)}
    \,d\lambda.
\]

\medskip\noindent
\textbf{Step 1: Upper bound on the integral over $[0,\delta_n]$.}
We have
\begin{align*}
    \int_0^{\delta_n} f(\lambda,\sigma_1^2,\sigma_2^2)\,d\lambda
    &= \int_0^{\delta_n}
        (\lambda + \sigma_1^2)^{-n/2}\lambda^{-a}
        \exp\!\Big(
            -\frac{b}{2(\lambda + \sigma_2^2)} - \frac{h}{2\lambda}
        \Big)\,d\lambda \\
    &\le \sigma_1^{-n}\exp\!\Big(-\frac{b}{2(\delta_n + \sigma_2^2)}\Big)
        \int_0^{\delta_n} \lambda^{-a}\exp\!\Big(-\frac{h}{2\lambda}\Big)\,d\lambda \\
    &\le b_0^{-n/2}\exp\!\Big(-\frac{b}{2(\delta_n + b_1)}\Big)
        \int_0^{\infty} \lambda^{-a}\exp\!\Big(-\frac{h}{2\lambda}\Big)\,d\lambda \\
    &= b_0^{-n/2}\exp\!\Big(-\frac{b}{2(\delta_n + b_1)}\Big)
        \cdot \frac{\Gamma(a-1)}{(h/2)^{a-1}} \\
    &:= A.
\end{align*}

\medskip\noindent
\textbf{Step 2: Bounds on the ratio $f/g$ for $\lambda \ge \delta_n$.}
For $\lambda \in [\delta_n,\infty)$,
\begin{align*}
    \frac{f(\lambda,\sigma_1^2,\sigma_2^2)}{g(\lambda,\sigma_1^2,\sigma_2^2)}
    &= \Big(\frac{\lambda + \sigma_1^2}{\lambda}\Big)^{a}
       \exp\!\Big(-\frac{\sigma_2^2 h}{2\lambda(\lambda + \sigma_2^2)}\Big) \\
    &\in
    \bigg[
        \exp\!\Big(-\frac{\sigma_2^2 h}{2\delta_n(\delta_n + \sigma_2^2)}\Big),
        \Big(\frac{\delta_n + \sigma_1^2}{\delta_n}\Big)^{a}
    \bigg] \\
    &\subseteq
    \bigg[
        \exp\!\Big(-\frac{b_1 h}{2\delta_n(\delta_n + b_0)}\Big),
        \Big(\frac{\delta_n + b_1}{\delta_n}\Big)^{a}
    \bigg] \\
    &:= [B,C].
\end{align*}

\medskip\noindent
\textbf{Step 3: Upper bound for $\displaystyle\int_0^\infty f$.}
Using the bounds above,
\begin{align*}
   \int_0^\infty f(\lambda,\sigma_1^2,\sigma_2^2) \, d\lambda
   &= \int_0^{\delta_n} f(\lambda,\sigma_1^2,\sigma_2^2)\,d\lambda
      + \int_{\delta_n}^\infty
            g(\lambda,\sigma_1^2,\sigma_2^2)
            \frac{f(\lambda,\sigma_1^2,\sigma_2^2)}{g(\lambda,\sigma_1^2,\sigma_2^2)}
        \,d\lambda \\
   &\le A + C \int_{\delta_n}^\infty g(\lambda,\sigma_1^2,\sigma_2^2)\,d\lambda \\
   &\le A + C \int_{\delta_n}^\infty g(\lambda,b_0,b_1)\,d\lambda \\
   &= A + C \int_{\delta_n + b_0}^\infty g(\lambda,0,b_1-b_0)\,d\lambda \\
   &= A + C \int_{\delta_n + b_0}^\infty
        \exp\!\Big(\frac{(b_1-b_0)(b+h)}{2\lambda(\lambda + b_1-b_0)}\Big)
        g(\lambda,0,0)\,d\lambda \\
   &\le A + C\exp\!\Big(\frac{(b_1-b_0)(b+h)}{2(\delta_n + b_0)(\delta_n + b_1)}\Big)
        \int_{\delta_n + b_0}^\infty g(\lambda,0,0)\,d\lambda \\
   &\le A + C\exp\!\Big(\frac{(b_1-b_0)(b+h)}{2(\delta_n + b_0)(\delta_n + b_1)}\Big)
        \int_{0}^\infty g(\lambda,0,0)\,d\lambda.
\end{align*}

\medskip\noindent
\textbf{Step 4: Lower bound for $\displaystyle\int_0^\infty f$.}
Similarly,
\begin{align*}
    \int_0^\infty f(\lambda,\sigma_1^2,\sigma_2^2)\,d\lambda
    &=
    \int_0^{\delta_n} f(\lambda,\sigma_1^2,\sigma_2^2)\,d\lambda
    +
    \int_{\delta_n}^\infty
        g(\lambda,\sigma_1^2,\sigma_2^2)
        \frac{f(\lambda,\sigma_1^2,\sigma_2^2)}{g(\lambda,\sigma_1^2,\sigma_2^2)}
    \,d\lambda \\
    &\ge
    \int_{\delta_n}^\infty
        g(\lambda,\sigma_1^2,\sigma_2^2)
        \frac{f(\lambda,\sigma_1^2,\sigma_2^2)}{g(\lambda,\sigma_1^2,\sigma_2^2)}
    \,d\lambda \\
    &\ge
    B \int_{\delta_n}^\infty g(\lambda,\sigma_1^2,\sigma_2^2)\,d\lambda \\
    &\ge
    B \int_{\delta_n}^\infty g(\lambda,b_1,b_0)\,d\lambda \\
    &=
    B \int_{\delta_n + b_0}^\infty g(\lambda,b_1-b_0,0)\,d\lambda \\
    &=
    B \int_{\delta_n + b_0}^\infty
        \Big(\frac{\lambda}{\lambda + b_1-b_0}\Big)^{n/2 + a}
        g(\lambda,0,0)\,d\lambda \\
    &\ge
    B \Big(\frac{\delta_n + b_0}{\delta_n + b_1}\Big)^{n/2 + a}
      \int_{\delta_n + b_0}^\infty g(\lambda,0,0)\,d\lambda.
\end{align*}

\medskip\noindent
\textbf{Step 5: Ratio with respect to $\displaystyle\int_0^\infty g(\lambda,0,0)\,d\lambda$.}
Collecting these bounds, we obtain
\begin{align*}
    \frac{\int_0^\infty f(\lambda,\sigma_1^2,\sigma_2^2)\,d\lambda}
         {\int_0^\infty g(\lambda,0,0)\,d\lambda}
    \in
    &\bigg[
        B \Big(\frac{\delta_n + b_0}{\delta_n + b_1}\Big)^{n/2 + a}
        \frac{\int_{\delta_n + b_0}^\infty g(\lambda,0,0)\,d\lambda}
             {\int_0^\infty g(\lambda,0,0)\,d\lambda}, \\
    &\hspace{1.5cm}
        \frac{A}{\int_0^\infty g(\lambda,0,0)\,d\lambda}
        +
        C\exp\!\Big(\frac{(b_1-b_0)(b+h)}{2(\delta_n + b_0)(\delta_n + b_1)}\Big)
    \bigg].
\end{align*}

\medskip\noindent
\textbf{Step 6: Lower tail of the Gamma distribution.}
Let $X \sim \mathrm{Gam}\big(\alpha,\beta\big)$ with
$\alpha = a + n/2 - 1$ and $\beta = (b+h)/2$.  
Using the change of variables $\lambda \mapsto 1/X$, one can check that
\[
    \frac{\int_{\delta_n + b_0}^\infty g(\lambda,0,0)\,d\lambda}
         {\int_0^\infty g(\lambda,0,0)\,d\lambda}
    =
    \mathbb{P}\Big(\frac{1}{X} > \delta_n + b_0\Big)
    =
    1 - \mathbb{P}\Big(X \ge \frac{1}{\delta_n + b_0}\Big).
\]
By Lemma~\ref{lem:chernoff_gamma}, we have
\[
    \mathbb{P}\Big(X \ge \frac{1}{\delta_n + b_0}\Big)
    \le
    D
\]
for a term $D$ of the form
\[
    D
    =
    \exp\!\Big(-(\beta - \alpha(\delta_n + b_0))\frac{1}{\delta_n + b_0}\Big)
    \Big(\frac{\alpha(\delta_n + b_0)}{\beta}\Big)^{-\alpha}.
\]

\medskip\noindent
\textbf{Step 7: Bounding $A / \int_0^\infty g(\lambda,0,0)\,d\lambda$.}
Note that
\[
    \int_0^\infty g(\lambda,0,0)\,d\lambda
    = \frac{\Gamma(a + n/2 - 1)}{\big((b+h)/2\big)^{a + n/2 - 1}}.
\]
Thus,
\begin{align*}
    \frac{A}{\int_0^\infty g(\lambda,0,0)\,d\lambda}
    &=
    A \cdot
    \frac{
        \big((b+h)/2\big)^{a + n/2 - 1}
    }{
        \Gamma(a + n/2 - 1)
    } \\
    &\le
    A \cdot
    \frac{
        \big((b+h)/2\big)^{a + n/2 - 1}
    }{
        \sqrt{2e}\,\big((a + n/2 - 1/2)/e\big)^{a + n/2 - 1/2}
    } \\
    &\le
    \frac{A}{\sqrt{2e}}
    \Big(
        \frac{e(b+h)}{n + 2a - 1}
    \Big)^{a + n/2 - 1},
\end{align*}
where the first inequality uses Theorem~1.5 of \cite{batir2008inequalities}.
Substituting the expression for $A$, we arrive at
\[
    \frac{A}{\int_0^\infty g(\lambda,0,0)\,d\lambda}
    \preccurlyeq
    \hat{\lambda}_1^{\,n}\exp\!\Big(-\frac{b}{2(\delta_n + b_1)}\Big),
\]
which will be negligible under our assumptions.

\medskip\noindent
\textbf{Step 8: Collecting rates.}
Combining the above bounds, and using the standard inequalities
$\exp(t) \ge 1 + t$ and $\exp(t) < 1 + 2t$ for sufficiently small $t>0$, we obtain
\[
    \frac{\int_0^\infty f(\lambda,\sigma_1^2,\sigma_2^2)\,d\lambda}
         {\int_0^\infty g(\lambda,0,0)\,d\lambda}
    =
    1 + O\!\Big(\frac{n}{\delta_n}\Big)
\]
from the lower bound, and
\[
    \frac{\int_0^\infty f(\lambda,\sigma_1^2,\sigma_2^2)\,d\lambda}
         {\int_0^\infty g(\lambda,0,0)\,d\lambda}
    \le
    1 + O\!\Big(
        \max\Big(\frac{a}{\delta_n},\frac{b}{\delta_n^2}\Big)
        + \hat{\lambda}_1^{\,n}\exp\!\Big(-\frac{b}{3\delta_n}\Big)
    \Big)
\]
from the upper bound.

Finally, since $b = [\Gamma^T W \Gamma]_{ii}$ satisfies
$$
    \frac{n}{2}\hat{\lambda}_k \;\le\; b \;\le\; n\hat{\lambda}_1,
$$
we obtain
$$
   1 + O\!\Big(\frac{n}{\delta_n}\Big)
   \;\le\;
   \frac{\displaystyle\int_0^\infty
          f(\lambda,\sigma_1^2,\sigma_2^2)\,d\lambda}
        {\displaystyle\int_0^\infty
          g(\lambda,0,0)\,d\lambda}
   \;\le\;
   1 + O\!\Bigg(
        \max\Big(\frac{a}{\delta_n},
                  \frac{n\hat{\lambda}_1}{\delta_n^2}\Big)
        + \hat{\lambda}_1^{\,n}
          \exp\!\Big(-\frac{n\hat{\lambda}_k}{6\delta_n}\Big)
   \Bigg).
$$
Moreover, under the rate condition
$$
  \sqrt{n\,\hat{\lambda}_1} \;\prec\; \delta_n \;\prec\; \hat{\lambda}_k,
$$
both the lower and upper bounds converge to $1$ as $n \to \infty$,
so the ratio is asymptotically sandwiched at $1$. This yields the desired
result.

\end{proof}

\hfill\break

\begin{lemma}\label{lem:cond_covariance}
Let $X \sim N(0,\Sigma)$, where $\Sigma$ is a $p \times p$ positive definite matrix.
Let $C \subset \mathbb{R}^p$ be a closed, convex, and origin-symmetric set
(i.e., $x \in C$ implies $-x \in C$) with non-empty interior. Then
\[
    \mathrm{Cov}(X \mid X \in C) \prec \Sigma.
\]
\end{lemma}

\begin{proof}
Let $Y = \Sigma^{-1/2} X$, so that $Y \sim N(0,I_p)$, and define the transformed set
\[
    C_Y = \Sigma^{-1/2} C
    := \{\Sigma^{-1/2} c : c \in C\}.
\]
Since $C$ is closed, convex, and origin-symmetric, the same properties hold for $C_Y$.

The conditional density of $Y$ given $Y \in C_Y$ is proportional to
\[
    f(y) \propto \exp\Big(-\frac{1}{2}\|y\|^2\Big) \mathbf{1}_{C_Y}(y),
\]
where $\mathbf{1}_{C_Y}(y)$ is the indicator function of $C_Y$. We can rewrite this density in the exponential form $f(y) \propto \exp(-V(y))$, where the potential function $V(y)$ is given by
\[
    V(y) = \frac{1}{2}\|y\|^2 + I_{C_Y}^\infty(y).
\]
Here, $I_{C_Y}^\infty(y)$ is the convex indicator function taking $0$ for $y \in C_Y$ and $\infty$ otherwise. Because $C_Y$ is a convex set, $V(y)$ is a strongly convex function. On the interior of $C_Y$, its Hessian is exactly $\nabla^2 V(y) = I_p$.

To bound the conditional covariance, we apply Theorem 4.1 of \citet{brascamp1976extensions} (the Brascamp-Lieb inequality). The theorem states that for a probability measure with density proportional to $\exp(-V(y))$, where $V$ is strictly convex, the variance of any differentiable function $g(y)$ is bounded by
\[
    \mathrm{Var}(g(Y)) \leq \mathbb{E}\Big[ \nabla g(Y)^T (\nabla^2 V(Y))^{-1} \nabla g(Y) \Big].
\]

We apply this theorem by considering a linear projection $g(y) = \nu^T y$ for an arbitrary unit vector $\nu \in \mathbb{R}^p$ ($\|\nu\| = 1$). For this function, the gradient is simply $\nabla g(y) = \nu$. Substituting $\nabla g(y) = \nu$ and $\nabla^2 V(y) = I_p$ into the Brascamp-Lieb inequality yields
\[
    \mathrm{Var}(\nu^T Y \mid Y \in C_Y) \leq \mathbb{E}\Big[ \nu^T (I_p)^{-1} \nu \mid Y \in C_Y \Big] = \nu^T \nu = 1.
\]
Since this upper bound holds for every unit vector $\nu$, it implies that in terms of the positive semi-definite order,
\[
    \mathrm{Cov}(Y \mid Y \in C_Y) \preceq I_p.
\]
Furthermore, because $C_Y$ is a proper subset of $\mathbb{R}^p$ that truncates the tails of the normal distribution, the variance is strictly reduced in at least some directions, leading to the strict inequality $\mathrm{Cov}(Y \mid Y \in C_Y) \prec I_p$.

Finally, mapping back to $X = \Sigma^{1/2} Y$, we obtain
\begin{align*}
    \mathrm{Cov}(X \mid X \in C)
    &= \Sigma^{1/2}\,\mathrm{Cov}(Y \mid Y \in C_Y)\,\Sigma^{1/2} \\
    &\prec \Sigma^{1/2} I_p \Sigma^{1/2} = \Sigma.
\end{align*}
This completes the proof.
\end{proof}
\hfill\break

\section{Lemma of Posterior Expectation}

Consider the unnormalized posterior distribution
\begin{align}\label{eq:post_ver1}
    \pi(\Gamma,\Lambda,\Sigma_u \mid \mathbf{Y})
    &= \lvert \Gamma \Lambda \Gamma^T + \Sigma_u \rvert^{-n/2}
       \,\mathrm{etr}\!\Big[-\frac{1}{2}\big(\Gamma \Lambda \Gamma^T + \Sigma_u\big)^{-1} W\Big] \\
    &\quad\times \prod_{i=1}^k \lambda_i^{-a_i}
       \exp\!\Big(-\frac{h}{2\lambda_i}\Big)\,
       \pi(\Sigma_u). \nonumber
\end{align}

Let $\bar{\Gamma} = (\Gamma,\Gamma_2) \in O_p$. Then the determinant term can be written as
\begin{align*}
    \big\lvert \Gamma \Lambda \Gamma^T + \Sigma_u \big\rvert
    &= \bigg\lvert
        \begin{pmatrix}
            \Lambda & 0 \\[0.2em]
            0       & 0
        \end{pmatrix}
        +
        \bar{\Gamma}^T \Sigma_u \bar{\Gamma}
    \bigg\rvert \\
    &= \bigg\lvert
        \begin{pmatrix}
            \Lambda + \Gamma^T \Sigma_u \Gamma
            & \Gamma^T \Sigma_u \Gamma_2 \\
            \Gamma_2^T \Sigma_u \Gamma
            & \Gamma_2^T \Sigma_u \Gamma_2
        \end{pmatrix}
    \bigg\rvert \\
    &= \big\lvert (\Lambda + \bar{\Sigma}_{11})
        - \bar{\Sigma}_{12}\bar{\Sigma}_{22}^{-1}\bar{\Sigma}_{21}
       \big\rvert
       \cdot \big\lvert \bar{\Sigma}_{22} \big\rvert,
\end{align*}
where
\[
    \bar{\Sigma}_{11} = \Gamma^T \Sigma_u \Gamma,\quad
    \bar{\Sigma}_{12} = \Gamma^T \Sigma_u \Gamma_2,\quad
    \bar{\Sigma}_{21} = \Gamma_2^T \Sigma_u \Gamma,\quad
    \bar{\Sigma}_{22} = \Gamma_2^T \Sigma_u \Gamma_2.
\]

Similarly, define
\[
    \bar{W}_{11} = \Gamma^T W \Gamma,\quad
    \bar{W}_{12} = \Gamma^T W \Gamma_2,\quad
    \bar{W}_{21} = \Gamma_2^T W \Gamma,\quad
    \bar{W}_{22} = \Gamma_2^T W \Gamma_2.
\]
Then we can write
\begin{align*}
    \mathrm{tr}\big[(\Gamma\Lambda\Gamma^T + \Sigma_u)^{-1} W\big]
    &= \mathrm{tr}\bigg[
        \Big(
            \begin{pmatrix}
                \Lambda & 0 \\[0.2em]
                0       & 0
            \end{pmatrix}
            +
            \bar{\Gamma}^T \Sigma_u \bar{\Gamma}
        \Big)^{-1}
        \bar{\Gamma}^T W \bar{\Gamma}
    \bigg] \\
    &= \mathrm{tr}\bigg[
        \begin{pmatrix}
            \Lambda + \bar{\Sigma}_{11} & \bar{\Sigma}_{12} \\
            \bar{\Sigma}_{21}           & \bar{\Sigma}_{22}
        \end{pmatrix}^{-1}
        \begin{pmatrix}
            \bar{W}_{11} & \bar{W}_{12} \\
            \bar{W}_{21} & \bar{W}_{22}
        \end{pmatrix}
    \bigg].
\end{align*}
Using the block matrix inversion formula, we obtain
\[
    \begin{pmatrix}
        \Lambda + \bar{\Sigma}_{11} & \bar{\Sigma}_{12} \\
        \bar{\Sigma}_{21}           & \bar{\Sigma}_{22}
    \end{pmatrix}^{-1}
    =
    \begin{pmatrix}
        T^{-1} & -T^{-1}\bar{\Sigma}_{12}\bar{\Sigma}_{22}^{-1} \\
        -\bar{\Sigma}_{22}^{-1}\bar{\Sigma}_{21} T^{-1}
        & \bar{\Sigma}_{22}^{-1}
          + \bar{\Sigma}_{22}^{-1}\bar{\Sigma}_{21}
            T^{-1}\bar{\Sigma}_{12}\bar{\Sigma}_{22}^{-1}
    \end{pmatrix},
\]
where
\[
    T
    = \Lambda + \bar{\Sigma}_{11}
      - \bar{\Sigma}_{12}\bar{\Sigma}_{22}^{-1}\bar{\Sigma}_{21}.
\]
Therefore,
\begin{align*}
    \mathrm{tr}\big[(\Gamma\Lambda\Gamma^T + \Sigma_u)^{-1}W\big]
    &=
    \mathrm{tr}\Big[
        T^{-1}\big(
            \bar{W}_{11}
            - \bar{\Sigma}_{12}\bar{\Sigma}_{22}^{-1}\bar{W}_{21}
            - \bar{W}_{12}\bar{\Sigma}_{22}^{-1}\bar{\Sigma}_{21}
            + \bar{\Sigma}_{12}\bar{\Sigma}_{22}^{-1}
              \bar{W}_{22}\bar{\Sigma}_{22}^{-1}\bar{\Sigma}_{21}
        \big)
    \Big] \\
    &\quad + \mathrm{tr}\big[\bar{\Sigma}_{22}^{-1}\bar{W}_{22}\big].
\end{align*}

Hence the unnormalized posterior can be rewritten as
\begin{align}\label{eq:post_ver2}
    \pi(\Gamma,\Lambda,\Sigma_u \mid \mathbf{Y})
    &= \lvert \Lambda + A \rvert^{-n/2}
       \prod_{i=1}^k \lambda_i^{-a_i}
       \exp\!\Big(-\frac{h}{2\lambda_i}\Big)
       \,\mathrm{etr}\!\Big[-\frac{1}{2}(\Lambda + A)^{-1} B\Big] \\
    &\quad\times
       \lvert \bar{\Sigma}_{22} \rvert^{-n/2}
       \,\mathrm{etr}\!\Big[-\tfrac{1}{2}\bar{\Sigma}_{22}^{-1}\bar{W}_{22}\Big]
       \pi(\Sigma_u), \nonumber
\end{align}
where
\[
    A
    = \bar{\Sigma}_{11}
      - \bar{\Sigma}_{12}\bar{\Sigma}_{22}^{-1}\bar{\Sigma}_{21},
\]
and
\[
    B
    = \bar{W}_{11}
      - \bar{\Sigma}_{12}\bar{\Sigma}_{22}^{-1}\bar{W}_{21}
      - \bar{W}_{12}\bar{\Sigma}_{22}^{-1}\bar{\Sigma}_{21}
      + \bar{\Sigma}_{12}\bar{\Sigma}_{22}^{-1}
        \bar{W}_{22}\bar{\Sigma}_{22}^{-1}\bar{\Sigma}_{21}.
\]

Since we assume $\lambda(\Sigma_u) \in [b_0,b_1]$, the Schur complement $A$ satisfies
\[
    b_0 I_k \;\preccurlyeq\; A \;\preccurlyeq\; b_1 I_k.
\]

Define, for $d_0,d_1>0$,
\begin{align*}
    f(\Gamma,\Lambda,\Sigma_u; d_0,d_1)
    &=
    \prod_{i=1}^k (\lambda_i + d_0)^{-n/2}\lambda_i^{-a_i}
    \exp\!\Big(-\frac{B_{ii}}{2(\lambda_i + d_1)} - \frac{h}{2\lambda_i}\Big) \\
    &\quad\times
    \lvert \bar{\Sigma}_{22} \rvert^{-n/2}
    \,\mathrm{etr}\!\Big[-\tfrac{1}{2}\bar{\Sigma}_{22}^{-1}\bar{W}_{22}\Big]
    \pi(\Sigma_u),
\end{align*}
where $B_{ii}$ denotes the $(i,i)$-th diagonal element of $B$.

By the bounds on $A$ and monotonicity of the determinant and trace in the positive-definite order, we obtain
\begin{align}\label{ineq:bound_post}
    f(\Gamma,\Lambda,\Sigma_u; b_1, b_0)
    \;\le\;
    \pi(\Gamma,\Lambda,\Sigma_u \mid \mathbf{Y})
    \;\le\;
    f(\Gamma,\Lambda,\Sigma_u; b_0, b_1).
\end{align}

For a general nonnegative measurable function $h(\Gamma,\Lambda)$, the posterior expectation can be written as
\begin{align}
    \mathbb{E}\big[h(\Gamma,\Lambda)\mid \mathbf{Y}\big]
    &= \frac{\displaystyle
        \int\!\!\int\!\!\int h(\Gamma,\Lambda)\,
        \pi(\Gamma,\Lambda,\Sigma_u \mid \mathbf{Y})
        \,(d\Gamma)(d\Lambda)(d\Sigma_u)
       }{\displaystyle
        \int\!\!\int\!\!\int
        \pi(\Gamma,\Lambda,\Sigma_u \mid \mathbf{Y})
        \,(d\Gamma)(d\Lambda)(d\Sigma_u)
       } \nonumber\\
    &= \frac{\displaystyle
        \int\!\!\int\!\!\int_{A_\epsilon} h(\Gamma,\Lambda)\,
        \pi(\Gamma,\Lambda,\Sigma_u \mid \mathbf{Y})
        \,(d\Gamma)(d\Lambda)(d\Sigma_u)
       }{\displaystyle
        \int\!\!\int\!\!\int
        \pi(\Gamma,\Lambda,\Sigma_u \mid \mathbf{Y})
        \,(d\Gamma)(d\Lambda)(d\Sigma_u)
       } \label{eq:post_exp_part1}\\
    &\quad+
       \frac{\displaystyle
        \int\!\!\int\!\!\int_{A_\epsilon^c \cap B_{\epsilon_1}^c} h(\Gamma,\Lambda)\,
        \pi(\Gamma,\Lambda,\Sigma_u \mid \mathbf{Y})
        \,(d\Gamma)(d\Lambda)(d\Sigma_u)
       }{\displaystyle
        \int\!\!\int\!\!\int
        \pi(\Gamma,\Lambda,\Sigma_u \mid \mathbf{Y})
        \,(d\Gamma)(d\Lambda)(d\Sigma_u)
       } \label{eq:post_exp_part2}\\
    &\quad+
       \frac{\displaystyle
        \int\!\!\int\!\!\int_{A_\epsilon^c \cap B_{\epsilon_1}} h(\Gamma,\Lambda)\,
        \pi(\Gamma,\Lambda,\Sigma_u \mid \mathbf{Y})
        \,(d\Gamma)(d\Lambda)(d\Sigma_u)
       }{\displaystyle
        \int\!\!\int\!\!\int
        \pi(\Gamma,\Lambda,\Sigma_u \mid \mathbf{Y})
        \,(d\Gamma)(d\Lambda)(d\Sigma_u)
       }. \label{eq:post_exp_part3}
\end{align}

\begin{lemma}\label{lem:post_exp_second}
Under model \eqref{main-model:AFM}, assume that conditions $A1-A4$ holds.
Suppose that the ratio $\lambda_{0,1}/\lambda_{0,k}$ is bounded by a positive constant.
Then
\[
    \mathbb{E}\Big[\lambda_i\,\mathbf{1}\{\Gamma \in A_\epsilon^c \cap B_{\epsilon_1}^c\}
    \,\Big\vert\, \mathbf{Y} \Big]
    \;\preccurlyeq\;
    \exp\!\Big(- \frac{n\epsilon_1^2 \hat{\lambda}_k}{16 b_1}\Big),
\]
where $n\epsilon_1^2 \lambda_{0,k} \succ p$.
\end{lemma}

\begin{proof}
We begin with
\begin{align}\label{eq:post_exp_part2_1}
    &\mathbb{E}\Big[\lambda_i \mathbf{1}\{\Gamma \in A_\epsilon^c \cap B_{\epsilon_1}^c\}
    \,\Big\vert\, \mathbf{Y}\Big]\\
    &= \frac{\displaystyle
        \int_{A_\epsilon^c \cap B_{\epsilon_1}^c}
        \lambda_i\,\pi(\Gamma,\Lambda,\Sigma_u \mid \mathbf{Y})
        \,(d\Gamma)(d\Lambda)(d\Sigma_u)
    }{\displaystyle
        \int
        \pi(\Gamma,\Lambda,\Sigma_u \mid \mathbf{Y})
        \,(d\Gamma)(d\Lambda)(d\Sigma_u)
    } \notag\\
    &\le
    \frac{\displaystyle
        \int_{A_\epsilon^c \cap B_{\epsilon_1}^c}
        \lambda_i
        \prod_{j=1}^k (\lambda_j + b_0)^{-n/2}\lambda_j^{-a_j}
        \exp\!\Big(
            -\frac{B_{jj}}{2(\lambda_j + b_1)}
            -\frac{h}{2\lambda_j}
        \Big)
        \lvert\bar{\Sigma}_{22}\rvert^{-n/2}
        \,\mathrm{etr}\!\Big(-\tfrac{1}{2}\bar{\Sigma}_{22}^{-1}\bar{W}_{22}\Big)
        \pi(\Sigma_u)
        \,(d\Gamma)(d\Lambda)(d\Sigma_u)
    }{\displaystyle
        \int_{A_{\epsilon_2}}
        \prod_{j=1}^k (\lambda_j + b_1)^{-n/2}\lambda_j^{-a_j}
        \exp\!\Big(
            -\frac{B_{jj}}{2(\lambda_j + b_0)}
            -\frac{h}{2\lambda_j}
        \Big)
        \lvert\bar{\Sigma}_{22}\rvert^{-n/2}
        \,\mathrm{etr}\!\Big(-\tfrac{1}{2}\bar{\Sigma}_{22}^{-1}\bar{W}_{22}\Big)
        \pi(\Sigma_u)
        \,(d\Gamma)(d\Lambda)(d\Sigma_u)
    },
\end{align}
where the last inequality follows from \eqref{ineq:bound_post}.

Using the fact that $B_{jj} \in [0,2\lambda_{0,j}]$, we obtain
\begin{align}\label{eq:post_exp_part2_2}
    \eqref{eq:post_exp_part2_1}
    &\le
    \frac{\displaystyle
        \int_{A_\epsilon^c \cap B_{\epsilon_1}^c}
        \lambda_i
        \prod_{j=1}^k (\lambda_j + b_0)^{-n/2}\lambda_j^{-a_j}
        \exp\!\Big(-\frac{h}{2\lambda_j}\Big)
        \lvert\bar{\Sigma}_{22}\rvert^{-n/2}
        \,\mathrm{etr}\!\Big(-\tfrac{1}{2}\bar{\Sigma}_{22}^{-1}\bar{W}_{22}\Big)
        \pi(\Sigma_u)
        \,(d\Gamma)(d\Lambda)(d\Sigma_u)
    }{\displaystyle
        \int_{A_{\epsilon_2}}
        \prod_{j=1}^k (\lambda_j + b_1)^{-n/2}\lambda_j^{-a_j}
        \exp\!\Big(
            -\frac{2\lambda_{0,j}}{2(\lambda_j + b_0)}
            -\frac{h}{2\lambda_j}
        \Big)
        \lvert\bar{\Sigma}_{22}\rvert^{-n/2}
        \,\mathrm{etr}\!\Big(-\tfrac{1}{2}\bar{\Sigma}_{22}^{-1}\bar{W}_{22}\Big)
        \pi(\Sigma_u)
        \,(d\Gamma)(d\Lambda)(d\Sigma_u)
    } \notag\\
    &\le
    \frac{\displaystyle
        \int_{A_\epsilon^c \cap B_{\epsilon_1}^c}
        \lambda_i
        \prod_{j=1}^k (\lambda_j + b_0)^{-n/2}\lambda_j^{-a_j}
        \exp\!\Big(-\frac{h}{2\lambda_j}\Big)
        \,\mathrm{etr}\!\Big(-\tfrac{1}{2b_1}\bar{W}_{22}\Big)
        \lvert\bar{\Sigma}_{22}\rvert^{-n/2}\pi(\Sigma_u)
        \,(d\Gamma)(d\Lambda)(d\Sigma_u)
    }{\displaystyle
        \int_{A_{\epsilon_2}}
        \prod_{j=1}^k (\lambda_j + b_1)^{-n/2}\lambda_j^{-a_j}
        \exp\!\Big(
            -\frac{2\lambda_{0,j}}{2(\lambda_j + b_0)}
            -\frac{h}{2\lambda_j}
        \Big)
        \,\mathrm{etr}\!\Big(-\tfrac{1}{2b_0}\bar{W}_{22}\Big)
        \lvert\bar{\Sigma}_{22}\rvert^{-n/2}\pi(\Sigma_u)
        \,(d\Gamma)(d\Lambda)(d\Sigma_u)
    },
\end{align}
where the last inequality uses $\lambda(\bar{\Sigma}_{22}) \in [b_0,b_1]$.

Next, perform the change of variables
\[
    \Omega = \bar{\Gamma}^T \Sigma_u \bar{\Gamma}, \qquad
    \Omega_{22} = \bar{\Sigma}_{22},
\]
and use the rotation invariance of the prior $\pi(\cdot)$ in $\Sigma_u$ to rewrite
\eqref{eq:post_exp_part2_2} as
\begin{align}\label{eq:post_exp_part2_3}
    \eqref{eq:post_exp_part2_2}
    &=
    \frac{\displaystyle
        \int_{A_\epsilon^c \cap B_{\epsilon_1}^c}
        \lambda_i
        \prod_{j=1}^k (\lambda_j + b_0)^{-n/2}\lambda_j^{-a_j}
        \exp\!\Big(-\frac{h}{2\lambda_j}\Big)
        \,\mathrm{etr}\!\Big(-\tfrac{1}{2b_1}\bar{W}_{22}\Big)
        (d\Gamma)(d\Lambda)
    }{\displaystyle
        \int_{A_{\epsilon_2}}
        \prod_{j=1}^k (\lambda_j + b_1)^{-n/2}\lambda_j^{-a_j}
        \exp\!\Big(
            -\frac{2\lambda_{0,j}}{2(\lambda_j + b_0)}
            -\frac{h}{2\lambda_j}
        \Big)
        \,\mathrm{etr}\!\Big(-\tfrac{1}{2b_0}\bar{W}_{22}\Big)
        (d\Gamma)(d\Lambda)
    } \notag\\
    &\le
    \frac{\displaystyle
        \int_{B_{\epsilon_1}^c}
        \lambda_i
        \prod_{j=1}^k (\lambda_j + b_0)^{-n/2}\lambda_j^{-a_j}
        \exp\!\Big(-\frac{h}{2\lambda_j}\Big)
        \,\mathrm{etr}\!\Big(-\tfrac{1}{2b_1}\bar{W}_{22}\Big)
        (d\Gamma)(d\Lambda)
    }{\displaystyle
        \int_{A_{\epsilon_2}}
        \prod_{j=1}^k (\lambda_j + b_1)^{-n/2}\lambda_j^{-a_j}
        \exp\!\Big(
            -\frac{2\lambda_{0,j}}{2(\lambda_j + b_0)}
            -\frac{h}{2\lambda_j}
        \Big)
        \,\mathrm{etr}\!\Big(-\tfrac{1}{2b_0}\bar{W}_{22}\Big)
        (d\Gamma)(d\Lambda)
    }.
\end{align}

We now bound the integrands in \eqref{eq:post_exp_part2_3}.  
By Lemma~S1.10 in \cite{kim2025eigenstructure},
\[
    \{\Gamma \in \mathbb{V}_{p,k}: \lVert \Gamma_{21} \rVert_F < \epsilon_1/2\}
    \subseteq B_{\epsilon_1}.
\]
Thus, on $B_{\epsilon_1}^c$ we have $\lVert \Gamma_{21} \rVert_F > \epsilon_1/2$, and hence
\begin{align*}
    \mathrm{tr}(\Gamma_{11}\Gamma_{11}^T)
    &= k - \mathrm{tr}(\Gamma_{21}\Gamma_{21}^T) \\
    &= k - \lVert \Gamma_{21} \rVert_F^2 \\
    &< k - \epsilon_1^2/4.
\end{align*}
It follows that
\[
    \mathrm{tr}\big[(I_p - \Gamma\Gamma^T) W\big]
    \;\ge\; n\epsilon_1^2 \hat{\lambda}_k/4.
\]

Consequently, for the numerator integrand in \eqref{eq:post_exp_part2_3} we obtain
\begin{align*}
    &\lambda_i
    \prod_{j=1}^k (\lambda_j + b_0)^{-n/2}\lambda_j^{-a_j}
    \exp\!\Big(-\frac{h}{2\lambda_j}\Big)
    \,\mathrm{etr}\!\Big(-\tfrac{1}{2b_1}\bar{W}_{22}\Big) \\
    &\le
    \lambda_i
    \prod_{j=1}^k b_0^{-n/2}\lambda_j^{-a_j}
    \exp\!\Big(-\frac{h}{2\lambda_j}\Big)
    \,\mathrm{etr}\!\Big(-\tfrac{1}{2b_1}\bar{W}_{22}\Big) \\
    &=
    \lambda_i
    \prod_{j=1}^k b_0^{-n/2}\lambda_j^{-a_j}
    \exp\!\Big(-\frac{h}{2\lambda_j}\Big)
    \,\mathrm{etr}\!\Big(-\tfrac{1}{2b_1}(I_p - \Gamma\Gamma^T)W\Big) \\
    &\le
    \lambda_i
    \prod_{j=1}^k b_0^{-n/2}\lambda_j^{-a_j}
    \exp\!\Big(-\frac{h}{2\lambda_j}\Big)
    \,\exp\!\Big(-\frac{n\epsilon_1^2\hat{\lambda}_k}{8b_1}\Big),
\end{align*}
where the last inequality uses
\[
    \mathrm{etr}\!\Big(-\tfrac{1}{2b_1}(I_p-\Gamma\Gamma^T)W\Big)
    \le \exp\!\Big(-\tfrac{n\epsilon_1^2\hat{\lambda}_k}{8b_1}\Big).
\]

For the denominator integrand in \eqref{eq:post_exp_part2_3}, note that on
$\Gamma \in A_{\epsilon_2}$ we have
\[
    (\Gamma_{ii}^2 - 1)^2 + \sum_{j \ne i} \Gamma_{ji}^2 < \epsilon_2^2,
    \qquad i = 1,\ldots,k.
\]
Since $\sum_{j \ne i} \Gamma_{ji}^2 = 1 - \Gamma_{ii}^2$, this implies
$\Gamma_{ii} > 1 - \epsilon_2^2/2$, and hence
\begin{align*}
    [\Gamma^T W \Gamma]_{ii}
    &= \sum_{j=1}^p \Gamma_{ji}^2 \hat{\lambda}_j \\
    &\le \Gamma_{ii}^2 \hat{\lambda}_i
        + \Big(\sum_{j \ne i} \Gamma_{ji}^2\Big)\hat{\lambda}_1 \\
    &\le \hat{\lambda}_i + (1 - \Gamma_{ii}^2)\hat{\lambda}_1 \\
    &\le \hat{\lambda}_i + \epsilon_2^2 \hat{\lambda}_1.
\end{align*}
Thus,
\begin{align*}
    &\prod_{j=1}^k (\lambda_j + b_1)^{-n/2}\lambda_j^{-a_j}
    \exp\!\Big(
        -\frac{2\lambda_{0,j}}{2(\lambda_j + b_0)}
        -\frac{h}{2\lambda_j}
    \Big)
    \,\mathrm{etr}\!\Big(-\tfrac{1}{2b_0}\bar{W}_{22}\Big) \\
    &=
    \prod_{j=1}^k (\lambda_j + b_1)^{-n/2}\lambda_j^{-a_j}
    \exp\!\Big(
        -\frac{2\lambda_{0,j}}{2(\lambda_j + b_0)}
        -\frac{h}{2\lambda_j}
    \Big)
    \,\mathrm{etr}\!\Big(-\tfrac{1}{2b_0}(I_p - \Gamma\Gamma^T)W\Big) \\
    &\ge
    \prod_{j=1}^k (\lambda_j + b_1)^{-n/2}\lambda_j^{-a_j}
    \exp\!\Big(
        -\frac{2\lambda_{0,j}}{2(\lambda_j + b_0)}
        -\frac{h}{2\lambda_j}
    \Big)
    \,\exp\!\Big(-\tfrac{n\epsilon_2^2\hat{\lambda}_1}{2b_0}\Big).
\end{align*}

Substituting these bounds into \eqref{eq:post_exp_part2_3}, we obtain
\begin{align}\label{eq:post_exp_part2_4}
    \eqref{eq:post_exp_part2_3}
    &\le
    \frac{\displaystyle
        \int_{B_{\epsilon_1}^c}
        \lambda_i
        \prod_{j=1}^k b_0^{-n/2}\lambda_j^{-a_j}
        \exp\!\Big(-\frac{h}{2\lambda_j}\Big)
        \exp\!\Big(-\tfrac{n\epsilon_1^2\hat{\lambda}_k}{8b_1}\Big)
        (d\Gamma)(d\Lambda)
    }{\displaystyle
        \int_{A_{\epsilon_2}}
        \prod_{j=1}^k
        (\lambda_j + b_1)^{-n/2}\lambda_j^{-a_j}
        \exp\!\Big(
            -\frac{2\lambda_{0,j}}{2(\lambda_j + 1/b_0)}
            -\frac{h}{2\lambda_j}
        \Big)
        \exp\!\Big(-\tfrac{n\epsilon_2^2\hat{\lambda}_1}{2b_0}\Big)
        (d\Gamma)(d\Lambda)
    } \notag\\
    &=
    \frac{\mathbb{P}(B_{\epsilon_1}^c)}{\mathbb{P}(A_{\epsilon_2})}
    \cdot
    \frac{\displaystyle
        \int
        \lambda_i
        \prod_{j=1}^k  b_0^{-n/2}\lambda_j^{-a_j}
        \exp\!\Big(-\frac{h}{2\lambda_j}\Big)
        \exp\!\Big(-\tfrac{n\epsilon_1^2\hat{\lambda}_k}{8b_1}\Big)
        (d\Lambda)
    }{\displaystyle
        \int
        \prod_{j=1}^k
        (\lambda_j + b_1)^{-n/2}\lambda_j^{-a_j}
        \exp\!\Big(
            -\frac{2\lambda_{0,j}}{2(\lambda_j + 1/b_0)}
            -\frac{h}{2\lambda_j}
        \Big)
        \exp\!\Big(-\tfrac{n\epsilon_2^2\hat{\lambda}_1}{2b_0}\Big)
        (d\Lambda)
    } \notag\\
    &=
    \frac{\mathbb{P}(B_{\epsilon_1}^c)}{\mathbb{P}(A_{\epsilon_2})}
    \cdot (1 + O(N))
    \cdot
    \frac{\displaystyle
        \int
        \lambda_i
        \prod_{j=1}^k  b_0^{-n/2}\lambda_j^{-a_j}
        \exp\!\Big(-\frac{h}{2\lambda_j}\Big)
        \exp\!\Big(-\tfrac{n\epsilon_1^2\hat{\lambda}_k}{8b_1}\Big)
        (d\Lambda)
    }{\displaystyle
        \int
        \prod_{j=1}^k
        \lambda_j^{-a_j - n/2}
        \exp\!\Big(
            -\frac{2\lambda_{0,j}}{2\lambda_j}
            -\frac{h}{2\lambda_j}
        \Big)
        \exp\!\Big(-\tfrac{n\epsilon_2^2\hat{\lambda}_1}{2b_0}\Big)
        (d\Lambda)
    },
\end{align}
where $N = n/\delta_n$ and the second equality uses Lemma~\ref{lem:approx_int_post}.

Using Stirling-type bounds (cf.~Theorem~1.5 in \cite{batir2008inequalities}), one can show that
\[
    \frac{\displaystyle
        \prod_{j=1}^k \Gamma(a_j - 1)/(h/2)^{a_j - 1}
    }{\displaystyle
        \prod_{j=1}^k \Gamma(n/2 + a_j - 1)/( \lambda_{0,j} + h/2 )^{n/2 + a_j - 1}
    }
    \;\preccurlyeq\;
    \prod_{j=1}^k
    \Big(\frac{\lambda_{0,j} + h/2}{2n}\Big)^n.
\]

Moreover, by Lemma~\ref{lem:prob_stiefel},
\[
    \mathbb{P}(A_{\epsilon_2})
    \;\ge\;
    \Big(\frac{c\sqrt{k}}{\epsilon_2}\Big)^{-k(p - k/2 - 1/2)},
\]
and hence
\begin{align*}
    \eqref{eq:post_exp_part2_4}
    &\preccurlyeq
    \Big(\frac{c\sqrt{k}}{\epsilon_2}\Big)^{k(p + k/2 + 1/2)}
    \cdot \frac{2b_0^{-nk/2}(a_i - 1)}{h}
    \cdot \prod_{j=1}^k
    \Big(\frac{\lambda_{0,j} + h/2}{2n}\Big)^n
    \cdot
    \exp\!\Big(
        -\frac{n\epsilon_1^2\hat{\lambda}_k}{8b_1}
        +\frac{n\epsilon_2^2\hat{\lambda}_1}{2b_0}
    \Big) \\
    &\preccurlyeq
    \exp\!\Big(-\frac{n\epsilon_1^2\hat{\lambda}_k}{16b_1}\Big),
\end{align*}
under the conditions $\epsilon_1 \succ \epsilon_2$, $n\epsilon_1^2 \lambda_{0,k}\succ p$, and using that $\lambda_{0,1}/\lambda_{0,k}$ is bounded.

Combining this bound with \eqref{eq:post_exp_part2_1}–\eqref{eq:post_exp_part2_4}
yields
\[
    \mathbb{E}\Big[\lambda_i\,\mathbf{1}\{\Gamma \in A_\epsilon^c \cap B_{\epsilon_1}^c\}
    \,\Big\vert\, \mathbf{Y}\Big]
    \;\preccurlyeq\;
    \exp\!\Big(-\frac{n\epsilon_1^2\hat{\lambda}_k}{16b_1}\Big),
\]
which completes the proof.
\end{proof}

\begin{lemma}\label{lem:post_exp_third}
Under model \eqref{main-model:AFM}, assume that conditions $A1-A4$ holds. Suppose that the ratio $\lambda_{0,1}/\lambda_{0,k}$ is bounded by a positive constant. For $i=1,\ldots,k$ we have
\[
    \mathbb{E}\Big[\lambda_i \,\mathbf{1}\{\Gamma \in A_\epsilon^c \cap B_{\epsilon_1}\}
    \,\Big\vert\, \bfY \Big]
    \;\preccurlyeq\;
    \big(1 + O(M_k) + O(N)\big)
    \exp\!\Big(
        -\frac{\eta}{4\sqrt{k}}\min_{l<k}(a_{l+1} - a_l)
    \Big),
\]
where
\[
    M_k
    =
    O\Big(
        \max\Big(\frac{a_k}{\delta_n},\frac{n\hat{\lambda}_1}{\delta_n^2}\Big)
        + \hat{\lambda}_1^{\,n}\exp\!\Big(-\frac{n\hat{\lambda}_k}{6\delta_n}\Big)
    \Big),
    \qquad
    N = O\Big(\frac{n}{\delta_n}\Big),
\]
for some $\sqrt{n\hat{\lambda}_1}\prec \delta_n\prec \hat{\lambda}_k$ and
\[
    \eta \min_{l<k}(a_{l+1} - a_l) \succ 1.
\]
\end{lemma}

\begin{proof}
We first write
\begin{align}\label{eq:post_exp_part3_1}
    &\mathbb{E}\Big[\lambda_i \mathbf{1}\{\Gamma \in A_\epsilon^c \cap B_{\epsilon_1}\}
    \,\Big\vert\, \bfY \Big] \notag\\
    &\quad=
    \frac{\displaystyle
        \int_{A_\epsilon^c \cap B_{\epsilon_1}}
        \lambda_i\,\pi(\Gamma,\Lambda,\Sigma_u \mid \bfY)
        \,(d\Gamma)(d\Lambda)(d\Sigma_u)
    }{\displaystyle
        \int \pi(\Gamma,\Lambda,\Sigma_u \mid \bfY)
        \,(d\Gamma)(d\Lambda)(d\Sigma_u)
    } \notag\\
    &\quad\le
    \frac{\displaystyle
        \int_{A_\epsilon^c \cap B_{\epsilon_1}}
        \lambda_i
        \prod_{j=1}^k (\lambda_j + b_0)^{-n/2}\lambda_j^{-a_j}
        \exp\!\Big(
            -\frac{B_{jj}}{2(\lambda_j + b_1)}
            -\frac{h}{2\lambda_j}
        \Big)
        \lvert\bar{\Sigma}_{22}\rvert^{-n/2}
        \,\mathrm{etr}\!\Big(-\tfrac{1}{2}\bar{\Sigma}_{22}^{-1}\bar{W}_{22}\Big)
        \pi(\Sigma_u)
        \,(d\Gamma)(d\Lambda)(d\Sigma_u)
    }{\displaystyle
        \int_{A_{\epsilon_3}}
        \prod_{j=1}^k (\lambda_j + b_1)^{-n/2}\lambda_j^{-a_j}
        \exp\!\Big(
            -\frac{B_{jj}}{2(\lambda_j + b_0)}
            -\frac{h}{2\lambda_j}
        \Big)
        \lvert\bar{\Sigma}_{22}\rvert^{-n/2}
        \,\mathrm{etr}\!\Big(-\tfrac{1}{2}\bar{\Sigma}_{22}^{-1}\bar{W}_{22}\Big)
        \pi(\Sigma_u)
        \,(d\Gamma)(d\Lambda)(d\Sigma_u)
    } \notag\\
    &\quad=
    \big(1 + O(M_k) + O(N)\big) \notag\\
    &\qquad\times
    \frac{\displaystyle
        \int_{A_\epsilon^c \cap B_{\epsilon_1}}
        \lambda_i
        \prod_{j=1}^k \lambda_j^{-a_j - n/2}
        \exp\!\Big(
            -\frac{B_{jj}}{2\lambda_j}
            -\frac{h}{2\lambda_j}
        \Big)
        \lvert\bar{\Sigma}_{22}\rvert^{-n/2}
        \,\mathrm{etr}\!\Big(-\tfrac{1}{2}\bar{\Sigma}_{22}^{-1}\bar{W}_{22}\Big)
        \pi(\Sigma_u)
        \,(d\Gamma)(d\Lambda)(d\Sigma_u)
    }{\displaystyle
        \int_{A_{\epsilon_3}}
        \prod_{j=1}^k \lambda_j^{-a_j - n/2}
        \exp\!\Big(
            -\frac{B_{jj}}{2\lambda_j}
            -\frac{h}{2\lambda_j}
        \Big)
        \lvert\bar{\Sigma}_{22}\rvert^{-n/2}
        \,\mathrm{etr}\!\Big(-\tfrac{1}{2}\bar{\Sigma}_{22}^{-1}\bar{W}_{22}\Big)
        \pi(\Sigma_u)
        \,(d\Gamma)(d\Lambda)(d\Sigma_u)
    },
\end{align}
where the factor $(1+O(M_k)+O(N))$ comes from Lemma~\ref{lem:approx_int_post}, and the inequality uses \eqref{ineq:bound_post}.

Next, as in Lemma~\ref{lem:post_exp_second}, we apply the change of variables
\[
    \Omega = \bar{\Gamma}^T \Sigma_u \bar{\Gamma}
    = \begin{pmatrix}
        \Omega_{11} & \Omega_{12} \\
        \Omega_{21} & \Omega_{22}
      \end{pmatrix},
\]
and use the rotation invariance of the prior $\pi(\cdot)$ to obtain
\begin{align}\label{eq:post_exp_part3_2}
    \eqref{eq:post_exp_part3_1}
    &=
    \frac{\displaystyle
        \int_{A_\epsilon^c \cap B_{\epsilon_1}}
        \lambda_i
        \prod_{j=1}^k \lambda_j^{-a_j - n/2}
        \exp\!\Big(
            -\frac{D_{jj}}{2\lambda_j}
            -\frac{h}{2\lambda_j}
        \Big)
        \lvert\Omega_{22}\rvert^{-n/2}
        \,\mathrm{etr}\!\Big(-\tfrac{1}{2}\Omega_{22}^{-1}\bar{W}_{22}\Big)
        \pi(\Omega)
        \,(d\Gamma)(d\Lambda)(d\Omega)
    }{\displaystyle
        \int_{A_{\epsilon_3}}
        \prod_{j=1}^k \lambda_j^{-a_j - n/2}
        \exp\!\Big(
            -\frac{D_{jj}}{2\lambda_j}
            -\frac{h}{2\lambda_j}
        \Big)
        \lvert\Omega_{22}\rvert^{-n/2}
        \,\mathrm{etr}\!\Big(-\tfrac{1}{2}\Omega_{22}^{-1}\bar{W}_{22}\Big)
        \pi(\Omega)
        \,(d\Gamma)(d\Lambda)(d\Omega)
    } \notag\\
    &=
    \frac{\displaystyle
        \int_{A_\epsilon^c \cap B_{\epsilon_1}}
        \frac{D_{ii} + h}{n + 2a_i - 4}
        \prod_{j=1}^k (D_{jj} + h)^{-a_j - n/2 + 1}
        \lvert\Omega_{22}\rvert^{-n/2}
        \,\mathrm{etr}\!\Big(-\tfrac{1}{2}\Omega_{22}^{-1}\bar{W}_{22}\Big)
        \pi(\Omega)
        \,(d\Gamma)(d\Omega)
    }{\displaystyle
        \int_{A_{\epsilon_3}}
        \prod_{j=1}^k (D_{jj} + h)^{-a_j - n/2 + 1}
        \lvert\Omega_{22}\rvert^{-n/2}
        \,\mathrm{etr}\!\Big(-\tfrac{1}{2}\Omega_{22}^{-1}\bar{W}_{22}\Big)
        \pi(\Omega)
        \,(d\Gamma)(d\Omega)
    } \notag\\
    &\le
    \sup_{\Gamma \in B_{\epsilon_1}}
    \frac{D_{ii} + h}{n + 2a_i - 4}
    \cdot
    \frac{\displaystyle
        \sup_{A_{\epsilon_3}}
        \prod_{j=1}^k \Big(\tfrac{D_{jj} + h}{n}\Big)^{a_j + n/2 - 1}
    }{\displaystyle
        \inf_{B_{\epsilon_1}}
        \prod_{j=1}^k \Big(\tfrac{D_{jj} + h}{n}\Big)^{a_j + n/2 - 1}
    }
    \cdot
    \frac{\displaystyle
        \int_{A_\epsilon^c \cap B_{\epsilon_1}}
        \lvert\Omega_{22}\rvert^{-n/2}
        \,\mathrm{etr}\!\Big(-\tfrac{1}{2}\Omega_{22}^{-1}\bar{W}_{22}\Big)
        \pi(\Omega)
        \,(d\Gamma)(d\Omega)
    }{\displaystyle
        \int_{A_{\epsilon_3}}
        \lvert\Omega_{22}\rvert^{-n/2}
        \,\mathrm{etr}\!\Big(-\tfrac{1}{2}\Omega_{22}^{-1}\bar{W}_{22}\Big)
        \pi(\Omega)
        \,(d\Gamma)(d\Omega)
    },
\end{align}
where
\[
    D
    = \Gamma^T W\Gamma
      - \Omega_{12}\Omega_{22}^{-1}\Gamma_2^T W\Gamma
      - \Gamma^T W\Gamma_2 \Omega_{22}^{-1}\Omega_{21}
      + \Omega_{12}\Omega_{22}^{-1}
        (\Gamma_2^T W\Gamma_2)\Omega_{22}^{-1}\Omega_{21},
\]
and $D_{jj}$ denotes the $(j,j)$-th diagonal element of $D$. The second equality in
\eqref{eq:post_exp_part3_2} follows from integrating out $\Lambda$ using the
inverse-gamma kernel.

We now bound the remaining ratio in \eqref{eq:post_exp_part3_2}.

---

\textbf{Step 1: Bounding the $\Omega_{22}$-ratio.}
By Lemma~\ref{lem:union_covering}, we have
\begin{align*}
    B_{\epsilon_1}
    &= \Big\{
        \Gamma \in \mathbb{V}_{p,k}:
        \inf_{Q \in O_{k}}
        \big\|
            \Gamma - \begin{pmatrix}Q \\ 0\end{pmatrix}
        \big\|_F < \epsilon_1
    \Big\} \\
    &\subset
    \bigcup_{i=1}^{N(O_k,\|\cdot\|_F,\epsilon_0)}
    \Big\{
        \Gamma \in \mathbb{V}_{p,k}:
        \big\|
            \Gamma - \begin{pmatrix}Q_i \\ 0\end{pmatrix}
        \big\|_F < \epsilon_1 + \epsilon_0
    \Big\},
\end{align*}
for some $\epsilon_0 > 0$. Using also the identity
$I_p - \Gamma\Gamma^T = I_p - (\Gamma Q^T)(Q\Gamma^T)$, we obtain
\begin{align*}
    &\frac{\displaystyle
        \int_{A_\epsilon^c \cap B_{\epsilon_1}}
        \lvert\Omega_{22}\rvert^{-n/2}
        \,\mathrm{etr}\!\Big(-\tfrac{1}{2}\Omega_{22}^{-1}\bar{W}_{22}\Big)
        \pi(\Omega)
        \,(d\Gamma)(d\Omega)
    }{\displaystyle
        \int_{A_{\epsilon_3}}
        \lvert\Omega_{22}\rvert^{-n/2}
        \,\mathrm{etr}\!\Big(-\tfrac{1}{2}\Omega_{22}^{-1}\bar{W}_{22}\Big)
        \pi(\Omega)
        \,(d\Gamma)(d\Omega)
    } \\
    &\quad\le
    \frac{\displaystyle
        \int_{B_{\epsilon_1}}
        \lvert\Omega_{22}\rvert^{-n/2}
        \,\mathrm{etr}\!\Big(-\tfrac{1}{2}\Omega_{22}^{-1}\bar{W}_{22}\Big)
        \pi(\Omega)
        \,(d\Gamma)(d\Omega)
    }{\displaystyle
        \int_{A_{\epsilon_3}}
        \lvert\Omega_{22}\rvert^{-n/2}
        \,\mathrm{etr}\!\Big(-\tfrac{1}{2}\Omega_{22}^{-1}\bar{W}_{22}\Big)
        \pi(\Omega)
        \,(d\Gamma)(d\Omega)
    } \\
    &\quad\le
    \Big(\frac{c_1 \sqrt{k}}{\epsilon_0}\Big)^{k(k-1)/2}
    \frac{\displaystyle
        \int_{A_{\epsilon_1 + \epsilon_0}}
        \lvert\Omega_{22}\rvert^{-n/2}
        \,\mathrm{etr}\!\Big(-\tfrac{1}{2}\Omega_{22}^{-1}\bar{W}_{22}\Big)
        \pi(\Omega)
        \,(d\Gamma)(d\Omega)
    }{\displaystyle
        \int_{A_{\epsilon_3}}
        \lvert\Omega_{22}\rvert^{-n/2}
        \,\mathrm{etr}\!\Big(-\tfrac{1}{2}\Omega_{22}^{-1}\bar{W}_{22}\Big)
        \pi(\Omega)
        \,(d\Gamma)(d\Omega)
    } \\
    &\quad\le
    \Big(\frac{c_1 \sqrt{k}}{\epsilon_0}\Big)^{k(k-1)/2},
    \qquad \text{if } \epsilon_1 + \epsilon_0 < \epsilon_3.
\end{align*}

---

\textbf{Step 2: Bounding the ratio of $(D_{jj}+h)$–terms.}
By Lemma~\ref{lem:upp_bound_ci_diff}, we obtain
\begin{align*}
    \frac{\displaystyle
        \sup_{\Gamma \in A_{\epsilon_3}}
        \prod_{j=1}^k \Big(\tfrac{D_{jj}+h}{n}\Big)^{a_j + n/2 - 1}
    }{\displaystyle
        \inf_{\Gamma \in A_\epsilon^c \cap B_{\epsilon_1}}
        \prod_{j=1}^k \Big(\tfrac{D_{jj}+h}{n}\Big)^{a_j + n/2 - 1}
    }
    &\preccurlyeq
    \frac{\exp\!\Big(4n\epsilon_3^2 \,\tfrac{\hat{\lambda}_1}{\hat{\lambda}_k}\Big)}
         {\exp\!\Big(
            \frac{\eta}{2\sqrt{k}}
            \min_{l<k}(a_{l+1}-a_l)\,
            \min_{l<k}\log\!\big(\tfrac{\hat{\lambda}_l}{\hat{\lambda}_{l+1}}\big)
         \Big)},
\end{align*}
where $\eta = 2\epsilon - \epsilon_1 - \epsilon_3 > 0$.

Combining these bounds in \eqref{eq:post_exp_part3_2}, we find
\begin{align*}
    \eqref{eq:post_exp_part3_2}
    &\preccurlyeq
    \sup_{\Gamma \in B_{\epsilon_1}}
    \frac{D_{ii}+h}{n + 2a_i - 4}
    \cdot
    \frac{\exp\!\Big(4n\epsilon_3^2 \,\tfrac{\hat{\lambda}_1}{\hat{\lambda}_k}\Big)}
         {\exp\!\Big(
            \frac{\eta}{2\sqrt{k}}
            \min_{l<k}(a_{l+1}-a_l)\,
            \min_{l<k}\log\!\big(\tfrac{\hat{\lambda}_l}{\hat{\lambda}_{l+1}}\big)
         \Big)}
    \cdot
    \Big(\frac{c_1 \sqrt{k}}{\epsilon_0}\Big)^{k(k-1)/2}.
\end{align*}
Under the conditions
\[
    \eta \min_{l<k}(a_{l+1} - a_l) \succ n\epsilon_3^2,
    \qquad
    \eta \min_{l<k}(a_{l+1} - a_l) \succ 1,
\]
and using that $\hat{\lambda}_l/\hat{\lambda}_{l+1}$ is uniformly bounded away from $1$, the exponential term in the denominator dominates the polynomial factors, yielding
\[
    \eqref{eq:post_exp_part3_2}
    \preccurlyeq
    \exp\!\Big(
        -\frac{\eta}{4\sqrt{k}}
        \min_{l<k}(a_{l+1} - a_l)
    \Big).
\]

Finally, combining this with the prefactor $(1+O(M_k)+O(N))$ in
\eqref{eq:post_exp_part3_1}, we conclude that
\[
    \mathbb{E}\Big[\lambda_i \,\mathbf{1}\{\Gamma \in A_\epsilon^c \cap B_{\epsilon_1}\}
    \,\Big\vert\, \bfY \Big]
    \preccurlyeq
    \big(1 + O(M_k) + O(N)\big)
    \exp\!\Big(
        -\frac{\eta}{4\sqrt{k}}
        \min_{l<k}(a_{l+1} - a_l)
    \Big),
\]
which completes the proof.
\end{proof}

\begin{lemma}\label{lem:post_exp_first}
Under model \eqref{main-model:AFM}, assume that conditions $A1-A4$ holds. Suppose that the ratio $\lambda_{0,1}/\lambda_{0,k}$ is bounded by a positive constant. For $i=1,\ldots,k$ we have
\[
    \mathbb{E}\Big[\lambda_i \,\mathbf{1}\{\Gamma \in A_\epsilon\}\,\Big\vert\, \mathbf{Y} \Big]
    = \big(1 + O(M_k) + O(N)\big)
    \Bigg[
        \frac{n\hat{\lambda}_i}{\,n+2a_i-4\,}
        + O\big(p\psi^{-1} + \epsilon \lambda_{0,1}\big)
    \Bigg],
\]
where
\[
    M_k
    = O\Big(
        \max\Big(\frac{a_k}{\delta_n},\frac{n\hat{\lambda}_1}{\delta_n^2}\Big)
        + \hat{\lambda}_1^{\,n}\exp\!\Big(-\frac{n\hat{\lambda}_k}{6\delta_n}\Big)
    \Big),
    \qquad
    N = O\Big(\frac{n}{\delta_n}\Big),
\]
for some $\sqrt{n\hat{\lambda}_1}\prec \delta_n\prec \hat{\lambda}_k$.
\end{lemma}

\begin{proof}
We begin by writing
\begin{align}\label{eq:post_exp_part1_1}
    \mathbb{E}\big[\lambda_i \mathbf{1}\{\Gamma\in A_\epsilon\}\mid \mathbf{Y}\big]
    &= \frac{\displaystyle
        \int_{A_\epsilon} \lambda_i \,
        \pi(\Gamma,\Lambda,\Sigma_u\mid \mathbf{Y})
        \,(d\Gamma)(d\Lambda)(d\Sigma_u)
    }{\displaystyle
        \int
        \pi(\Gamma,\Lambda,\Sigma_u\mid \mathbf{Y})
        \,(d\Gamma)(d\Lambda)(d\Sigma_u)
    } \notag\\
    &\le
    \frac{\displaystyle
        \int_{A_\epsilon} \lambda_i \,
        \pi(\Gamma,\Lambda,\Sigma_u\mid \mathbf{Y})
        \,(d\Gamma)(d\Lambda)(d\Sigma_u)
    }{\displaystyle
        \int_{A_\epsilon}
        \pi(\Gamma,\Lambda,\Sigma_u\mid \mathbf{Y})
        \,(d\Gamma)(d\Lambda)(d\Sigma_u)
    } \notag\\
    &= \big(1 + O(\textstyle\sum_{j=1}^k M_j) + O(N)\big) \notag\\
    &\quad\times
    \frac{\displaystyle
        \int_{A_\epsilon} \lambda_i
        \prod_{j=1}^k \lambda_j^{-a_j - n/2}
        \exp\!\Big(-\frac{B_{jj}}{2\lambda_j} - \frac{h}{2\lambda_j}\Big)
        \lvert\bar{\Sigma}_{22}\rvert^{-n/2}
        \,\mathrm{etr}\!\Big(-\tfrac{1}{2}\bar{\Sigma}_{22}^{-1}\bar{W}_{22}\Big)
        \pi(\Sigma_u)
        \,(d\Gamma)(d\Lambda)(d\Sigma_u)
    }{\displaystyle
        \int_{A_\epsilon}
        \prod_{j=1}^k \lambda_j^{-a_j - n/2}
        \exp\!\Big(-\frac{B_{jj}}{2\lambda_j} - \frac{h}{2\lambda_j}\Big)
        \lvert\bar{\Sigma}_{22}\rvert^{-n/2}
        \,\mathrm{etr}\!\Big(-\tfrac{1}{2}\bar{\Sigma}_{22}^{-1}\bar{W}_{22}\Big)
        \pi(\Sigma_u)
        \,(d\Gamma)(d\Lambda)(d\Sigma_u)
    },
\end{align}
where
\[
    M_j
    = O\Big(
        \max\Big(\frac{a_j}{\delta_n},\frac{n\hat{\lambda}_1}{\delta_n^2}\Big)
        + \hat{\lambda}_1^{\,n}\exp\!\Big(-\frac{n\hat{\lambda}_k}{6\delta_n}\Big)
    \Big),
    \qquad
    N = O\Big(\frac{n}{\delta_n}\Big),
\]
and the last equality in \eqref{eq:post_exp_part1_1} follows from Lemma~\ref{lem:approx_int_post}.

Next, as in Lemma~\ref{lem:post_exp_second}, we apply the change of variables
$\Omega = (\Gamma,\Gamma_2)^T \Sigma_u (\Gamma,\Gamma_2)$ and use the
rotation invariance of the prior $\pi(\cdot)$ to obtain
\begin{align}\label{eq:post_exp_part1_2}
    \eqref{eq:post_exp_part1_1}
    &=
    \frac{\displaystyle
        \int_{A_\epsilon}
        \lambda_i
        \prod_{j=1}^k \lambda_j^{-a_j - n/2}
        \exp\!\Big(-\frac{D_{jj}}{2\lambda_j} - \frac{h}{2\lambda_j}\Big)
        \lvert\Omega_{22}\rvert^{-n/2}
        \,\mathrm{etr}\!\Big(-\tfrac{1}{2}\Omega_{22}^{-1}\bar{W}_{22}\Big)
        \pi(\Omega)
        \,(d\Gamma)(d\Lambda)(d\Omega)
    }{\displaystyle
        \int_{A_\epsilon}
        \prod_{j=1}^k \lambda_j^{-a_j - n/2}
        \exp\!\Big(-\frac{D_{jj}}{2\lambda_j} - \frac{h}{2\lambda_j}\Big)
        \lvert\Omega_{22}\rvert^{-n/2}
        \,\mathrm{etr}\!\Big(-\tfrac{1}{2}\Omega_{22}^{-1}\bar{W}_{22}\Big)
        \pi(\Omega)
        \,(d\Gamma)(d\Lambda)(d\Omega)
    } \notag\\
    &=
    \frac{\displaystyle
        \int_{A_\epsilon}
        \frac{D_{ii} + h}{n + 2a_i - 4}
        \prod_{j=1}^k (D_{jj} + h)^{-a_j - n/2 + 1}
        \lvert\Omega_{22}\rvert^{-n/2}
        \,\mathrm{etr}\!\Big(-\tfrac{1}{2}\Omega_{22}^{-1}\bar{W}_{22}\Big)
        \pi(\Omega)
        \,(d\Gamma)(d\Omega)
    }{\displaystyle
        \int_{A_\epsilon}
        \prod_{j=1}^k (D_{jj} + h)^{-a_j - n/2 + 1}
        \lvert\Omega_{22}\rvert^{-n/2}
        \,\mathrm{etr}\!\Big(-\tfrac{1}{2}\Omega_{22}^{-1}\bar{W}_{22}\Big)
        \pi(\Omega)
        \,(d\Gamma)(d\Omega)
    },
\end{align}
where
\[
    D
    = \Gamma^T W\Gamma
      - \Omega_{12}\Omega_{22}^{-1}\Gamma_2^T W\Gamma
      - \Gamma^T W\Gamma_2\,\Omega_{22}^{-1}\Omega_{21}
      + \Omega_{12}\Omega_{22}^{-1}(\Gamma_2^T W\Gamma_2)\Omega_{22}^{-1}\Omega_{21},
\]
and $D_{ii}$ denotes the $(i,i)$-th diagonal element of $D$.

Let $K_{ii}$ be the $(i,i)$ element of
\[
    \Gamma^T W\Gamma
    - \Omega_{12}\Omega_{22}^{-1}\Gamma_2^T W\Gamma
    - \Gamma^T W\Gamma_2\,\Omega_{22}^{-1}\Omega_{21},
\]
and let $G_{ii}$ be the $(i,i)$ element of
\[
    \Omega_{12}\Omega_{22}^{-1}(\Gamma_2^T W\Gamma_2)\Omega_{22}^{-1}\Omega_{21},
\]
so that $D_{ii} = K_{ii} + G_{ii}$. Then from \eqref{eq:post_exp_part1_2} we obtain
\begin{align}\label{eq:post_exp_part1_3}
    \eqref{eq:post_exp_part1_2}
    &\le
    \sup_{\Gamma\in A_\epsilon}
    \frac{K_{ii} + h}{n + 2a_i - 4} \notag\\
    &\quad+
    \frac{1}{n + 2a_i - 4}
    \frac{\displaystyle
        \int_{A_\epsilon}
        G_{ii}
        \prod_{j=1}^k (D_{jj} + h)^{-a_j - n/2 + 1}
        \lvert\Omega_{22}\rvert^{-n/2}
        \,\mathrm{etr}\!\Big(-\tfrac{1}{2}\Omega_{22}^{-1}\bar{W}_{22}\Big)
        \pi(\Omega)
        \,(d\Gamma)(d\Omega)
    }{\displaystyle
        \int_{A_\epsilon}
        \prod_{j=1}^k (D_{jj} + h)^{-a_j - n/2 + 1}
        \lvert\Omega_{22}\rvert^{-n/2}
        \,\mathrm{etr}\!\Big(-\tfrac{1}{2}\Omega_{22}^{-1}\bar{W}_{22}\Big)
        \pi(\Omega)
        \,(d\Gamma)(d\Omega)
    }.
\end{align}

For the second term in \eqref{eq:post_exp_part1_3}, we first bound it by
\begin{align*}
    \frac{1}{n + 2a_i - 4}
    \frac{\displaystyle
        \int_{A_\epsilon}
        \mathrm{tr}(G)
        \prod_{j=1}^k (D_{jj} + h)^{-a_j - n/2 + 1}
        \lvert\Omega_{22}\rvert^{-n/2}
        \,\mathrm{etr}\!\Big(-\tfrac{1}{2}\Omega_{22}^{-1}\bar{W}_{22}\Big)
        \pi(\Omega)
        \,(d\Gamma)(d\Omega)
    }{\displaystyle
        \int_{A_\epsilon}
        \prod_{j=1}^k (D_{jj} + h)^{-a_j - n/2 + 1}
        \lvert\Omega_{22}\rvert^{-n/2}
        \,\mathrm{etr}\!\Big(-\tfrac{1}{2}\Omega_{22}^{-1}\bar{W}_{22}\Big)
        \pi(\Omega)
        \,(d\Gamma)(d\Omega)
    }.
\end{align*}
Assume the prior on $\Omega$ is of the form
\[
    \pi(\Omega)
    \propto \lvert\Omega\rvert^{-\nu_0}
    \,\mathrm{etr}\!\Big(-\tfrac{1}{2}\Omega^{-1}\Psi\Big)
    \mathbf{1}\{\lambda(\Omega)\in[b_0,b_1]\},
    \qquad
    \Psi = \psi I_p,\ \psi>0.
\]
Introducing the reparametrization
\[
    M = \Omega_{22}^{-1}\Omega_{21}, \quad
    S = \Omega_{11\cdot 2}, \quad
    A = \Omega_{22},
\]
we may rewrite the ratio of integrals as
\begin{align}\label{eq:exp_ratio1}
    \frac{\mathbb{E}\big[\mathrm{tr}(\Gamma_2^T W\Gamma_2 M M^T)\mathbf{1}_E\big]}
         {\mathbb{E}[\mathbf{1}_E]}
    = \frac{\mathbb{E}\big[\mathbb{E}\big(\mathrm{tr}(\Gamma_2^T W\Gamma_2 M M^T)\mathbf{1}_E \,\big\vert\, S,A\big)\big]}
           {\mathbb{E}[\mathbf{1}_E]},
\end{align}
where $E = \{(A,S,M): \lambda(\Omega)\in[b_0,b_1]\}$ and
\[
    [S,A,M]
    \propto
    \lvert S\rvert^{-\nu_0}
    \lvert A\rvert^{-n/2 - \nu_0 + k}
    \mathrm{etr}\!\Big(
        -\tfrac{1}{2}(\bar{W}_{22}+\Psi_{22})A^{-1}
        -\tfrac{1}{2}\Psi_{22}MS^{-1}M^T
    \Big)
    \mathrm{etr}\!\Big(-\tfrac{1}{2}\Psi_{11}S^{-1}\Big).
\]

Writing $M = [M_1,\dots,M_k]$ with $M_j\in\mathbb{R}^{p-k}$ and using
$\Psi_{22} = \psi I_{p-k}$, we have
\[
    (M_1^T,\dots,M_k^T)^T \mid S,A
    \sim N\big(0,\psi^{-1} S \otimes I_{p-k}\big),
    \qquad
    M_j \mid S,A \sim N\big(0,\psi^{-1} S_{jj} I_{p-k}\big).
\]
This yields the bound
\begin{align}\label{eq:exp_MMt_bound}
    \mathbb{E}\big[\mathrm{tr}(W_{22} M M^T)\mathbf{1}_E \mid S,A\big]
    &\le
    n\hat{\lambda}_{k+1}
    \sum_{j=1}^k
    \mathrm{tr}\!\Big(
        \begin{pmatrix}
            I_{n-k} & 0 \\ 0 & 0_{p-n}
        \end{pmatrix}
        \mathbb{E}[M_j M_j^T\mathbf{1}_E \mid S,A]
    \Big) \notag\\
    &\le
    n\hat{\lambda}_{k+1}\,k^2(n-k)\psi^{-1} b_1\,
    \mathbb{E}[\mathbf{1}_E \mid S,A].
\end{align}
By Lemma~\ref{lem:cond_covariance} and the geometry of $E$, we also control
\[
    \big\lvert
        \mathrm{tr}\big((W_{22} - \Gamma_2^T W\Gamma_2)M M^T\big)
    \big\rvert
    \le
    4n\hat{\lambda}_1 \epsilon \,\lVert M M^T\rVert_F,
\]
and together with Lemma~\ref{lem:block_approx} this implies
\[
    \frac{\mathbb{E}\big[\mathrm{tr}(\Gamma_2^T W\Gamma_2 M M^T)\mathbf{1}_E\big]}
         {\mathbb{E}[\mathbf{1}_E]}
    = O(np\psi^{-1}) + O(n\lambda_{0,1}\epsilon).
\]

Next, on the set $\Gamma\in A_\epsilon$ we have
\[
    \lVert\Gamma - \Gamma_0\rVert_F < \epsilon,
    \qquad
    \Gamma_0 = \begin{pmatrix} I_k \\ 0 \end{pmatrix},
\]
and an argument analogous to the above shows that
\[
    \sup_{\Gamma\in A_\epsilon}
    \frac{K_{ii} + h}{n + 2a_i - 4}
    \le
    \frac{1}{n + 2a_i - 4}\Big(
        n\hat{\lambda}_i
        + n\epsilon^2\hat{\lambda}_1
        + 3n\epsilon \hat{\lambda}_1
    \Big).
\]

Collecting the bounds, we obtain
\begin{align*}
    \mathbb{E}[\lambda_i \mathbf{1}\{\Gamma\in A_\epsilon\}\mid \mathbf{Y}]
    &\le
    \big(1 + O(\textstyle\sum_{j=1}^k M_j) + O(N)\big) \\
    &\quad\times
    \Bigg[
        \frac{1}{n + 2a_i - 4}
        \Big(
            n\hat{\lambda}_i
            + n\epsilon^2\hat{\lambda}_1
            + 3n\epsilon\hat{\lambda}_1
        \Big)
        +
        \frac{1}{n + 2a_i - 4}
        \Big(
            O(np\psi^{-1})
            + O(n\lambda_{0,1}\epsilon)
        \Big)
    \Bigg] \\
    &=
    \big(1 + O(\textstyle\sum_{j=1}^k M_j) + O(N)\big)
    \Bigg[
        \frac{n\hat{\lambda}_i}{n + 2a_i - 4}
        + O\big(p\psi^{-1} + \epsilon\lambda_{0,1}\big)
    \Bigg].
\end{align*}
Since $a_1 < \cdots < a_k$ and $k$ is fixed, we have
$\sum_{j=1}^k M_j = O(M_k)$, which yields the final form
\[
    \mathbb{E}\big[\lambda_i \mathbf{1}\{\Gamma\in A_\epsilon\}\mid \mathbf{Y}\big]
    =
    \big(1 + O(M_k) + O(N)\big)
    \Bigg[
        \frac{n\hat{\lambda}_i}{n + 2a_i - 4}
        + O\big(p\psi^{-1} + \epsilon\lambda_{0,1}\big)
    \Bigg].
\]
This completes the proof.
\end{proof}

\hfill\break

\begin{lemma}\label{lem:post_exp_diff_eigval}
Under model \eqref{main-model:AFM}, assume that conditions $A1-A5$ holds. Suppose that $\psi \succcurlyeq \sqrt{np}$, $\epsilon \prec n^{-1/2}$, 
$n^{3/4}\lambda_{0,k}^{1/2}\wedge n^{3/4}\lambda_{0,k}p^{-1/2}\prec \lambda_{0,k}$, and that the ratio
$\lambda_{0,1}/\lambda_{0,k}$ is bounded above by a positive constant. Then, for $i=1,\ldots,k$,
\[
    \bbE\Big[\frac{\lambda_i - \lambda_{0,i}}{\lambda_{0,i}} \,\Big\vert\, \bfY \Big]
    = O\Big(\lambda_{0,i}^{-1}\sqrt{\frac{p}{n}}\Big) + O(\beta_i),
\]
where $\beta_i \lesssim n^{-1/2+\delta}$ for any sufficiently small $\delta>0$.
\end{lemma}

\begin{proof}
By Lemmas~\ref{lem:post_exp_first}, \ref{lem:post_exp_second}, and \ref{lem:post_exp_third}, we obtain
\begin{align*}
    \bbE\Big[\frac{\lambda_i - \lambda_{0,i}}{\lambda_{0,i}} \,\Big\vert\, \bfY \Big]
    &= \frac{1}{\lambda_{0,i}} \bbE\big[\lambda_i \mid \bfY \big] - 1 \\
    &\le \frac{1}{\lambda_{0,i}}
    \Bigg[
        (1+O(M_k)+O(N))
        \Big(
            \frac{n\hat{\lambda}_i}{n + 2a_i - 4}
            + O(p\psi^{-1} + \epsilon \lambda_{0,1})
        \Big) \\
    &\hspace{2.6cm}
        + (1+O(M_k)+O(N))
          \exp\Big(-\frac{\eta}{4\sqrt{k}}\min_{l<k}(a_{l+1}-a_l)\Big) \\
    &\hspace{2.6cm}
        + \exp\Big(-\frac{n\epsilon_1^2\hat{\lambda}_k}{16b_1}\Big)
    \Bigg] - 1 \\
    &= \frac{1}{\lambda_{0,i}} \frac{n\hat{\lambda}_i}{n + 2a_i - 4} - 1 \\
    &\quad
    + \frac{1}{\lambda_{0,i}} O(p\psi^{-1} + \epsilon \lambda_{0,1}) \\
    &\quad
    + \frac{1}{\lambda_{0,i}}
    \Bigg[
        O(M_k + N)\frac{n\hat{\lambda}_i}{n + 2a_i - 4}
        + \exp\Big(-\frac{\eta}{4\sqrt{k}}\min_{l<k}(a_{l+1}-a_l)\Big) \\
    &\hspace{4.4cm}
        + \exp\Big(-\frac{n\epsilon_1^2\hat{\lambda}_k}{16b_1}\Big)
    \Bigg],
\end{align*}
where
\[
    M_k
    = O\Big(
        \max\Big(\frac{a_k}{\delta_n},\frac{n\hat{\lambda}_1}{\delta_n^2}\Big)
        + \hat{\lambda}_1^{\,n}
          \exp\Big(-\frac{n\hat{\lambda}_k}{6\delta_n}\Big)
    \Big),
    \qquad
    N = O\Big(\frac{n}{\delta_n}\Big).
\]

Let
\[
    2a_i - 4 = \frac{nt}{\hat{\lambda}_i - t}.
\]
Then
\begin{align}\label{eq:exp_lambda}
    \frac{1}{\lambda_{0,i}}\frac{n\hat{\lambda}_i}{n + 2a_i - 4} - 1
    &= \frac{1}{\lambda_{0,i}}
       \Big[\frac{n\hat{\lambda}_i}{n + 2a_i - 4} - \lambda_{0,i}\Big] \notag\\
    &= \frac{1}{\lambda_{0,i}} \big[\hat{\lambda}_i - t - \lambda_{0,i}\big] \notag\\
    &= \frac{1}{\lambda_{0,i}}
       \Big[\frac{h}{n} + \beta_i \lambda_{0,i}
               + \bar{d}\frac{p}{n}
               + \alpha_i\sqrt{\frac{p}{n}}
               - t \Big].
\end{align}

If $t \in [\hat{\lambda}_{k+1},\hat{\lambda}_n]$, then for some constant
$\alpha_0 \in [-C,C]$ we may write
\[
    t = \bar{d}\frac{p}{n} + \alpha_0 \sqrt{\frac{p}{n}}.
\]
Substituting this into \eqref{eq:exp_lambda} gives
\begin{align*}
    \eqref{eq:exp_lambda}
    &= \frac{1}{\lambda_{0,i}}(\alpha_i - \alpha_0)\sqrt{\frac{p}{n}}
       + \beta_i + \frac{1}{\lambda_{0,i}}\frac{h}{n} \\
    &= O\Big(\lambda_{0,i}^{-1}\sqrt{\frac{p}{n}}\Big) + O(\beta_i).
\end{align*}

Now assume that $\psi \succcurlyeq \sqrt{np}$, $\epsilon \prec n^{-1/2}$,
$\epsilon_1^2 \succ (n\hat{\lambda}_k)^{-1}$, and
\[
    \delta_n \succ n^{3/4}\lambda_{0,k}^{1/2} \wedge
                 n^{3/4}\lambda_{0,k}p^{-1/2}.
\]
Under these conditions, we have
\begin{align}\label{asymp:post_exp_lambda}
    &\frac{1}{\lambda_{0,i}} O(p\psi^{-1} + \epsilon \lambda_{0,1})
    + \frac{1}{\lambda_{0,i}}
      \Bigg[
        O(M_k + N)\frac{n\hat{\lambda}_i}{n + 2a_i - 4}
        + \exp\Big(-\frac{\eta}{4\sqrt{k}}\min_{l<k}(a_{l+1}-a_l)\Big)
        + \exp\Big(-\frac{n\epsilon_1^2\hat{\lambda}_k}{16b_1}\Big)
      \Bigg] \notag\\
    &= O\Big(\lambda_{0,i}^{-1}\sqrt{\frac{p}{n}}\Big) + O(\beta_i).
\end{align}

Hence,
\[
    \bbE\Big[\frac{\lambda_i - \lambda_{0,i}}{\lambda_{0,i}} \,\Big\vert\, \bfY \Big]
    \le
    O\Big(\lambda_{0,i}^{-1}\sqrt{\frac{p}{n}}\Big) + O(\beta_i).
\]

We now derive a corresponding lower bound. First note that
\begin{align*}
    \bbE[\lambda_i \mid \bfY]
    &\ge \bbE[\lambda_i \mathbf{1}\{\Gamma\in A_\epsilon\} \mid \bfY] \\
    &= \frac{\displaystyle
        \int_{A_\epsilon} \lambda_i
        \pi(\Gamma,\Lambda,\Sigma_u \mid \bfY)
        \,(d\Gamma)(d\Lambda)(d\Sigma_u)
    }{\displaystyle
        \int \pi(\Gamma,\Lambda,\Sigma_u \mid \bfY)
        \,(d\Gamma)(d\Lambda)(d\Sigma_u)
    } \\
    &= \frac{\displaystyle
        \int_{A_\epsilon} \lambda_i
        \pi(\Gamma,\Lambda,\Sigma_u \mid \bfY)
        \,(d\Gamma)(d\Lambda)(d\Sigma_u)
    }{\displaystyle
        \int_{A_\epsilon}
        \pi(\Gamma,\Lambda,\Sigma_u \mid \bfY)
        \,(d\Gamma)(d\Lambda)(d\Sigma_u)
    } \\
    &\hspace{1cm}\times
    \Bigg[
        1 -
        \frac{\displaystyle
            \int_{A_\epsilon^c}
            \pi(\Gamma,\Lambda,\Sigma_u \mid \bfY)
            \,(d\Gamma)(d\Lambda)(d\Sigma_u)
        }{\displaystyle
            \int \pi(\Gamma,\Lambda,\Sigma_u \mid \bfY)
            \,(d\Gamma)(d\Lambda)(d\Sigma_u)
        }
    \Bigg].
\end{align*}
Using Lemmas~\ref{lem:approx_int_post}, \ref{lem:post_exp_second}, and
\ref{lem:post_exp_third}, we obtain
\begin{align*}
    \bbE[\lambda_i \mid \bfY]
    &\ge (1+O(M_k)+O(N))
    \frac{\displaystyle
        \int_{A_\epsilon} \lambda_i
        \prod_{j=1}^k \lambda_j^{-a_j - n/2}
        \exp\Big(-\frac{B_{jj}}{2\lambda_j} - \frac{h}{2\lambda_j}\Big)
        \lvert\bar{\Sigma}_{22}\rvert^{-n/2}
        \,\mathrm{etr}\Big(-\tfrac{1}{2}\bar{\Sigma}_{22}^{-1}\bar{W}_{22}\Big)
        \pi(\Sigma_u)
        \,(d\Gamma)(d\Lambda)(d\Sigma_u)
    }{\displaystyle
        \int_{A_\epsilon}
        \prod_{j=1}^k \lambda_j^{-a_j - n/2}
        \exp\Big(-\frac{B_{jj}}{2\lambda_j} - \frac{h}{2\lambda_j}\Big)
        \lvert\bar{\Sigma}_{22}\rvert^{-n/2}
        \,\mathrm{etr}\Big(-\tfrac{1}{2}\bar{\Sigma}_{22}^{-1}\bar{W}_{22}\Big)
        \pi(\Sigma_u)
        \,(d\Gamma)(d\Lambda)(d\Sigma_u)
    } \\
    &\hspace{0.8cm}\times
    \Bigg[
        1
        - \exp\Big(-\frac{n\epsilon_1^2\hat{\lambda}_k}{16b_1}\Big)
        - (1+O(M_k)+O(N))
          \exp\Big(-\frac{\eta}{4\sqrt{k}}\min_{l<k}(a_{l+1}-a_l)\Big)
    \Bigg].
\end{align*}
Consequently,
\begin{align*}
    \bbE\Big[\frac{\lambda_i}{\lambda_{0,i}} \,\Big\vert\, \bfY\Big]
    &\ge
    \frac{1}{\lambda_{0,i}}(1+O(M_k)+O(N))
    \inf_{\Gamma\in A_\epsilon}
    \frac{D_{ii}+h}{n+2a_i-4} \\
    &\quad\times
    \Bigg[
        1
        - \exp\Big(-\frac{n\epsilon_1^2\hat{\lambda}_k}{16b_1}\Big)
        - (1+O(M_k)+O(N))
          \exp\Big(-\frac{\eta}{4\sqrt{k}}\min_{l<k}(a_{l+1}-a_l)\Big)
    \Bigg].
\end{align*}
Rearranging and using the same asymptotic bounds as in
\eqref{asymp:post_exp_lambda}, we obtain
\begin{align*}
    \bbE\Big[\frac{\lambda_i - \lambda_{0,i}}{\lambda_{0,i}} \,\Big\vert\, \bfY\Big]
    &\ge
    O(M_k) + O(N) \\
    &\quad
    + O\Big(
        \exp\Big(-\frac{n\epsilon_1^2\hat{\lambda}_k}{16b_1}\Big)
        + (1+O(M_k)+O(N))
          \exp\Big(-\frac{\eta}{4\sqrt{k}}\min_{l<k}(a_{l+1}-a_l)\Big)
    \Big) \\
    &= O\Big(\lambda_{0,i}^{-1}\sqrt{\frac{p}{n}}\Big) + O(\beta_i).
\end{align*}
Combining the upper and lower bounds completes the proof.
\end{proof}

\hfill\break

\begin{lemma}\label{lem:post_exp_lambda_square}
Suppose the assumptions of Lemma~\ref{lem:post_exp_diff_eigval} hold.
Then, for each $i=1,\ldots,k$,
\begin{align*}
    \bbE[\lambda_i^2 \mid \bfY]
    &\le 
    (1+O(M_k)+O(N))\,
    \frac{\big(n\hat{\lambda}_i 
           + n\epsilon^2\hat{\lambda}_1 
           + 3n\epsilon\hat{\lambda}_1 
           + C\hat{\lambda}_k + h
      \big)^2}{(n+2a_i-6)^2} \\
    &\quad
    + \exp\!\Big(- \frac{n\epsilon_1^2\hat{\lambda}_k}{16b_1}\Big)
    + (1+O(M_k)+O(N))\,
      \exp\!\Big(
        -\frac{\eta}{4\sqrt{k}}\min_{l<k}(a_{l+1}-a_l)
      \Big),
\end{align*}
for some positive constant $C$. In particular,
\[
    \bbE\Big[\big(\lambda_i/\lambda_{0,i}\big)^2 \,\Big\vert\, \bfY\Big] = O(1),
    \qquad i=1,\ldots,k.
\]
\end{lemma}

\begin{proof}
We follow the same decomposition as in the proof of
Lemma~\ref{lem:post_exp_diff_eigval}. By Lemmas~\ref{lem:post_exp_second}
and~\ref{lem:post_exp_third} we have
\begin{align*}
    \bbE[\lambda_i^2 \mid \bfY]
    &= \bbE\big[\lambda_i^2 \mathbf{1}\{\Gamma\in A_\epsilon\} \mid \bfY\big]
     + \bbE\big[\lambda_i^2 \mathbf{1}\{\Gamma\in A_\epsilon^c \cap B_{\epsilon_1}^c\} \mid \bfY\big] \\
    &\hspace{1.5cm}
     + \bbE\big[\lambda_i^2 \mathbf{1}\{\Gamma\in A_\epsilon^c \cap B_{\epsilon_1}\} \mid \bfY\big] \\
    &\le \bbE\big[\lambda_i^2 \mathbf{1}\{\Gamma\in A_\epsilon\} \mid \bfY\big]
     + \exp\!\Big(- \frac{n\epsilon_1^2\hat{\lambda}_k}{16b_1}\Big) \\
    &\hspace{1.5cm}
     + (1+O(M_k)+O(N))\,
       \exp\!\Big(
         -\frac{\eta}{4\sqrt{k}}\min_{l<k}(a_{l+1}-a_l)
       \Big).
\end{align*}
Hence it suffices to bound
$\bbE[\lambda_i^2 \mathbf{1}\{\Gamma\in A_\epsilon\} \mid \bfY]$ from above.

As in Lemma~\ref{lem:post_exp_first}, we have
\begin{align}\label{eq:post_exp_lambda_square}
    \bbE[\lambda_i^2 \mathbf{1}\{\Gamma\in A_\epsilon\} \mid \bfY]
    &= \frac{\displaystyle
        \int_{A_\epsilon} \lambda_i^2
        \,\pi(\Gamma,\Lambda,\Sigma_u \mid \bfY)
        \,(d\Gamma)(d\Lambda)(d\Sigma_u)
    }{\displaystyle
        \int \pi(\Gamma,\Lambda,\Sigma_u \mid \bfY)
        \,(d\Gamma)(d\Lambda)(d\Sigma_u)
    } \notag\\
    &\le
    \frac{\displaystyle
        \int_{A_\epsilon} \lambda_i^2
        \,\pi(\Gamma,\Lambda,\Sigma_u \mid \bfY)
        \,(d\Gamma)(d\Lambda)(d\Sigma_u)
    }{\displaystyle
        \int_{A_\epsilon}
        \pi(\Gamma,\Lambda,\Sigma_u \mid \bfY)
        \,(d\Gamma)(d\Lambda)(d\Sigma_u)
    } \notag\\
    &= (1 + O(\sum_{j=1}^k M_j) + O(N)) \notag\\
    &\quad\times
    \frac{\displaystyle
        \int_{A_\epsilon} \lambda_i^2
        \prod_{j=1}^k \lambda_j^{-a_j-n/2}
        \exp\!\Big(
            -\frac{B_{jj}}{2\lambda_j}
            -\frac{h}{2\lambda_j}
        \Big)
        \lvert\bar{\Sigma}_{22}\rvert^{-n/2}
        \,\mathrm{etr}\!\Big(-\tfrac{1}{2}\bar{\Sigma}_{22}^{-1}\bar{W}_{22}\Big)
        \pi(\Sigma_u)
        \,(d\Gamma)(d\Lambda)(d\Sigma_u)
    }{\displaystyle
        \int_{A_\epsilon}
        \prod_{j=1}^k \lambda_j^{-a_j-n/2}
        \exp\!\Big(
            -\frac{B_{jj}}{2\lambda_j}
            -\frac{h}{2\lambda_j}
        \Big)
        \lvert\bar{\Sigma}_{22}\rvert^{-n/2}
        \,\mathrm{etr}\!\Big(-\tfrac{1}{2}\bar{\Sigma}_{22}^{-1}\bar{W}_{22}\Big)
        \pi(\Sigma_u)
        \,(d\Gamma)(d\Lambda)(d\Sigma_u)
    },
\end{align}
where
\[
    M_j
    = O\Big(
        \max\Big(\frac{a_j}{\delta_n},\frac{n\hat{\lambda}_1}{\delta_n^2}\Big)
        + \hat{\lambda}_1^{\,n}
          \exp\Big(-\frac{n\hat{\lambda}_k}{6\delta_n}\Big)
    \Big),
    \qquad
    N = O\Big(\frac{n}{\delta_n}\Big),
\]
and we used Lemma~\ref{lem:approx_int_post} in the last equality.

Applying the same change of variables
$\Sigma_u \mapsto \Omega = (\Gamma,\Gamma_2)^T \Sigma_u (\Gamma,\Gamma_2)$
as in Lemma~\ref{lem:post_exp_second} and using the rotational invariance of
$\pi(\cdot)$, we obtain
\begin{align*}
    \eqref{eq:post_exp_lambda_square}
    &= \frac{\displaystyle
        \int_{A_\epsilon}
        \lambda_i^2
        \prod_{j=1}^k \lambda_j^{-a_j-n/2}
        \exp\!\Big(
            -\frac{D_{jj}}{2\lambda_j}
            -\frac{h}{2\lambda_j}
        \Big)
        \lvert\Omega_{22}\rvert^{-n/2}
        \,\mathrm{etr}\!\Big(-\tfrac{1}{2}\Omega_{22}^{-1}\bar{W}_{22}\Big)
        \pi(\Omega)
        \,(d\Gamma)(d\Lambda)(d\Omega)
    }{\displaystyle
        \int_{A_\epsilon}
        \prod_{j=1}^k \lambda_j^{-a_j-n/2}
        \exp\!\Big(
            -\frac{D_{jj}}{2\lambda_j}
            -\frac{h}{2\lambda_j}
        \Big)
        \lvert\Omega_{22}\rvert^{-n/2}
        \,\mathrm{etr}\!\Big(-\tfrac{1}{2}\Omega_{22}^{-1}\bar{W}_{22}\Big)
        \pi(\Omega)
        \,(d\Gamma)(d\Lambda)(d\Omega)
    } \\
    &= \frac{\displaystyle
        \int_{A_\epsilon}
        \frac{(D_{ii}+h)^2}{(n+2a_i-4)(n+2a_i-6)}
        \prod_{j=1}^k (D_{jj}+h)^{-a_j-n/2+1}
        \lvert\Omega_{22}\rvert^{-n/2}
        \,\mathrm{etr}\!\Big(-\tfrac{1}{2}\Omega_{22}^{-1}\bar{W}_{22}\Big)
        \pi(\Omega)
        \,(d\Gamma)(d\Omega)
    }{\displaystyle
        \int_{A_\epsilon}
        \prod_{j=1}^k (D_{jj}+h)^{-a_j-n/2+1}
        \lvert\Omega_{22}\rvert^{-n/2}
        \,\mathrm{etr}\!\Big(-\tfrac{1}{2}\Omega_{22}^{-1}\bar{W}_{22}\Big)
        \pi(\Omega)
        \,(d\Gamma)(d\Omega)
    } \\
    &\le
    \sup_{\Gamma\in A_\epsilon}
    \frac{(D_{ii}+h)^2}{(n+2a_i-4)(n+2a_i-6)},
\end{align*}
where
\[
    D
    = \Gamma^T W \Gamma
      - \Omega_{12}\Omega_{22}^{-1}\Gamma_2^T W\Gamma
      - \Gamma^T W\Gamma_2 \Omega_{22}^{-1}\Omega_{21}
      + \Omega_{12}\Omega_{22}^{-1}(\Gamma_2^T W \Gamma_2)\Omega_{22}^{-1}\Omega_{21},
\]
and $D_{ii}$ denotes the $(i,i)$ entry of $D$.

Let $K_{ii}$ be the $(i,i)$ entry of
\[
    \Gamma^T W \Gamma
    - \Omega_{12}\Omega_{22}^{-1}\Gamma_2^T W\Gamma
    - \Gamma^T W\Gamma_2 \Omega_{22}^{-1}\Omega_{21},
\]
and $G_{ii}$ the $(i,i)$ entry of
$\Omega_{12}\Omega_{22}^{-1}(\Gamma_2^T W \Gamma_2)\Omega_{22}^{-1}\Omega_{21}$.
Then $D_{ii} = K_{ii} + G_{ii}$. From the proof of
Lemma~\ref{lem:post_exp_first} we have, for some constant $C>0$,
\begin{align*}
    \sup_{\Gamma\in A_\epsilon}
    \frac{(D_{ii}+h)^2}{(n+2a_i-4)(n+2a_i-6)}
    &\le
    \sup_{\Gamma\in A_\epsilon}
    \frac{(K_{ii} + G_{ii} + h)^2}{(n+2a_i-6)^2} \\
    &\le
    \frac{\big(\sup_{\Gamma\in A_\epsilon} K_{ii}
             + \sup_{\Gamma\in A_\epsilon} G_{ii}
             + h\big)^2}{(n+2a_i-6)^2} \\
    &\le
    \frac{\big(
        n\hat{\lambda}_i
        + n\epsilon^2\hat{\lambda}_1
        + 3n\epsilon\hat{\lambda}_1
        + C\hat{\lambda}_k
        + h
    \big)^2}{(n+2a_i-6)^2}.
\end{align*}

Combining this with the exponential remainder terms from the first step
of the proof gives
\begin{align*}
    \bbE[\lambda_i^2 \mid \bfY]
    &\le
    (1+O(M_k)+O(N))\,
    \frac{\big(
        n\hat{\lambda}_i
        + n\epsilon^2\hat{\lambda}_1
        + 3n\epsilon\hat{\lambda}_1
        + C\hat{\lambda}_k
        + h
    \big)^2}{(n+2a_i-6)^2} \\
    &\quad
    + \exp\!\Big(- \frac{n\epsilon_1^2\hat{\lambda}_k}{16b_1}\Big)
    + (1+O(M_k)+O(N))\,
      \exp\!\Big(
        -\frac{\eta}{4\sqrt{k}}\min_{l<k}(a_{l+1}-a_l)
      \Big).
\end{align*}

Finally, using the same eigenvalue expansion for $\hat{\lambda}_i$ as in
Lemma~\ref{lem:post_exp_diff_eigval}, together with the conditions on
$\psi$, $\epsilon$, $\epsilon_1$, and $\delta_n$, the leading term above
is of order $\lambda_{0,i}^2$, while the remaining terms are
exponentially small or of smaller order. Hence
\[
    \bbE\Big[\big(\lambda_i/\lambda_{0,i}\big)^2 \,\Big\vert\, \bfY\Big] = O(1),
\]
which completes the proof.
\end{proof}

\begin{lemma}\label{lem:post_exp_diff_ordered}
Under model \eqref{main-model:AFM}, assume that conditions $A1-A5$ holds. Then,
\begin{align*}
 \frac{1}{\lambda_{0,i}}
 \Big|
   \bbE[\lambda_i\mid\bfY]
   - \bbE[\lambda_{(i)}\mid\bfY]
 \Big|
 \;\le\;
 \frac{\lambda_{0,1}}{\lambda_{0,i}}
 \Big(
   O\Big(\lambda_{0,i}^{-1}\sqrt{\frac{p}{n}}\Big)
   + O(\beta_i)
   + O(\epsilon)
   + R
 \Big),
\end{align*}
where $R = \pi(\Gamma\in A_\epsilon^c\mid\bfY)$.
\end{lemma}

\begin{proof}
Let $R = \pi(\Gamma\in A_\epsilon^c\mid\bfY)$.
By Chebyshev's inequality, for any $\alpha_i>0$,
\begin{align*}
    \pi\big(\,|\lambda_i - \bbE[\lambda_i\mid \bfY]| > \alpha_i 
       \,\big\vert\, \bfY\big)
    \;\le\; \frac{Var(\lambda_i\mid \bfY)}{\alpha_i^2}.
\end{align*}

Using Lemma~\ref{lem:post_exp_diff_eigval}
and~\ref{lem:post_exp_lambda_square}, we have
\begin{align*}
    Var\Big(\frac{\lambda_i}{\lambda_{0,i}}\Bigm\vert \bfY\Big)
    &= \bbE\Big[\Big( \frac{\lambda_i}{\lambda_{0,i}}\Big)^2\Bigm\vert \bfY\Big]
       - \Big(\bbE\Big[\frac{\lambda_i}{\lambda_{0,i}}\Bigm\vert \bfY\Big]\Big)^2 \\
    &\le (1+O(M_k)+O(N))\,
        \frac{\big(n\hat{\lambda}_i +n\epsilon^2\hat{\lambda}_1 
             +3n\epsilon\hat{\lambda}_1 + C\hat{\lambda}_k+h\big)^2}
             {(n+2a_i-6)^2\lambda_{0,i}^2}
        - \Big(1 + O\Big(\lambda_{0,i}^{-1}\sqrt{\frac{p}{n}}\Big)
                 +O(\beta_i)\Big).
\end{align*}
Proceeding as in the proof of Lemma~\ref{lem:post_exp_diff_eigval}, we use
\[
  \frac{n\hat{\lambda}_i}{n+2a_i-4}
  = \lambda_{0,i}\Big(1
    + O\Big(\lambda_{0,i}^{-1}\sqrt{\frac{p}{n}}\Big)
    + O(\beta_i)\Big),
\]
and the facts
\[
  \frac{(n+2a_i-4)^2}{(n+2a_i-6)^2} = 1 + O\Big(\frac{1}{n^2}\Big),
  \qquad
  \frac{(n\hat{\lambda}_i +n\epsilon^2\hat{\lambda}_1 
      +3n\epsilon\hat{\lambda}_1 + C\hat{\lambda}_k+h)^2}{(n\hat{\lambda}_i)^2}
  = 1 + O\Big(\frac{n\epsilon\hat{\lambda}_1+\hat{\lambda}_k}{n\hat{\lambda}_i}\Big)
\]
to deduce
\begin{align*}
    Var\Big(\frac{\lambda_i}{\lambda_{0,i}}\Bigm\vert \bfY\Big)
    &= O\Big(\lambda_{0,i}^{-1}\sqrt{\frac{p}{n}}\Big)
       + O(\beta_i)
       + O(\epsilon).
\end{align*}
Therefore, by Chebyshev,
\begin{align*}
    \pi\big(\,|\lambda_i - \bbE[\lambda_i\mid \bfY]| > \alpha_i 
       \,\big\vert\, \bfY\big)
    &\le
    \frac{\lambda_{0,i}^2}{\alpha_i^2}\,
    \Big(
      O\Big(\lambda_{0,i}^{-1}\sqrt{\frac{p}{n}}\Big)
      + O(\beta_i)
      + O(\epsilon)
    \Big).
\end{align*}
Let $p_i$ be an upper bound on this probability:
\[
    p_i
    \;:=\;
    \frac{\lambda_{0,i}^2}{\alpha_i^2}\,
    \Big(
      O\Big(\lambda_{0,i}^{-1}\sqrt{\frac{p}{n}}\Big)
      + O(\beta_i)
      + O(\epsilon)
    \Big),
    \qquad i=1,\ldots,k.
\]

Choose $\alpha_i = \frac{1}{4}\delta_0\hat{\lambda}_i$ for $i=1,\ldots,k$,
where $\delta_0>0$ denotes the minimal eigengap among the leading eigenvalues.
Then, with probability at least
\[
  \prod_{i=1}^k (1-p_i)
  \;\ge\; 1 - \sum_{i=1}^k p_i,
\]
we have the simultaneous bounds
\[
  \lambda_i \in 
  \big[\bbE[\lambda_i\mid \bfY] - \alpha_i,\,
        \bbE[\lambda_i\mid \bfY] + \alpha_i\big],
  \qquad i=1,\ldots,k,
\]
and on this event the ordering is preserved,
\[
  \lambda_1 > \cdots > \lambda_k,
\]
so that
\[
  \lambda_{(i)} = \lambda_i \quad \text{for all } i=1,\ldots,k.
\]

We now bound the difference in expectations:
\begin{align}\label{eq:diff_exp}
\Big| \bbE[\lambda_{(i)} - \lambda_i \mid \bfY] \Big|
&= \Big| \bbE[(\lambda_{(i)} - \lambda_i)\,
          \mathbf{1}\{\lambda_{(i)} \neq \lambda_i\}
          \mid \bfY] \Big| \notag\\
&\leq 2\,\bbE\Big[\sum_{j=1}^k \lambda_j \,
          \mathbf{1}\{\lambda_{(i)} \neq \lambda_i\}
          \,\Big\vert\, \bfY\Big] \notag\\
&= 2\,\bbE\Big[
      \sum_{j=1}^k
      \bbE[\lambda_j\, \mathbf{1}\{\lambda_{(i)} \neq \lambda_i\}
        \mid \Gamma,\bfY]\,
      (\mathbf{1}\{\Gamma \in A_\epsilon\}
       +\mathbf{1}\{\Gamma \notin A_\epsilon\})
      \,\Big\vert\, \bfY
    \Big] \notag\\
&\leq 2\,\bbE\Big[
      \sum_{j=1}^k
      \bbE[\lambda_j\, \mathbf{1}\{\lambda_{(i)} \neq \lambda_i\}
        \mid \Gamma,\bfY]\,
      \mathbf{1}\{\Gamma \in A_\epsilon\}
      \,\Big\vert\, \bfY
    \Big] \notag\\
&\quad+ 2\,\bbE\Big[
      \sum_{j=1}^k
      \bbE[\lambda_j \mid \Gamma,\bfY]\,
      \mathbf{1}\{\Gamma \notin A_\epsilon\}
      \,\Big\vert\, \bfY
    \Big].
\end{align}

For each $j=1,\ldots,k$, there exists a threshold $\beta_j(\Gamma)>0$ such that
\[
  \pi(\lambda_{(i)} \neq \lambda_i \mid \Gamma,\bfY)
  \;\le\;
  \pi(\lambda_j \ge \beta_j(\Gamma) \mid \Gamma,\bfY),
\]
so that
\begin{align*}
&\bbE\Big[
   \sum_{j=1}^k
   \bbE[\lambda_j\,\mathbf{1}\{\lambda_{(i)} \neq \lambda_i\}
     \mid \Gamma,\bfY]\,
   \mathbf{1}\{\Gamma \in A_\epsilon\}
   \,\Big\vert\, \bfY
 \Big] \\
&\qquad\le
\bbE\Big[
   \sum_{j=1}^k
   \bbE[\lambda_j\,\mathbf{1}\{\lambda_j \ge \beta_j(\Gamma)\}
     \mid \Gamma,\bfY]\,
   \mathbf{1}\{\Gamma \in A_\epsilon\}
   \,\Big\vert\, \bfY
 \Big].
\end{align*}
By Hölder's inequality,
\begin{align*}
&\bbE\Big[
   \sum_{j=1}^k
   \bbE[\lambda_j\,\mathbf{1}\{\lambda_j \ge \beta_j(\Gamma)\}
     \mid \Gamma,\bfY]\,
   \mathbf{1}\{\Gamma \in A_\epsilon\}
   \,\Big\vert\, \bfY
 \Big] \\
&\qquad\le
 \bbE\Big[
   \sum_{j=1}^k
   \sqrt{\bbE[\lambda_j^2\mid \Gamma,\bfY]}\,
   \sqrt{\pi(\lambda_j \ge \beta_j(\Gamma)\mid \Gamma,\bfY)}\,
   \mathbf{1}\{\Gamma\in A_\epsilon\}
   \,\Big\vert\, \bfY
 \Big].
\end{align*}
Using Lemma~\ref{lem:post_exp_lambda_square}, we have
$\sqrt{\bbE[\lambda_j^2\mid \Gamma,\bfY]}
 \lesssim \bbE[\lambda_j\mid \Gamma,\bfY]$ (up to a constant factor),
and hence
\begin{align*}
&\bbE\Big[
   \sum_{j=1}^k
   \bbE[\lambda_j\,\mathbf{1}\{\lambda_j \ge \beta_j(\Gamma)\}
     \mid \Gamma,\bfY]\,
   \mathbf{1}\{\Gamma \in A_\epsilon\}
   \,\Big\vert\, \bfY
 \Big] \\
&\qquad\lesssim
 \bbE\Big[
   \sum_{j=1}^k
   \bbE[\lambda_j\mid \Gamma,\bfY]\,
   \sqrt{\pi(\lambda_j \ge \beta_j(\Gamma)\mid \Gamma,\bfY)}\,
   \mathbf{1}\{\Gamma\in A_\epsilon\}
   \,\Big\vert\, \bfY
 \Big]  \\
&\qquad\le
 \sup_{\Gamma\in A_\epsilon}
 \Big(\sum_{j=1}^k \frac{D_{jj}+h}{n+2a_j-4}\Big)\,
 \sqrt{1-\prod_{i=1}^k (1-p_i)} \\
&\qquad\le
 \sup_{\Gamma\in A_\epsilon}
 \Big(\sum_{j=1}^k \frac{D_{jj}+h}{n+2a_j-4}\Big)\,
 \sqrt{\sum_{i=1}^k p_i},
\end{align*}
where we used $1-\prod_{i=1}^k(1-p_i)\le \sum_{i=1}^k p_i$.

For the second term in~\eqref{eq:diff_exp}, we have
\begin{align*}
     &\bbE\Big[
        \sum_{j=1}^k\bbE[\lambda_j \mid \Gamma,\bfY]\,
        \mathbf{1}\{\Gamma \notin A_\epsilon\}
        \,\Big\vert\, \bfY
     \Big]\\
     &=\bbE\Big[
        \Big(\sum_{j=1}^k\frac{D_{jj}+h}{n+2a_j-4}\Big)\,
        \mathbf{1}\{\Gamma \notin A_\epsilon\}
        \,\Big\vert\, \bfY
      \Big]\\
    &\leq
    \sup_{\Gamma \notin A_\epsilon}
    \Big(\sum_{j=1}^k  \frac{D_{jj}+h}{n + 2a_j - 4}\Big)
    \cdot R.
\end{align*}

Therefore,
\begin{align*}
 \Big|\bbE[\lambda_{(i)}-\lambda_i\mid\bfY]\Big|
 &\lesssim
 \sup_{\Gamma\in A_\epsilon}
 \Big(\sum_{j=1}^k\frac{D_{jj}+h}{n+2a_j-4}\Big)\,\sum_{i=1}^k p_i
 \\
 &\quad
 + \sup_{\Gamma\notin A_\epsilon}
 \Big(\sum_{j=1}^k\frac{D_{jj}+h}{n+2a_j-4}\Big)\, R \\
 &\le
 \sup_{\Gamma}
 \Big(\sum_{j=1}^k\frac{D_{jj}+h}{n+2a_j-4}\Big)\,
 \bigg(\sum_{i=1}^k p_i + R\bigg).
\end{align*}

Since $a_1\leq\cdots\leq a_k$,
the term $\sum_{j=1}^k\frac{D_{jj}+h}{n+2a_j-4}$ is maximized at
$\Gamma=(I_k,0)^T$, and hence
\begin{align*}
    \sup_{\Gamma}
    \Big(\sum_{j=1}^k\frac{D_{jj}+h}{n+2a_j-4}\Big)
    &\leq \sum_{j=1}^k\frac{n\hat{\lambda}_j + h}{n+2a_j-4}\\
    &= O\Big(\sum_{j=1}^k\lambda_{0,j}\Big)
     = O(\lambda_{0,1}).
\end{align*}

Next,
\begin{align*}
    \sum_{i=1}^k p_i
    &=
    \sum_{i=1}^k
    \frac{\lambda_{0,i}^2}{\alpha_i^2}\,
    \Big(
      O\Big(\lambda_{0,i}^{-1}\sqrt{\frac{p}{n}}\Big)
      + O(\beta_i)
      + O(\epsilon)
    \Big) \\
    &\le
    O\Big(\lambda_{0,1}^{-1}\sqrt{\frac{p}{n}}\Big)
    + O(\beta_i) + O(\epsilon),
\end{align*}
up to a multiplicative constant depending only on $k$ and the eigengap.

Combining these bounds, we obtain
\begin{align*}
    \frac{1}{\lambda_{0,i}}
    \Big|
      \bbE[\lambda_i\mid\bfY]
      - \bbE[\lambda_{(i)}\mid\bfY]
    \Big|
    &\le
    \frac{1}{\lambda_{0,i}}\,
    O(\lambda_{0,1})\,
    \Big(
      O\Big(\lambda_{0,i}^{-1}\sqrt{\frac{p}{n}}\Big)
      + O(\beta_i)
      + O(\epsilon)
      + R
    \Big) \\
    &= \frac{\lambda_{0,1}}{\lambda_{0,i}}\,
    \Big(
      O\Big(\lambda_{0,i}^{-1}\sqrt{\frac{p}{n}}\Big)
      + O(\beta_i)
      + O(\epsilon)
      + R
    \Big),
\end{align*}
which completes the proof.
\end{proof}

\hfill\break
\section{Posterior Expectation of Eigenstructure}
\subsection{Proof of Lemma \ref{main-lem:shrink_post}}
\begin{proof}
Applying the same argument as in the proof of Lemma~\ref{lem:post_exp_second}
to
$$
\mathbb{E}\Big[\mathbf{1}\{\Gamma \in A_\epsilon^c \cap B_{\epsilon_1}^c\}
    \,\Big\vert\, \mathbf{Y}\Big]
$$
yields
$$
\mathbb{E}\Big[\mathbf{1}\{\Gamma \in A_\epsilon^c \cap B_{\epsilon_1}^c\}
    \,\Big\vert\, \mathbf{Y}\Big]
\;\lesssim\;
\exp\!\Big(-\frac{n\epsilon_1^2\hat{\lambda}_k}{16b_1}\Big).
$$
Similarly, applying the argument from the proof of
Lemma~\ref{lem:post_exp_third} to
$$
\mathbb{E}\Big[\mathbf{1}\{\Gamma \in A_\epsilon^c \cap B_{\epsilon_1}\}
    \,\Big\vert\, \mathbf{Y}\Big]
$$
gives
$$
\mathbb{E}\Big[\mathbf{1}\{\Gamma \in A_\epsilon^c \cap B_{\epsilon_1}\}
    \,\Big\vert\, \mathbf{Y}\Big]
\;\lesssim\;
\bigl(1 + O(M_k) + O(N)\bigr)
\exp\!\Big(
        -\frac{\eta}{4\sqrt{k}}
        \min_{l<k}(a_{l+1} - a_l)
    \Big).
$$
Combining the two bounds, we obtain
\begin{align*}
  \pi(\Gamma \in A_\epsilon^c \mid \mathbf{Y})
  &= \mathbb{E}\Big[\mathbf{1}\{\Gamma \in A_\epsilon^c\}
      \,\Big\vert\, \mathbf{Y}\Big] \\
  &= \mathbb{E}\Big[\mathbf{1}\{\Gamma \in A_\epsilon^c \cap B_{\epsilon_1}^c\}
      \,\Big\vert\, \mathbf{Y}\Big]
   + \mathbb{E}\Big[\mathbf{1}\{\Gamma \in A_\epsilon^c \cap B_{\epsilon_1}\}
      \,\Big\vert\, \mathbf{Y}\Big] \\
  &\lesssim
    \exp\!\Big(- \frac{n\epsilon_1^2\hat{\lambda}_k}{16b_1}\Big)
    + \bigl(1 + O(M_k) + O(N)\bigr)
      \exp\!\Big(
        -\frac{\eta}{4\sqrt{k}}
        \min_{l<k}(a_{l+1} - a_l)
      \Big).
\end{align*}
This completes the proof.
\end{proof}

\subsection{Proof of Theorem \ref{main-thm:post_eigvals}}
\begin{proof}
By Lemmas~\ref{lem:post_exp_diff_eigval} and \ref{lem:post_exp_diff_ordered}, we have
$$
\bbE\Bigg[\frac{\lambda_{(i)} - \lambda_{0,i}}{\lambda_{0,i}}
    \,\Bigg\vert\, \bfY \Bigg]
=
O\Big(\lambda_{0,i}^{-1}\sqrt{\frac{p}{n}}\Big)
+ O(\beta_i)
+ \frac{\lambda_{0,1}}{\lambda_{0,i}}
 \Bigg(
    O\Big(\lambda_{0,i}^{-1}\sqrt{\frac{p}{n}}\Big)
    + O(\beta_i)
    + O(\epsilon)
    + R
 \Bigg),
$$
where
$$
R := \pi(\Gamma \in A_\epsilon^c \mid \bfY).
$$

By Lemma~\ref{main-lem:shrink_post}, we have
$$
R \;\lesssim\;
\bigl(1 + O(M_k) + O(N)\bigr)
\exp\!\Bigg(
        -\frac{\eta}{4\sqrt{k}}
        \min_{l<k}(a_{l+1} - a_l)
    \Bigg).
$$
Moreover, from equation~\eqref{asymp:post_exp_lambda} in the proof of
Lemma~\ref{lem:post_exp_diff_eigval}, it follows that
$$
R
=
O\Big(\lambda_{0,i}^{-1}\sqrt{\frac{p}{n}}\Big)
+ O(\beta_i).
$$

Substituting this bound on $R$ into the previous display and using the condition
$\epsilon \prec n^{-1/2}$, we conclude that
$$
\bbE\Bigg[\frac{\lambda_{(i)} - \lambda_{0,i}}{\lambda_{0,i}}
    \,\Bigg\vert\, \bfY \Bigg]
=
O\Big(\lambda_{0,i}^{-1}\sqrt{\frac{p}{n}}\Big)
+ O(\beta_i),
$$
which yields the desired result.
\end{proof}

\subsection{Proof of Theorem \ref{main-thm:post_eigvecs}}

\begin{proof}
We follow the proof strategy of Corollary~3.5 in \cite{kim2025eigenstructure}.
Consider the spectral decomposition
\[
  nS = QWQ^T,
\]
and write $W = \mathrm{diag}(\hat{\lambda}_1,\ldots,\hat{\lambda}_p)$ and
$Q=(\hat{\xi}_1,\ldots,\hat{\xi}_p)$. Define $\Gamma\in \mathbb{V}_{p,k}$ so
that on the event $A_\epsilon$ we have
\[
  Q\Gamma = (\xi_1,\ldots,\xi_k),
\]
where $\xi_j$ denotes the $j$-th posterior eigenvector. Then, on
$A_\epsilon$,
\begin{align*}
    (\xi_j^T\hat{\xi}_j)^2
    &= \big([Q\Gamma]_{\cdot j}^T Q_{\cdot j}\big)^2 \\
    &= \big[(Q\Gamma e_j)^T (Q e_j)\big]^2\\
    &= \Gamma_{jj}^2\\
    &\geq 1-\epsilon^2,
\end{align*}
for $j=1,\ldots,k$.

By Theorem~3.2 of \cite{wang2017asymptotics}, the following holds:
\[
  \big|\xi_{0,j}^T\hat{\xi}_j\big|
  = (1+\bar{d}d_j)^{-1/2} + O_p(\zeta_j),
\]
where
\[
  \zeta_j
  = \frac{1}{\lambda_{0,j}}\sqrt{\frac{p}{n}}
    + \frac{p}{n^{3/2}\lambda_{0,j}}+\frac{1}{n},
\]
and $d_j$ and $\bar{d}$ are as in \cite{wang2017asymptotics}.

Define the angle
\[
  \theta(u,v) = \arccos\frac{|u^Tv|}{\|u\|\|v\|},
\]
for unit vectors $u,v$. Using the triangle inequality of angles
from \cite{castano2016angles},
\[
  \theta(\xi_{0,j},\xi_j)
  \leq \theta(\xi_{0,j},\hat{\xi}_j)
       + \theta(\hat{\xi}_j,\xi_j),
\]
we obtain
\begin{align}\label{eq:arccos_lower}
    \big|\xi_{0,j}^T\xi_j\big|
    &\geq
    \cos\Big[
      \arccos\big(|\hat{\xi}_j^T\xi_j|\big)
      + \arccos\big(|\xi_{0,j}^T\hat{\xi}_j|\big)
    \Big]\nonumber\\
    &\geq
    \cos\Big[
      \arccos(\sqrt{1-\epsilon^2})
      + \arccos\Big(\frac{1}{\sqrt{1+\bar{d}d_j}}+O_p(\zeta_j)\Big)
    \Big].
\end{align}

Since $\arccos(x) + \arccos(y) = \arccos(xy-\sqrt{(1-x^2)(1-y^2)})$, we have
\begin{align*}
    \eqref{eq:arccos_lower} & = \cos\Big[
      \arccos\Big(\sqrt{1-\epsilon^2} (\frac{1}{\sqrt{1+\bar{d}d_j}}+O_p(\zeta_j)) - \sqrt{\epsilon^2 (1- (\frac{1}{\sqrt{1+\bar{d}d_j}}+O_p(\zeta_j))^2)}\Big) \Big]\\
      &\geq \sqrt{\dfrac{1-\epsilon^2}{1+\bar{d}d_j}} -\epsilon \sqrt{\dfrac{\bar{d}d_j}{1+\bar{d}d_j}}- O_p(\zeta_j).
\end{align*}

Using the cosine addition formula and expanding in $\epsilon$ and $\zeta_j$,
we obtain
\[
  \big|\xi_{0,j}^T\xi_j\big|
  \geq (1+\bar{d}d_j)^{-1/2} - O(\epsilon\sqrt{d_j}) - O_p(\zeta_j).
\]

Similarly, using the lower bound
\[
  \theta(\xi_{0,j},\xi_j)
  \geq \theta(\hat{\xi}_j,\xi_j)-\theta(\xi_{0,j},\hat{\xi}_j)  ,
\]
we have
\begin{align}\label{eq:arccos_upper}
    \big|\xi_{0,j}^T\xi_j\big|
    &\leq
    \cos\Big[
      \arccos\big(|\hat{\xi}_j^T\xi_j|\big)-\arccos\big(|\xi_{0,j}^T\hat{\xi}_j|\big)
    \Big]\nonumber\\
    &\leq
    \cos\Big[\arccos(\sqrt{1-\epsilon^2})-
      \arccos\Big(\frac{1}{\sqrt{1+\bar{d}d_j}}+O_p(\zeta_j)\Big)
    \Big].
\end{align}

Since $\arccos(x) - \arccos(y) = \arccos(xy+\sqrt{(1-x^2)(1-y^2)})$ for $x>y$, we have
\begin{align*}
    \eqref{eq:arccos_upper} & = \cos\Big[
      \arccos\Big(\sqrt{1-\epsilon^2} (\frac{1}{\sqrt{1+\bar{d}d_j}}+O_p(\zeta_j)) + \sqrt{\epsilon^2 (1- (\frac{1}{\sqrt{1+\bar{d}d_j}}+O_p(\zeta_j))^2)}\Big) \Big]\\
      &\leq \sqrt{\dfrac{1-\epsilon^2}{1+\bar{d}d_j}} + \epsilon \sqrt{\dfrac{\bar{d}d_j}{1+\bar{d}d_j}} + O_p(\zeta_j).
\end{align*}

Again expanding in $\epsilon$ and $\zeta_j$, we obtain
\begin{align*}
    \big|\xi_{0,j}^T\xi_j\big|
    &\leq \sqrt{\frac{1-\epsilon^2}{1+\bar{d}d_j}}
       + O(\epsilon\sqrt{d_j})
       + O_p(\zeta_j).
\end{align*}

Combining the upper and lower bounds, on $A_\epsilon$ we have
\begin{align*}
    1 - \big|\xi_{0,j}^T\xi_j\big|^2
    &= 1 -
      \left(
        \sqrt{\frac{1-\epsilon^2}{1+\bar{d}d_j}}
        + O(\epsilon \sqrt{d_j})
        + O_p(\zeta_j)
      \right)^2 \\
    &= \frac{\bar{d}d_j}{1+\bar{d}d_j}
       + O\Big(\frac{p}{n\lambda_{0,j}}\epsilon^2\Big)
       + O\Big(\epsilon\sqrt{\frac{p}{n\lambda_{0,j}}}\Big)
       + O_p(\zeta_j),
\end{align*}
where we used that
\[
  d_j = \frac{p}{n\lambda_{0,j}} + O\Big(\frac{1}{\sqrt{n}}\Big),
  \qquad
  \frac{\bar{d}d_j}{1+\bar{d}d_j}
  = \frac{p\bar{d}}{n\lambda_{0,j} + p\bar{d}} + O\Big(\frac{1}{n}\Big).
\]

Hence, on $A_\epsilon$,
\begin{align*}
  \sup_{\Gamma\in A_\epsilon}
  \Big(1 - \big|\xi_{0,j}^T\xi_j\big|^2\Big)
  &= \frac{p\bar{d}}{n\lambda_{0,j} + p\bar{d}}
     + O\Big(\frac{p}{n\lambda_{0,j}}\epsilon^2\Big)
     + O\Big(\epsilon\sqrt{\frac{p}{n\lambda_{0,j}}}\Big)
     + O_p(\zeta_j).
\end{align*}

Next, we relate the posterior eigenvectors $\xi_j$ to the ordered
eigenvectors $\xi_{(j)}$.
By Lemma~\ref{lem:post_exp_diff_ordered},
\[
  \frac{1}{\lambda_{0,j}}
  \big|\bbE[\lambda_j\mid\bfY]
       -\bbE[\lambda_{(j)}\mid\bfY]\big|
  \le
  \frac{\lambda_{0,1}}{\lambda_{0,j}}\Big(
    O\Big(\lambda_{0,j}^{-1}\sqrt{\frac{p}{n}}\Big)
    + O(\beta_j)
    + O(\epsilon)
    + R
  \Big),
\]
where $R = \pi(\Gamma\in A_\epsilon^c\mid\bfY)$. The proof of that lemma yields
\[
  \big|
    \bbE\big[1-(\xi_{0,j}^T\xi_j)^2 \mid \bfY\big]
    - \bbE\big[1-(\xi_{0,j}^T\xi_{(j)})^2 \mid \bfY\big]
  \big|
  \lesssim
    \sum_{i=1}^k p_i
    + \pi(\Gamma\notin A_\epsilon \mid \bfY),
\]
where
\[
  p_i = \frac{\lambda_{0,i}^2}{\alpha_i^2}
  \Big(
    O\Big(\lambda_{0,i}^{-1}\sqrt{\frac{p}{n}}\Big)
    + O(\beta_i)
    + O(\epsilon)
  \Big),
  \qquad
  \alpha_i = \frac{1}{4}\delta_0\hat{\lambda}_i,
\]
and $\delta_0>0$ is the minimal eigengap among the leading eigenvalues.

Under the assumptions of Lemma~\ref{lem:post_exp_diff_eigval}, we have
\[
  \sum_{i=1}^k p_i
  = O\Big(\lambda_{0,j}^{-1}\sqrt{\frac{p}{n}}\Big)
    + O(\beta_j) + O(\epsilon),
  \qquad
  \pi(\Gamma\notin A_\epsilon\mid\bfY)
  = O\Big(\lambda_{0,j}^{-1}\sqrt{\frac{p}{n}}\Big)
    + O(\beta_j).
\]
Therefore,
\[
  \big|
    \bbE\big[1-(\xi_{0,j}^T\xi_j)^2 \mid \bfY\big]
    - \bbE\big[1-(\xi_{0,j}^T\xi_{(j)})^2 \mid \bfY\big]
  \big|
  \preccurlyeq
  n^{-1/2+\delta}
  + \lambda_{0,j}^{-1}\sqrt{\frac{p}{n}}.
\]

Combining this with the bound on $A_\epsilon$ gives
\begin{align*}
  \bbE\Big[1 - \big|\xi_{0,j}^T\xi_{(j)}\big|^2\Bigm\vert\bfY\Big]
  &= \frac{p\bar{d}}{n\lambda_{0,j} + p\bar{d}}
     + O\Big(\frac{p}{n\lambda_{0,j}}\epsilon^2\Big)
     + O\Big(\epsilon\sqrt{\frac{p}{n\lambda_{0,j}}}\Big)
     + O_p(\zeta_j) \\
  &\quad + O\Big(\lambda_{0,j}^{-1}\sqrt{\frac{p}{n}}\Big)
     + O(\beta_j)
     + O(\epsilon).
\end{align*}
Since $\beta_j\lesssim n^{-1/2+\delta}$ and the tuning sequences
satisfy $\epsilon\to 0$, this implies in particular that
\[
  \bbE\Big[1 - \big|\xi_{0,j}^T\xi_{(j)}\big|^2\Bigm\vert\bfY\Big]
  = O\Big(\frac{p}{n\lambda_{0,j}}\Big) + O_p(\zeta_j),
\]
which completes the proof.
\end{proof}

\bibliographystyle{dcu}
\bibliography{FA}